%% file: arxiv.tex
\documentclass{jpp}

\usepackage{amsmath}
\usepackage{natbib}
\usepackage{graphicx}
\usepackage{pdflscape}
\usepackage{tablefootnote}
\usepackage{hyperref}

\newcommand\physrep{{Phys.~Rep.}}
\newcommand\apj{{ApJ}}
\newcommand\apjl{{ApJ}}
\newcommand\apjs{{ApJS}}
\newcommand\aap{{A\&A}}
\newcommand\aapr{{A\&A~Rev.}}

\newcommand\mnras{{MNRAS}}
\newcommand\araa{{ARA\&A}}
\newcommand\procspie{{Proc.~SPIE}}

\newcommand{\simgt}{\,\hbox{\lower0.6ex\hbox{$\sim$}\llap{\raise0.6ex\hbox{$>$}}}\,}
\newcommand{\simlt}{\,\hbox{\lower0.6ex\hbox{$\sim$}\llap{\raise0.6ex\hbox{$<$}}}\,}
\newcommand{\code}[1]{\texttt{#1}}

\title[Cold Fronts]{Cold Fronts: Probes of Plasma Astrophysics in Galaxy Clusters}

\author[J.~A. ZuHone and E. Roediger]
{J.\ns A.\ns Z\ls U\ls H\ls O\ls N\ls E$^{1,2}$\thanks{Email address for correspondence: john.zuhone@cfa.harvard.edu} \and
E.\ns R\ls O\ls E\ls D\ls I\ls G\ls E\ls R$^3$\thanks{Email address for correspondence: e.roediger@hull.ac.uk}}

\affiliation{$^1$Kavli Institute for Astrophysics and Space Research,
Massachusetts Institute of Technology, 77 Massachusetts Avenue, Cambridge, MA 02139, USA\\
$^2$Smithsonian Astrophysical Observatory, 60 Garden St., Cambridge, MA 02138, USA\\
$^3$E.A. Milne Centre for Astrophysics, Department of Physics and Mathematics, University of Hull, Hull, HU6 7RX, United Kingdom}

\begin{document}

\maketitle

\begin{abstract}
The most massive baryonic component of galaxy clusters is the ``intracluster medium'' (ICM), a diffuse, hot, weakly magnetized plasma that is most easily observed in the X-ray band. Despite being observed for decades, the macroscopic transport properties of the ICM are still not well-constrained. A path to determine macroscopic ICM properties opened up with the discovery of ``cold fronts''. These were observed as sharp discontinuities in surface brightness and temperature in the ICM, with the property that the brighter (and denser) side of the discontinuity is the colder one. The high spatial resolution of the {\it Chandra X-ray Observatory} revealed two puzzles about the cold fronts. First, they should be subject to Kelvin-Helmholtz instabilites, yet in many cases they appear relatively smooth and undisturbed. Second, the width of the interface between the two gas phases is typically narrower than the mean free path of the particles in the plasma, indicating negligible thermal conduction. From the time of their discovery, it was realized that these special characteristics of cold fronts may be used to probe the physical properties of the cluster plasma. In this review, we will discuss the recent simulations of cold front formation and evolution in galaxy clusters, with a focus on those which have attempted to use these features to constrain the physics of the ICM. In particular, we will focus on the effects of magnetic fields, viscosity, and thermal conductivity on the stability properties and long-term evolution of cold fronts. We conclude with a discussion on what important questions remain unanswered, and the future role of simulations and the next generation of X-ray observatories.
\end{abstract}

\tableofcontents

\section{Introduction}\label{sec:intro}

Galaxy clusters are the largest gravitationally bound objects in the current universe. They reach total masses of $\sim{10^{13}-10^{15}}~M_\odot$, only about $\sim$3-5\% of which is in the member galaxies. The other two major components are the ``intracluster medium'' (ICM), a hot, ionized plasma that comprises $\sim$15-17\% of a cluster's mass, and the majority of the mass ($\sim$80\%) exists in the form of ``dark matter'' (DM), the precise nature of which is at present unknown, but is likely comprised of weakly-interacting non-baryonic particles.

The ICM is one of the largest-scale examples of a space plasma, and was discovered in early X-ray observations of clusters \citep[e.g.,][]{cav71,mee71,kel72,for72,ser77}. It consists of fully ionized hydrogen and helium with traces of highly ionized heavy elements at roughly solar to one-third solar abundances. With electron densities of $n_e \sim 10^{-4}-10^{-1}$~cm$^{-3}$ and temperatures of $T \sim 10^{6}-10^{8}$~K, the ICM radiates in X-rays predominantly through bremsstrahlung, 2-photon emission, photo and dielectronic recombinations, and line emission due to collisional excitation processes. Except for the brightest emission lines in the cores of clusters, the gas is optically thin to such X-rays. Inverse Compton scattering of the cosmic microwave background (CMB) off of the thermal electrons in the ICM leads to observable distortions in the CMB spectrum known as the Sunyaev-Zeldovich effect \citep{sun72}. Observations of radio sources in clusters and Faraday rotation measurements reveal that the ICM is weakly magnetized \citep[see][for reviews]{car02,fer12}, with magnetic field strengths of $B \sim 0.1-10$~$\mu$G, corresponding to a magnetic pressure which is roughly a few percent of the thermal pressure.

Table \ref{tab:length_scales} shows the relevant length scales of the ICM for typical values of the density, temperature, and magnetic field strength. First, we see that the ICM is weakly collisional, since the typical dynamical scales of interest range from $\sim$100-1000~kpc and the electron and ion mean free paths are on the order of less than a tenth of a kpc (as in the case of the core of the Virgo cluster) up to tens of kpc (as in cluster outskirts). Under these assumptions, the ICM is typically modeled as a collisional fluid using the equations of hydrodynamics (HD) or magnetohydrodynamics (MHD), though structure on scales below several kpc are observable and thus of interest, where, formally, the assumption of collisionality is violated. Second, the relevant scales for plasma physics (e.g., the inertial lengths and Larmor radii of the electrons and ions) are many orders of magnitude smaller than these scales, and are not directly observable. It should also be noted that though in one sense the effect of the magnetic field is ``strong'' due to the fact that the electron and ion Larmor radii are orders of magnitude smaller than their mean free paths, in another sense the magnetization is ``weak'' since over most of the cluster volume the magnetic pressure is much weaker than the thermal pressure. As we discuss below, accounting for these effects in physical models of the ICM has important and potentially observable consequences.

\input tab1.tex

The launch of the {\it Chandra X-ray Observatory} in 1999 provided a unique window into the physics of the ICM, due to its $\sim$0.5-arcsecond spatial resolution. However, even for deep observations of the diffuse ICM emission, realistically the resolution is reduced to a few arcseconds due to the need for sufficient photon statistics. Nevertheless, for nearby clusters at distances of 16 to 300~Mpc from us, an angular resolution of 3 arcsec still corresponds to a projected spatial resolution on the sky of about 0.25~kpc to 4~kpc, a scale comparable to the particle mean free path. Among the first cluster discoveries by {\it Chandra} were sharp edges in surface brightness, first seen in the clusters A2142 \citep{mar00} and A3667 \citep[][]{vik01b}. In the latter case, the edge had been previously seen by {\it ROSAT}, but not with the same detail \citep[][]{mar99}. Such edges may be expected in colliding clusters, which should drive shocks into the ICM. However, spectroscopic temperature measurements revealed these edges were not shocks, since the denser side of the discontinuity is the cooler one. Because of this difference, these edges were dubbed ``cold fronts''.

An extended review of the properties and theory of cold fronts in clusters was first given by \citet[][hereafter MV07]{MV07}. Since that review, many more cold fronts have been discovered in galaxy clusters and similar objects, e.g., galaxy groups and elliptical galaxies. All of these systems are similar in the sense that in the context of this review all of them can be thought of as DM potentials (of masses $\sim{10^{13}-10^{15}}~M_\odot$) filled with a hot plasma atmospheres (of temperatures $T \sim 10^{6}-10^{8}$~K). Despite this large range in mass, for the remainder of the review we will refer to cold fronts in ``clusters'' as a shorthand.

\subsection{Classes of Cold Fronts}\label{sec:cold_front_classes}

\begin{figure}
\begin{center}
\includegraphics[width=0.98\textwidth]{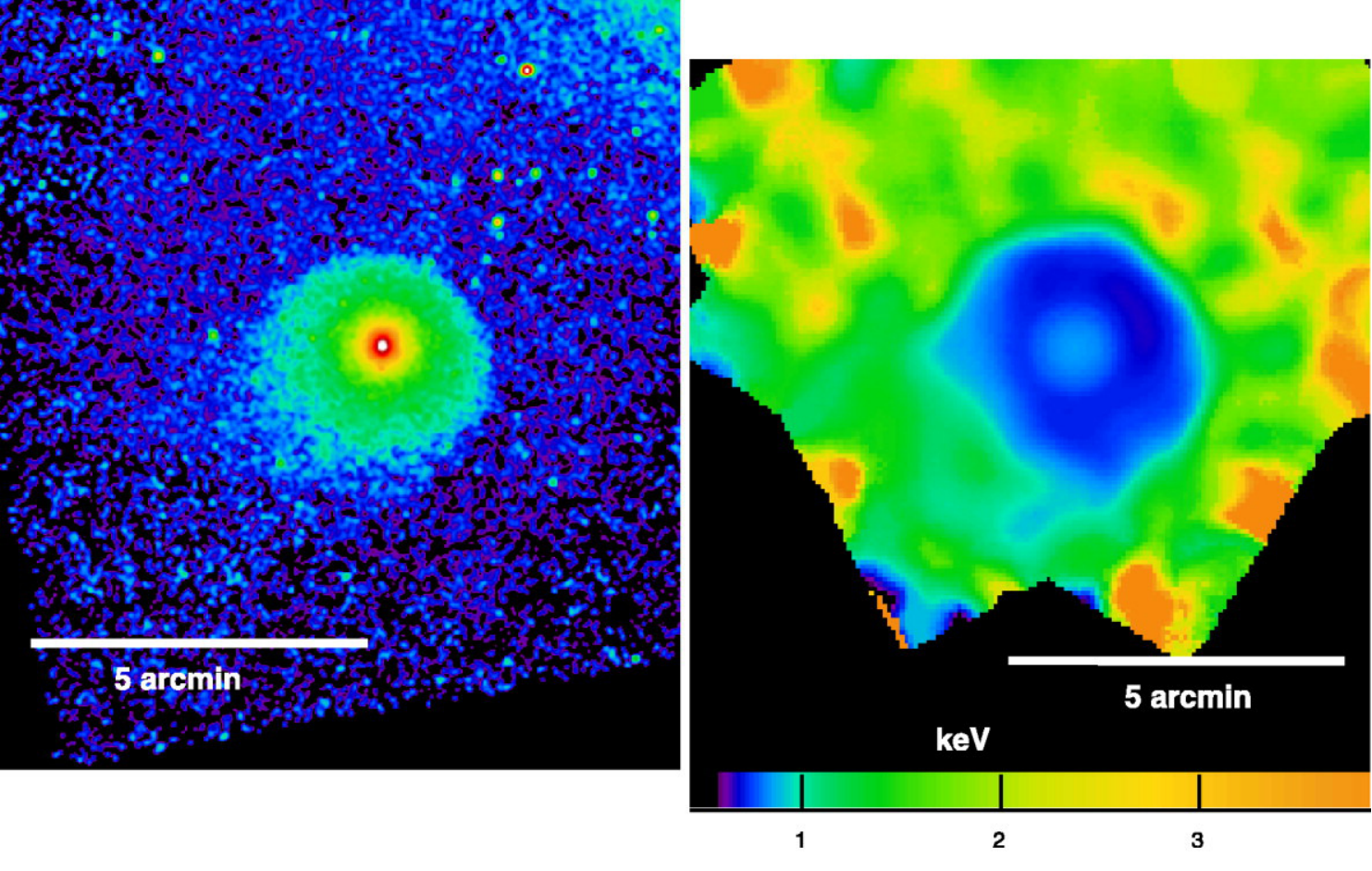}
\caption{X-ray surface brightness (left) and spectroscopic temperature (right) maps for the elliptical galaxy NGC~1404, which is falling into the Fornax cluster. The galaxy moves towards the NW to the Fornax center. NGC 1404 is a classic example of a remnant core cold front, which stretches around the galaxy's atmosphere from NE through NW to SW. Reproduced from \citet{mac05}.\label{fig:ngc1404}}
\end{center}
\end{figure}

Cosmological structure formation and hence the growth of galaxy clusters is still ongoing; thus, clusters are dynamic objects. The motion of gas within or between clusters and the presence of entropy gradients, both of which are conditions that are prevalent in clusters, provide the conditions for the formation of cold fronts. There are two classes of cold fronts, ``remnant core'' cold fronts and ``sloshing'' cold fronts, which were first distinguished by \citet{tit05}. In the most general terms, remnant core cold fronts form when a galaxy, group, or cluster moves through a hotter ambient plasma and the resulting head wind strips off the outer layer of its atmosphere. This situation arises when a galaxy or group falls into a larger cluster, and even when two clusters are passing through each other. This process has also been described as ``ram-pressure stripping'' in the context of galaxies moving through the ICM of their host clusters (\citealt{gun72}, see also \citealt{rod15a} and references therein). The formation of the cold front is also aided by the fact that the area-proportional drag force on the gas will decelerate the less dense gas of the core more, bringing the densest gas forward (see Figure 23 of MV07 for an illustration). These processes lead to an upstream contact discontinuity between the cooler, denser atmosphere of the infalling object and the hotter, less dense ambient gas. This leading edge is the remnant core cold front (Figure \ref{fig:ngc1404}). It stretches from the stagnation point to the sides of the infalling atmosphere. Prominent examples are 1E0657-56 \citep[][the ``Bullet Cluster'']{mar02}, the infalling elliptical galaxy NGC~1404 in the Fornax cluster \citep[][shown in Figure \ref{fig:ngc1404}]{dos02,scha05,mac05}, and A3667 \citep{vik01a,vik01b,dat14}. In the literature, this type of cold front has also been referred to as a ``merger'' cold front.

\begin{figure}
\begin{center}
\includegraphics[width=0.48\textwidth]{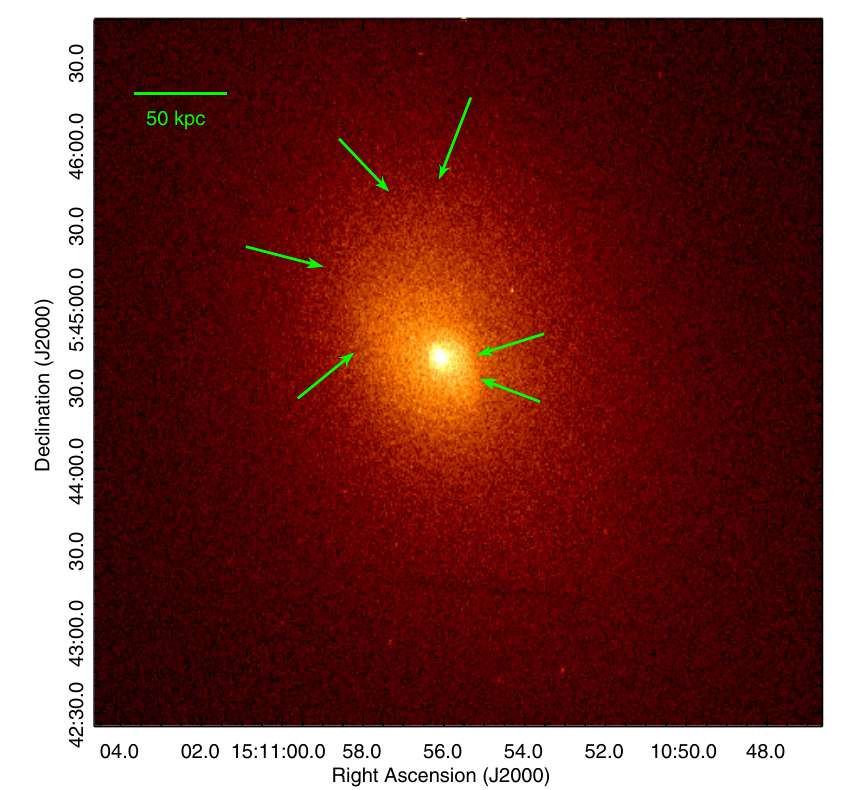}
\includegraphics[width=0.48\textwidth]{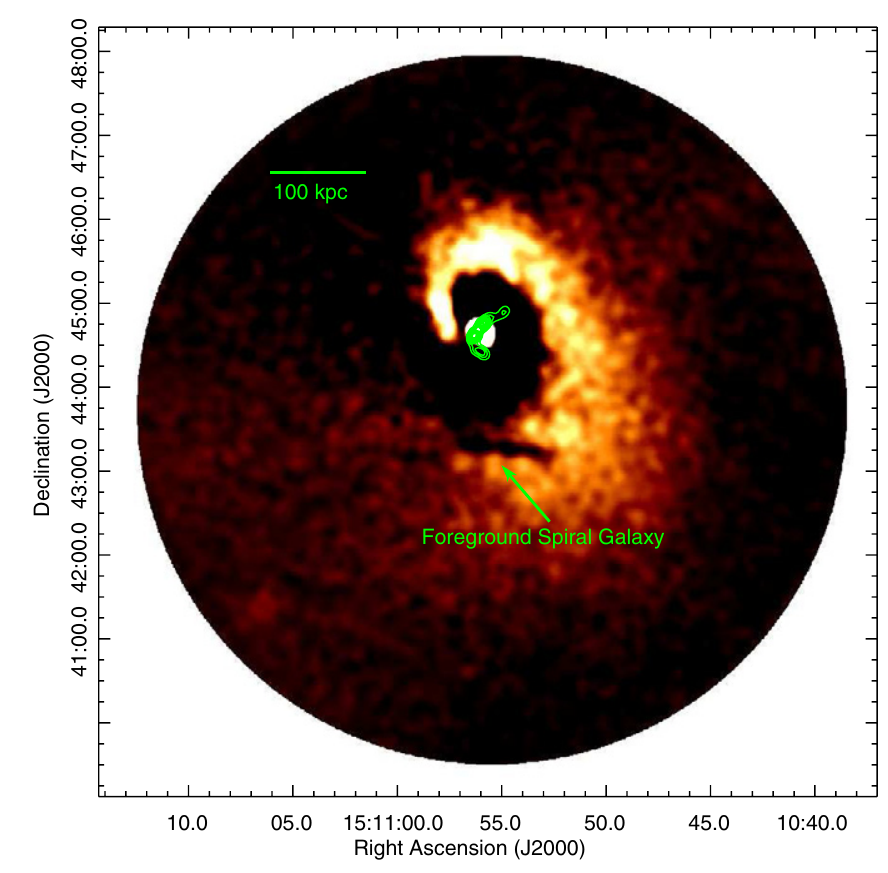}
\caption{Sloshing in A2029. Left: X-ray surface brightness image of the central part of the cluster, showing sloshing cold fronts arranged in a spiral pattern. Right: Surface brightness residual image, highlighting the associated spiral-shaped surface brightness enhancement, and demonstrating the extent of the sloshing motions out to 400~kpc. Reproduced from \citet{pat13}.\label{fig:a2029}}
\end{center}
\end{figure}

The second class of cold front, sloshing cold fronts, arises when some process offsets the bulk of the central ICM in a cluster from its hydrostatic equilibrium in the cluster potential. The ICM then slowly oscillates--or sloshes--around its hydrostatic equilibrium configuration \citep[][AM06]{mar00,tit05}. The variation of sloshing frequency with cluster radius leads to the formation of arc-shaped contact discontinuities staggered on opposite sides of the cluster core (Nulsen et al., in prep.). They are particularly visible if the cluster is of the ``cool-core'' variety, where the cluster has been relatively undisturbed for some time and the temperature in the center has slowly decreased via radiative cooling, causing the central density to go up to maintain hydrostatic equilibrium. In such clusters, the entropy profile is a steep function of radius, increasing the likelihood of producing contact discontinuities. Again, since the denser side of the contact discontinuity is the cooler one, the feature is a cold front. The required global perturbation of the ICM is achieved easily in a minor merger, i.e., the passage of a subcluster through the main cluster causes the ICM offset and subsequent sloshing (AM06), though \citep{chu03} demonstrated that similar effects could be caused by the passage of a sufficiently strong shock. An off-center passage of a subcluster introduces some additional angular momentum into the ICM, resulting in sloshing cold fronts wrapped around the cluster core in a one-armed spiral-like pattern (Figure \ref{fig:a2029}). The exact combination of linear sloshing, angular momentum, and viewing angle determines whether sloshing cold fronts are arranged in a clear spiral or rather in staggered arcs on opposite sides of the cluster core. Sloshing lasts for several Gyrs, such that the sloshing cluster can appear quite relaxed except for the sloshing signatures. Prominent examples of relaxed clusters with sloshing cold fronts are A2029 (\citealt{cla04,pat13}, see Figure \ref{fig:a2029}), Virgo \citep{sim10,wer16}, A2319 \citep{gov04,oha04}, A2142 \citep{mar00,ros13}, and A1795 \citep{mar01,ehl15}. By some estimates, sloshing cold fronts may exist in the cores of $\sim$1/2~-~2/3 of relaxed galaxy clusters \citep{mvf03,ghi10}.

\subsection{Cold Fronts and ICM Plasma Physics}\label{sec:cfs_and_plasma_physics}

From their initial discovery, it was noted that cold fronts have special properties. First, all of the formation scenarios for cold fronts predict that there should be a velocity shear across the sharp interface. Simple perturbation analysis assuming pure inviscid hydrodynamics shows that such shear flow interfaces should develop Kelvin-Helmholtz instabilities (KHI) on a relatively short timescale (\citealt{chandrasekhar}), which is a few Myr for cold fronts in the ICM. Thus, a major question in relating cold fronts to ICM properties is whether or not cold fronts are stable or unstable to KHIs.

KHIs at a given shear flow interface can be suppressed in a variety of ways, by, e.g., a sufficiently strong magnetic field aligned with the shear flow interface (\citealt{chandrasekhar,vik02}), or by a sufficient ICM viscosity (\citealt{rod13b}). In the case of magnetic fields, KHIs are suppressed if the magnetic energy in the component of the field oriented parallel to the interface is roughly in excess of the kinetic energy in the shear flow \citep[][]{lan60,vik02}. Even if the field is too weak to completely suppress KHI, it may decrease its growth rate. Viscosity suppresses the growth of KHI below a certain length scale by damping out perturbations and smoothing out the velocity gradient at the interface layer by momentum diffusion \citep{kai05,jun10,rod13b}, depending on the Reynolds number of the plasma. It should also be noted in this context that a small initial width of the cold front interface may be sufficient to suppress KHIs for a restricted range of angles along the surface \citet{chu04}, though this is not expected to be true for all observed cold fronts.

At remnant core cold fronts the shear flow has a specific structure along the cold front. The shear velocity is zero at the stagnation point, i.e., the most upstream point of the remnant core, and increases along the upstream edge towards the sides of the remnant core. Shear flows also arise along large stretches of sloshing cold fronts. For this reason, we should expect to see evidence of KHIs at the sides of remnant cores and along many sloshing cold fronts, or a washing out of the interface by the KHI. Nevertheless, many (though not all) cold fronts appear relatively undisturbed. This may indicate that the KHI are being suppressed, or that they are not as easy to recognize.

The second interesting property of cold fronts is the fact of the sharp interface itself. In many cases, the width of the interface is unresolved even by {\it Chandra}, and is often smaller than the electron and ion mean free paths. Diffusion of particles and heat should be very efficient across such an interface. In fact, the temperature and density of the gas on either side of the front imply that the thermal conduction timescale across the interface should be very fast, in which case the temperature and density gradients would be smeared out. Nevertheless, sharp cold front interfaces are present in many clusters, indicating that both heat flux and particle diffusion across the interface are both strongly suppressed.

These special properties of cold fronts provide an opportunity to use the combination of numerical simulations with X-ray observations to investigate the physical properties of the cluster plasma. This involves going beyond a simple HD description of the plasma to explicitly including the effects of magnetic fields and transport processes such as viscosity and thermal conduction. Simulations can provide predictions for observable signatures of specific ICM properties, e.g., the appearance or absence of KHIs in X-ray observations, or specific shapes for profiles of density and temperature across the fronts. By modeling ICM magnetic fields and transport processes explicitly, simulations may determine what plasma physics of the ICM is responsible for maintaining the smooth and sharp appearance of the cold fronts after their formation, and provide a more accurate description of the physics of the ICM. As more simulation works are incorporating these processes to understand their effect on ICM turbulence, their association with AGN feedback, and their effects on the thermal state of the cluster as a whole, it is vital to use detailed comparisions between simulations and observations to understand the nature of these processes in the ICM.

In this vein, a number of simulation works have been carried out over the past several years which incorporate various physics beyond pure HD in an attempt to explain the apparent long-term stability properties of cold fronts. These works have provided a number of key insights into ICM physics, but some questions still remain. The purpose of this paper is to review the results of these investigations. These simulations have approached the study of the ICM from the framework of a magnetized fluid, and so the focus of this review is on the constraints that can be placed on parameterized transport coefficients in this framework rather than detailed treatment of the underlying kinetics (see, however, Section \ref{sec:plasma_instabilities} for some comments on the latter).

In Section \ref{sec:observations}, we review a number of key observations of nearby cold fronts. In Section \ref{sec:simulations}, we review the simulations of cold fronts in clusters which have incorporated physics beyond HD. In Section \ref{sec:disc} we describe the remaining open questions, and the role of future observations and simulations. Lastly, in Section \ref{sec:summary} we provide a summary.

\section{Observations of Cold Fronts}\label{sec:observations}

Since the launch of {\it Chandra}, many cold fronts have been discovered in galaxies, galaxy groups, and galaxy clusters. In this section we provide an overview of recent X-ray observations of nearby cold fronts that are most relevant to the question of constraining the properties of the cluster plasma. We will first discuss examples of remnant core cold fronts, followed by sloshing cold fronts.

\subsection{NGC~1404/Fornax}\label{sec:ngc1404}

An excellent example of a remmant-core cold front is found in the elliptical galaxy NGC~1404 (Figure \ref{fig:ngc1404}), which is falling toward the galaxy NGC~1399 at the center of the Fornax cluster. Its gas tail, seen already in {\it ROSAT} data (\citealt{jon97,pao02}) shows that the galaxy is undergoing ram-pressure stripping. Early {\it Chandra} observations revealed a sharp leading edge \citep{dos02,scha05,mac05}, and recent deep {\it Chandra} observations will allow the width of the remnant core cold front to be measured in several wedges from the stagnation point towards the sides (Su et al., in prep.).

Using stripped galaxies or groups as probes for ICM plasma properties can be complicated by a variety of additional processes superimposed on the infall and stripping, such as central AGN activity as in e.g., M89 (\citealt{mac06}), a strongly elliptical potential and interactions with other galaxies as in M86 (\citealt{ran08}), or an early stage of infall as for the stripped groups observed in A2142 (\citealt{eck14}) and A85 (\citealt{ich15}). None of these complications apply to NGC~1404, making it a particularly clean probe of ICM plasma properties.

\subsection{A3667}\label{sec:a3667}

\begin{figure}
\begin{center}
\includegraphics[width=0.42\textwidth]{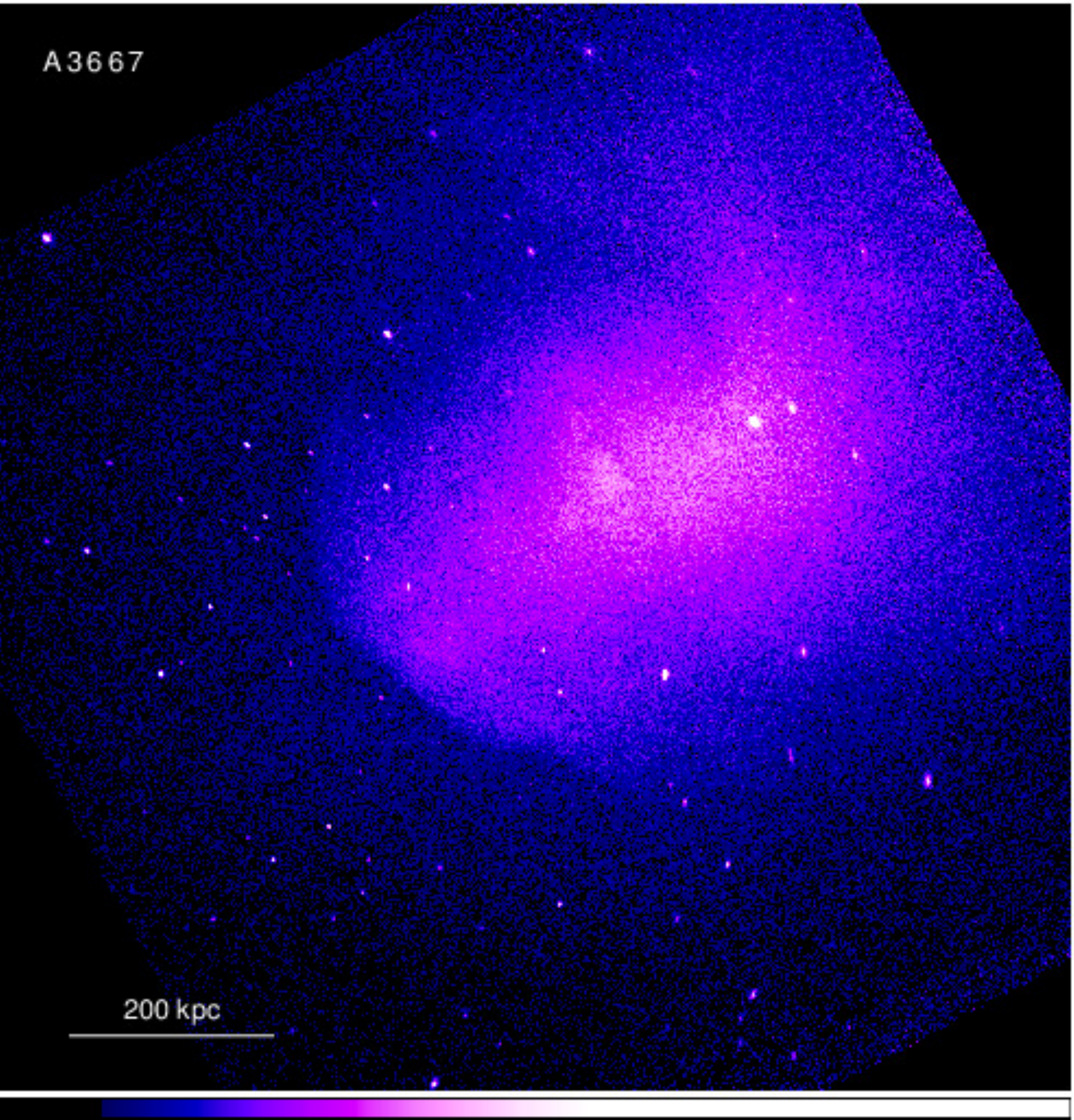}
\includegraphics[width=0.48\textwidth]{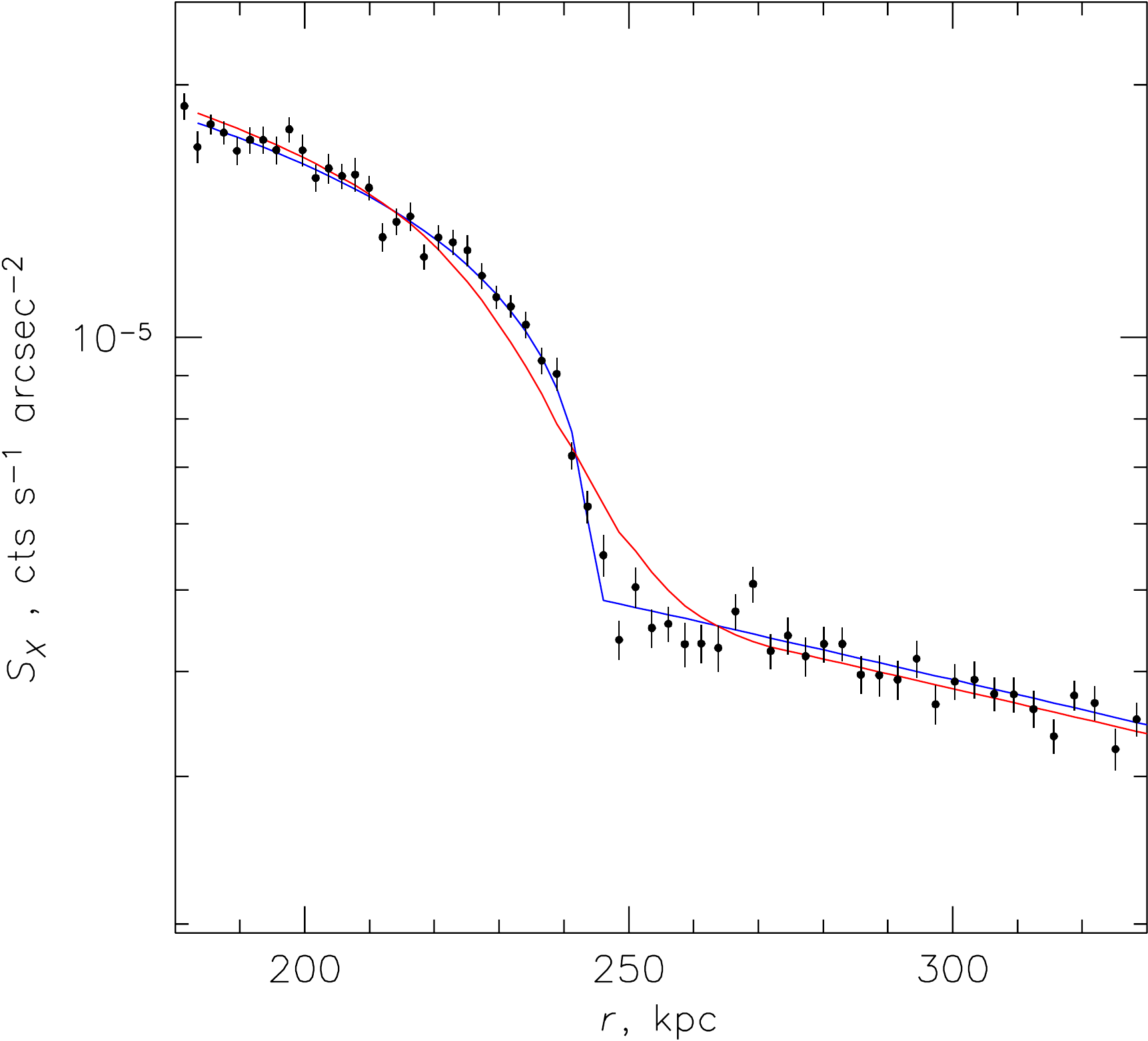}
\caption{{\it Chandra} observation of the cold front in A3667 \citep[from a longer exposure than][$\sim$500~ks]{vik01b}. Left: X-ray surface brightness image. The cold front is the sharp edge feature to the SE. Right: X-ray surface brightness profile of the cold front. The blue line shows the best-fit projected spherical density discontinuity, with an infinitely small width. The red line shows the same model smoothed with a Gaussian of width $\sigma$ = 13~kpc, corresponding to the collisional mean free path $\lambda_{\rm in-out}$, which is inconsistent with the data (Reproduced from MV07).\label{fig:a3667}}
\end{center}
\end{figure}

Another nearby example of a remnant core cold front is in the merging cluster A3667 (Figure \ref{fig:a3667}). Line-of-sight velocity measurements of the cluster galaxies show no evidence for multiple components in redshift space, indicating that A3667 is a merger occurin in the plane of the sky \citep{joh08,owe09}. A surface brightness edge in the ICM was identified already in {\it ROSAT} and {\it ASCA} observations \citep{kno96,mar98,mar99}, but only {\it Chandra} data had the sufficient resolution to allow the edge to be identified as a cold front \citep{vik01b}.

Due to the high spatial resolution of the {\it Chandra} data, \citet{vik01b} were able to show that the surface brightness profiles across the edge were well fit by a projected density discontinuity. They measured an upper limit on the width of the front interface of $\sim$4~kpc, which is comparable to or smaller than the particle mean free paths on both sides of the cold front, and of those particles moving across the interface (\citealt{vik01b}). Particle diffusion across the interface proceeds mostly from the inside of the front (i.e., the denser side) to the outside, since the particle flux through the unit area is proportional to $nT^{1/2}$. Therefore, the most relevant mean free path to consider is $\lambda_{\rm in\rightarrow{out}}$ \citep{spi62,vik01b}:
\begin{equation}
\lambda_{\rm in\rightarrow{out}} = \lambda_{\rm out}\frac{T_{\rm in}}{T_{\rm out}}\frac{G(1)}{G(\sqrt{T_{\rm in}/T_{\rm out}})}
\end{equation}
where $G(x) = [\Phi(x)-x\Phi'(x)]/2x^2$ and $\Phi(x)$ is the error function. For the cold front in A3667, $\lambda_{\rm in\rightarrow{out}} \approx 10-13$~kpc (MV07). If Coloumb diffusion is not suppressed across the interface, the front width should be at least several times $\lambda_{\rm in\rightarrow{out}}$, yet this is not the case (Figure \ref{fig:a3667}), implying that there is a mechanism for such suppression. More recently, \citet{dat14} estimated a larger width for the cold front of $\Delta{R} = 24 \pm 12$~kpc, but their estimate appears to neglect the fact that the projected X-ray emissivity a zero-width density jump superficially appears as a surface brightness jump with a finite width. Nevertheless, even with this larger estimate for the width, they were still able to conclude that particle diffusion and thermal conductivity must be strongly suppressed across the interface.

\begin{figure}
\begin{center}
\includegraphics[width=0.98\textwidth]{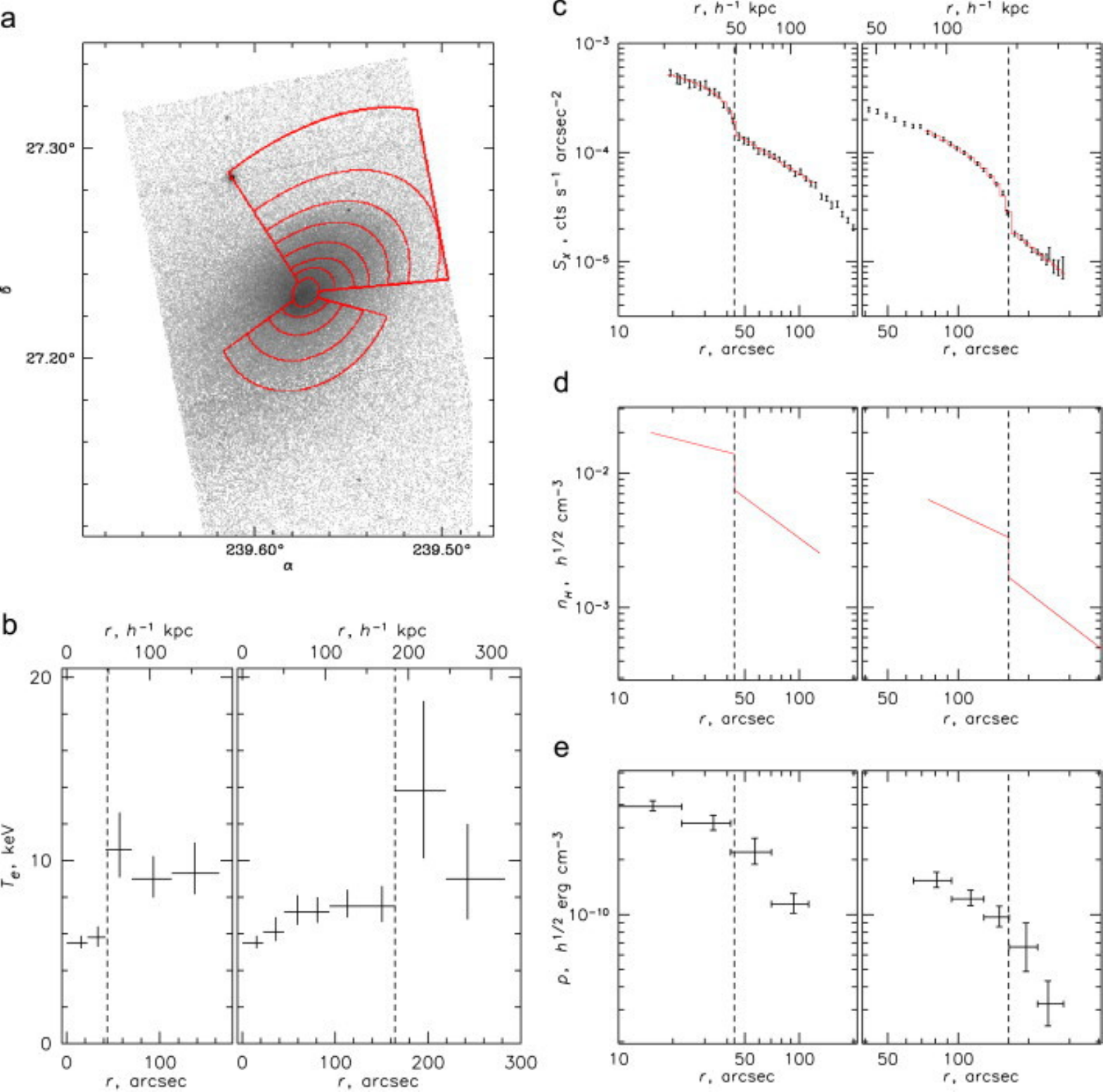}
\caption{Analysis of cold fronts discoverd in A2142 by \citet{mar00}. Panel (a): X-ray image with red regions overlaid showing the regions used for derivation of the temperature profiles shown in panel (b). Panels (b-e) show the profiles of temperature, surface brightness, density, and pressure across the southern (left) and northwestern (right) edges. Vertical dashed lines show the positions of density jumps. Reproduced from MV07.\label{fig:a2142}}
\end{center}
\end{figure}

The front surface itself appears very smooth, lacking evidence of pertubations within an angular sector of $\phi \approx 30^\circ$ from the direction of the front motion. Outside of this sector, the front becomes less sharp and eventually blends in with the surrounding emission. However, such an interface should be susceptible to the effects of KHI even closer to the axis defined by the direction of motion than this. \citet{vik02} determined that perturbations with wavelengths of $\approx$~20-30~kpc would be unstable across the front surface even if the stabilizing effect of the cluster's gravity is taken into account. However, \citet{chu04} argued that the upstream part of a remnant core cold front could be stable to the KHI if the front has a small initial intrinsic width. \citet{maz02} pointed out that the sides of the remnant core in A3667 show a ragged appearance which may indeed be due to the KHIs expected at this location.

Assuming that the front is stabilized against KHI by a magnetic field layer, \citet{vik01a,vik02} estimated the minimum magnetic field strength that would be required to suppress KHI at small angles from the direction of the front's motion. They determined that the tangential magnetic field strength at the front surface must be within $6~{\rm \mu{G}} < B < 14~{\rm \mu{G}}$, which though likely stronger than in most regions of the cluster is still only a small fraction of the thermal pressure, with $p_m/p_{\rm th} \sim 0.1-0.2$. Together with the observed sharpness of the front, this is powerful indirect evidence for a magnetic field draping layer at the cold front in A3667.

\subsection{A2142}\label{sec:a2142}

Cold fronts were first discovered by \citet{mar00} in A2142. They identified two arc-shaped fronts on opposite sides of the cluster core (Figure \ref{fig:a2142}). They first proposed that the edges arose from dense subcluster cores which were ram-pressure stripped during a cluster merger--in other words, remnant-core cold fronts. As more cold fronts wrapped around the  cores of other clusters were found, it became clear that the cold fronts in A2142 are of the sloshing type (MV07). \citealt{ros13} reported another cold front $\sim$1~Mpc to the SE, suggesting that sloshing extends to large scales beyond the core region. The hypothesis of sloshing triggered by a minor merger is supported by the observations of \citet{owe11}, who identified substructures in the galaxy distribution and their kinematics in A2142 which are candidates for the perturbing subclusters which initiate the sloshing motions.

\citet{ett00} noted that the existence of the sharp temperature discontinuities in the A2142 cold fronts indicated that heat flux across the interface must be suppressed by a factor of $\sim$250-2500, since otherwise they would have been smeared out by thermal conduction on a very short timescale. This was the first indication that cold fronts may be used to constrain transport processes in the ICM.

\subsection{Virgo}\label{sec:virgo}

\begin{figure}
\begin{center}
\includegraphics[width=0.50\textwidth]{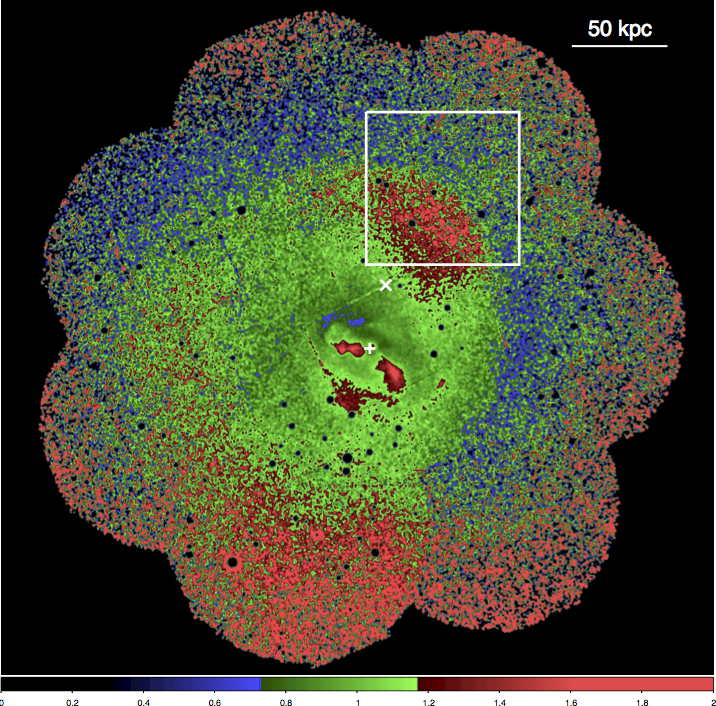}
\caption{{\it XMM-Newton} EPIC/MOS mosaic image of the central r$\sim$150~kpc region of the Virgo cluster \citep{sim10}. The white square indicates the {\it Chandra} ACIS-I pointing of \citet{wer16} shown in \ref{fig:virgo_chandra}. Reproduced from \citet{wer16}.\label{fig:virgo_xmm}}
\end{center}
\end{figure}

\begin{figure}
\begin{center}
\includegraphics[width=0.48\textwidth]{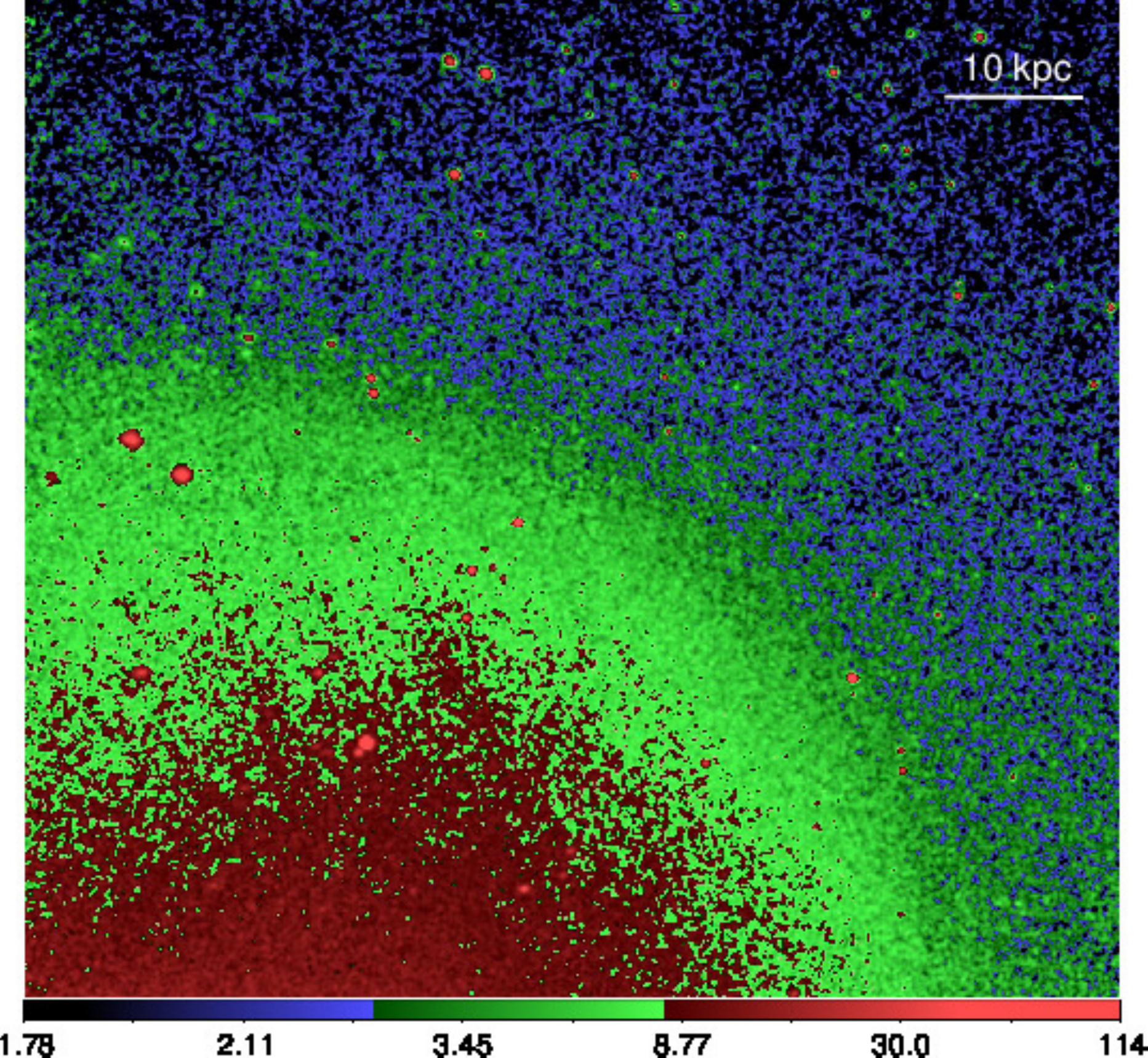}
\includegraphics[width=0.48\textwidth]{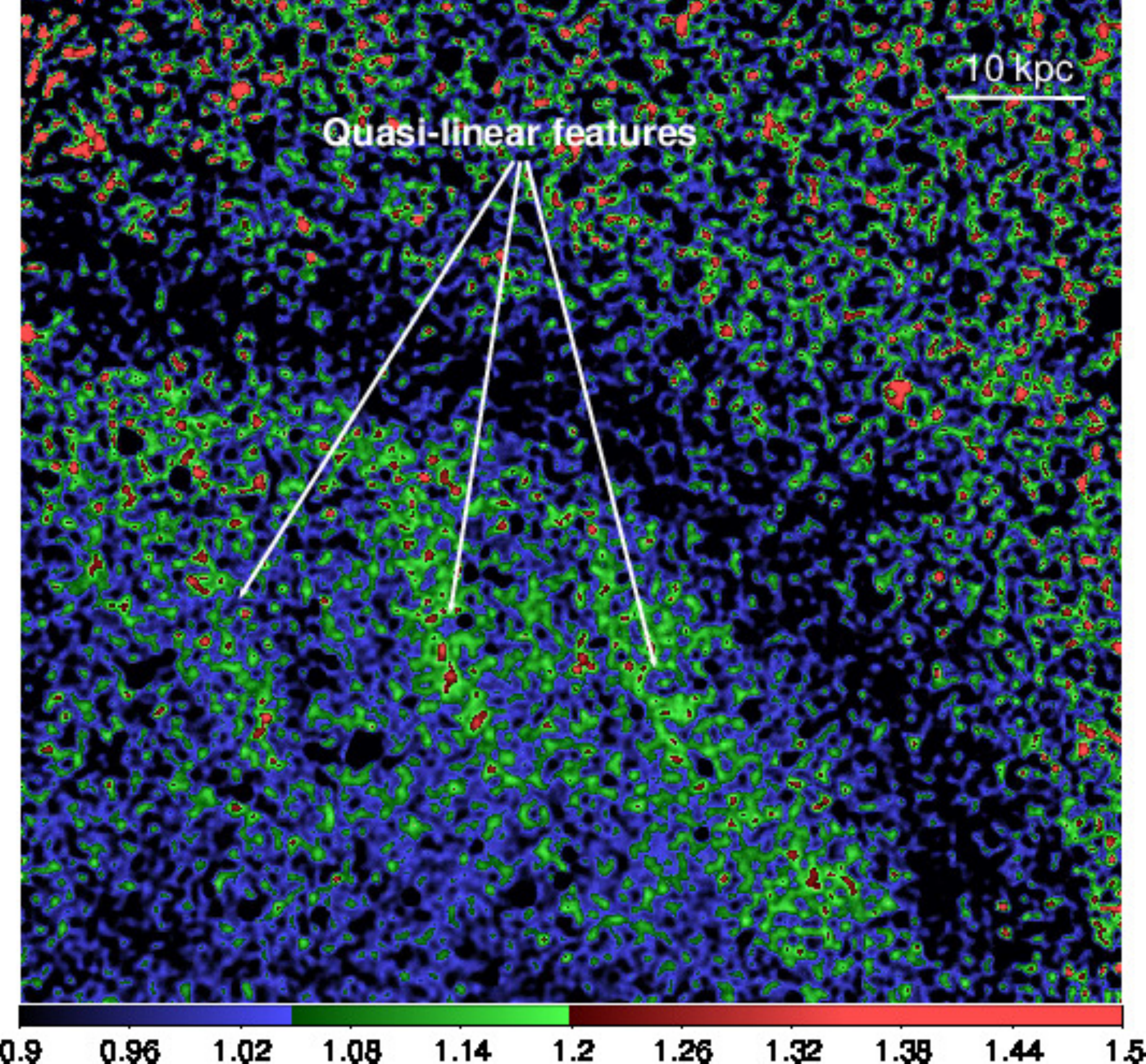}
\caption{{\it Left panel:} {\it Chandra} image of the Virgo cold front in the 0.8--4.0~keV energy band from \citet{wer16}. The image was smoothed with a Gaussian function with a 1.5 arcsec window. {\it Right panel:} The residual image, which reveals three X-ray bright quasi-linear features separated from each other by $\sim15$~kpc in projection. The dark band outside the cold front is an image processing artifact due to the sharp surface brightness discontinuity.\label{fig:virgo_chandra}}
\end{center}
\end{figure}

The very nearby Virgo cluster of galaxies (with its central galaxy M87) possesses a pair of sloshing cold fronts (Figure \ref{fig:virgo_xmm}, \citealt{sim07,sim10}). At a distance of only 16.1~Mpc \citep{ton01}, the arcsecond resolution of {\it Chandra} corresponds to a projected spatial resolution of 78~pc, making the Virgo cold fronts an ideal target to study cold fronts at the highest resolution. In a recent 500~ks {\it Chandra} observation focused on the northern Virgo cold front, \citet{wer16} measured the width of the interface at a number of locations along the front surface. In the northwestern part of the front, which appears to be the sharpest, they found an upper limit on the width of the interface to be $\sim$2.5~kpc at 99\% confidence level, which is approximately 1.5 times the local Coulomb mean free path $\lambda_{\rm in{\rightarrow}out}$ for particles diffusing across the interface. In the northern part of the front, the upper limit on the width is found to be somewhat wider, $\sim$4.3~kpc. They argued that at this location the front may be smeared by the effects of KHI, as had been suggested by the Virgo cluster simulations of \citet[][hereafter R13; see Section \ref{sec:viscosity}]{rod13a}. In several locations, narrow-wedge surface brightness profiles across the Virgo cold front display a multi-step structure, a signature for the presence of KHIs, predicted by R13 from mock observations of their simulations. In either case, the Virgo cold front width is much narrower than would be expected if particle diffusion across the front were unsuppressed, which would smooth the front by several $\lambda_{\rm in{\rightarrow}out}$. They also argued from these estimates that thermal conduction is also highly suppressed across the cold front surface.

These observations also revealed a somewhat unexpected feature associated with the cold front. The residual image in the right panel of Figure \ref{fig:virgo_chandra} shows the presence of three X-ray bright quasi-linear features, extending away from the front surface at an angle which roughly aligns them with the direction of the curvature of the front to the southwest. The features are separated from each other by $\sim$15~kpc in projection, and are not associated with any instrumental artifacts (e.g., chip gaps or readout streaks). Therefore, these features correspond to density enhancements underneath the cold front seen in projection. \citet{wer16} argued for an origin of these features that was related to magnetic fields, a possibility we discuss in Section \ref{sec:bfields}.

\subsection{NGC 7618/UGC 12491}\label{sec:n12491}

\begin{figure}
\begin{center}
\includegraphics[width=0.42\textwidth]{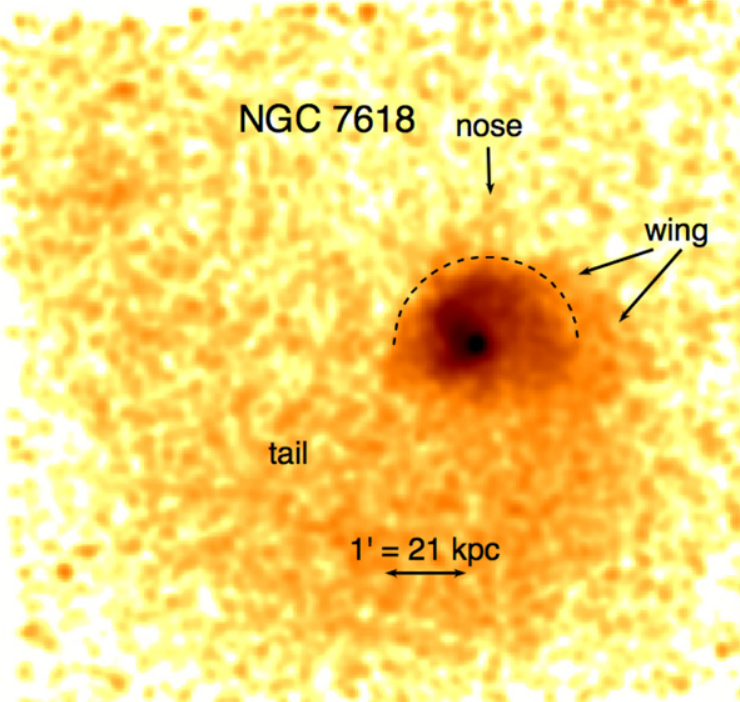}
\includegraphics[width=0.53\textwidth]{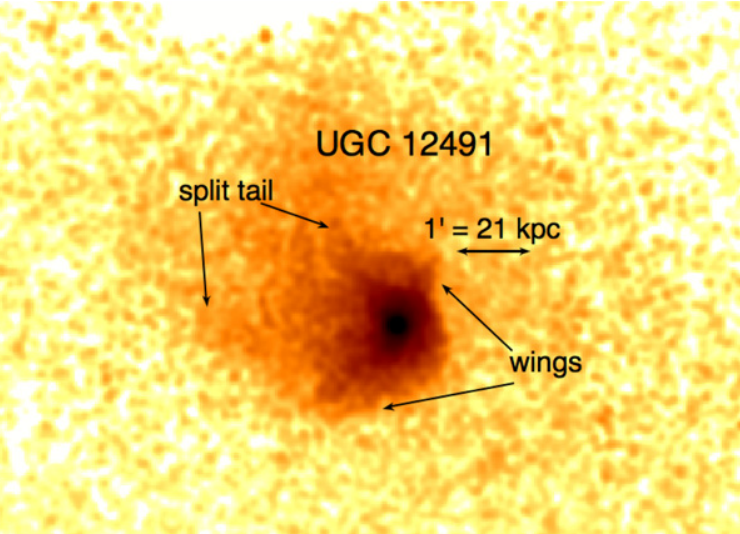}
\caption{The two galaxy groups around NGC 7618 (left) and UGC 12491 (right) passed each other a few 100~Myr ago, and caused sloshing in each other’s ICM atmospheres. The spiral-shaped cold fronts wrapped around each group center show distortions in the form of wings and noses, resembling KHIs. Reproduced from \citet{rod12a}.\label{fig:n7618}}
\end{center}
\end{figure}

The galaxy groups around the elliptical galaxies NGC 7618 and UGC 12491 show  classic spiral-shaped sloshing cold fronts, arising from their recent mutual encounter (\citealt{kra06}). However, the cold fronts in these groups are not as smooth as in, e.g., A2142, but display distortions in the form of noses and wings (Figure \ref{fig:n7618}, \citealt{rod12a}), that resemble the appearance of KHIs as seen in simulations. Cold fronts with similar features exist also in A496 \citep[][R12]{dup07}.

\section{Simulations of Cold Fronts}\label{sec:simulations}

Since the discovery of cold fronts, a large number of simulations have been performed to model their formation and evolution in merging galaxy clusters. In this review, we focus on those recent simulations which have attempted to place constraints on the detailed physical properties of the underlying plasma. A table listing the major simulations of cold fronts referenced in this paper, along with the numerical models employed and other properties, can be found in Appendix \ref{sec:appendix}.

\subsection{Basic Physics}\label{sec:physics}

We will first outline the basic physics underlying these simulations. As previously mentioned, due the fact that the electron and ion mean free paths are smaller than the length scales under consideration, the ICM is typically modeled as a thermal, magnetized plasma in the fluid approximation. However, contrary to the usual MHD ordering of scales, the Larmor radii of the particles are much smaller than their mean free paths, where the Larmor radii are on the order of npc compared to the kpc scale of the mean free path. Therefore, a given particle gyrates many times around the local magnetic field line before encountering another particle. Due to the small Larmor radii and the relative weakness of collisions, we may deduce that the perpendicular and parallel pressures of the particles and the transport coefficients of diffusive processes are anisotropic with respect to the local direction of the magnetic field line. In this situation, the anisotropy of the perpendicular and parallel pressures is produced by the conservation of the first and second adiabatic invariants of the particles.

Under these assumptions, the fluid description of the ICM can be written down in terms of the Braginskii-MHD equations \citep[][presented here in conservative form, and in Gaussian units]{bra65}. The equations for conservation of mass, momentum, energy, and magnetic field are:
\begin{eqnarray}
\frac{\partial{\rho}}{\partial{t}} + \nabla \cdot (\rho{\bf v}) &=& 0, \label{eqn:mass}\\
\frac{\partial{(\rho{\bf v})}}{\partial{t}} + \nabla \cdot
(\rho{\bf vv} + {\sf P}) &=& \rho{\bf g} \label{eqn:momentum}, \\
\frac{\partial{(\rho{e})}}{\partial{t}} + \nabla \cdot [(\rho{e}{\sf I} + {\sf P}) \cdot {\bf v} +
{\bf Q}] &=& \rho{\bf g \cdot v}\label{eqn:energy}, \\
\frac{\partial{\bf B}}{\partial{t}} + \nabla \cdot ({\bf vB} - {\bf
 Bv}) &=& 0, \label{eqn:bfield}
\end{eqnarray}
where the total (thermal and magnetic) pressure tensor in Equation \ref{eqn:momentum} and \ref{eqn:energy} is
\begin{equation}
{\sf P} = p_\perp{\sf I}-(p_\perp-p_\parallel)\hat{\textbf{b}}\hat{\textbf{b}} + \frac{B^2}{8\pi}{\sf I}-\frac{\bf BB}{4\pi}.\label{eqn:ptensor}
\end{equation}
As usual, $\hat{\textbf{b}} = {\bf B}/B$ is the unit vector in the direction of the local magnetic field, and the total thermal pressure satisfies
\begin{equation}
p = \frac{2}{3} \,p_\perp+\frac{1}{3} \,p_\parallel,
\end{equation}
The terms $B^2{\sf I}/8\pi$ and $-{\bf B}{\bf B}/4\pi$ in Equation \ref{eqn:ptensor} are the familiar ``magnetic pressure'' and ``magnetic tension'', respectively. In Equation \ref{eqn:energy}, the total energy per unit volume is
\begin{equation}
\rho{e} = \frac{{\rho}v^2}{2} + \rho\epsilon + \frac{B^2}{8\pi}.
\end{equation}
In these equations, $\rho$ is the gas density, $\bf{v}$ is the fluid velocity, $\epsilon$ is the gas thermal energy per unit volume, $\bf{B}$ is the magnetic field vector, ${\bf g}$ is the gravitational acceleration, ${\bf Q}$ is the heat flux vector, and $\sf{I}$ is the unit dyad. Generally, an adiabatic equation of state of an ideal gas $p = (\gamma - 1)\rho\epsilon$ with $\gamma = 5/3$ is assumed, along with equal electron and ion temperatures, and primordial abundances of H and He with trace amounts of metals, for an average molecular weight of $\bar{A} \approx 0.6$.

Most of the astrophysical literature on galaxy clusters does not present the MHD equations with a pressure anisotropy, but instead describes this effect in terms of a viscous flux. We will now show how these two descriptions are equivalent. Differences in the two components of the thermal pressure arise from the conservation of the first and second adiabatic invariants for each particle on timescales much greater than the inverse of the ion gyrofrequency, $\Omega_g^{-1}$ \cite{che56}. When the ion-ion collision frequency $\nu_{\rm ii}$ is much larger than the rates of change of all fields, an equation for the pressure anisotropy can be obtained by balancing its production by adiabatic invariance with its relaxation via collisions \cite[cf.][]{sch05}:
\begin{equation}
p_\perp-p_\parallel = 0.960\,\frac{p_{\rm i}}{\nu_{\rm ii}}\frac{d}{dt}\ln{\frac{B^3}{\rho^2}},
\label{eqn:pressure_anisotropy}
\end{equation}
where $p_{\rm i}$ is the thermal pressure of the ions (due to their much smaller mass, the effect of the electrons on the viscosity may be neglected).

The factor in front of the time derivative on the right-hand side of Equation \ref{eqn:pressure_anisotropy} is simply the dynamic viscosity coefficient for the ions, or the ``Spitzer'' viscosity, which is given by \citep{spi62,bra65,sar88}:
\begin{eqnarray}\label{eqn:spitzer_viscosity}
\mu_{\rm Sp} &=& 0.960\,\frac{p_i}{\nu_{\rm ii}} \\
\nonumber &\approx& 2.2 \times 10^{-15}\frac{T^{5/2}}{\ln\Lambda_{\rm i}}~{\rm g~cm^{-1}~s^{-1}},
\end{eqnarray}
where the temperature $T$ is in Kelvin, $\nu_{\rm ii}$ is the ion-ion collision frequency, and $\ln\Lambda_{\rm i}$ is the ion Coulomb logarithm, which is a weak function of $\rho$ and $T$; for conditions in the ICM, it may be approximated as $\ln\Lambda_{\rm i} \approx 40$. Using Equations (\ref{eqn:mass}) and (\ref{eqn:bfield}) to replace the time derivatives of density and magnetic field strength with velocity gradients, Equation \ref{eqn:pressure_anisotropy} may be written as
\begin{equation}
p_\perp-p_\parallel = \mu_{\rm Sp}[3\hat{\textbf{b}}\cdot(\hat{\textbf{b}}\cdot\nabla{\bf v})-\nabla \cdot {\bf v}].
\end{equation}
which, upon plugging back into Equation \ref{eqn:momentum}, shows that the effect of the pressure anisotropy manifests itself as an anisotropic viscous flux.

Next, we consider the effect of thermal conduction. Due to the effect of the small Larmor radii of the electrons, heat only flows along the field lines, and the only component of the temperature gradient which affects the heat flux is that which is parallel to the field direction:
\begin{equation}
{\bf Q} = -\kappa_{\rm Sp}\hat{\textbf{b}}\hat{\textbf{b}} \cdot \nabla{T},
\end{equation}
where $\kappa_{\rm Sp}$ is the familiar ``Spitzer'' thermal conductivity coefficient \citep{spi62,bra65,sar88}:
\begin{eqnarray}
\kappa_{\rm Sp} &=& \frac{3.2 k_{\rm B} p_e}{m_{\rm e}\nu_{\rm ee}} \\
\nonumber &\approx& 1.84 \times 10^{-5}\frac{T^{5/2}}{\ln\Lambda_{\rm e}}~{\rm erg~cm^{-1}~s^{-1}~K^{-1}},
\end{eqnarray}
where $p_e$ is the thermal pressure of the electrons, $\nu_{\rm ee}$ is the electron-electron collision frequency, and $\ln\Lambda_{\rm e}$ is the electron Coulomb logarithm, which may also may be approximated as $\ln\Lambda_e \approx 40$. Due to their much larger mass, the heat flux due to ion collisions is negligible, so it is neglected.

So far, most simulations of the ICM have not worked with the full set of these equations, but rather in certain physically-motivated limits (though \citealt{don09}, \citealt{kun12}, \citealt{par12}, \citealt{suz13}, and \citealt{zuh15a} are notable exceptions). In particular, there have been two approximations of these equations in common use. In the limit that the pressure anisotropy $p_\perp-p_\parallel$ is small (in other words, collisions are relatively strong and the viscosity is low), we recover standard MHD (or simply HD if it is assumed that the magnetic field can also be neglected). In this review we will describe a number of simulations that work in this MHD limit, which emphasize the effect of the magnetic field on cold fronts. Since the Prandtl number of the Spitzer description of the ICM is very small, the thermal diffusivity dominates over the momentum diffusivity:
\begin{equation}
{\rm Pr} \equiv \frac{\nu}{\chi} = 0.5 \,
\frac{\ln\Lambda_{\rm e}}{\ln\Lambda_{\rm i}}\left(\frac{2m_{\rm e}}{m_{\rm i}}\right)^{1/2} \simeq 0.02.
\end{equation}
Therefore, another limit of these equations considers MHD with thermal conductivity added. We will also consider a number of simulations that have operated in this limit.

A second limit of these equations that is often modeled is that in which the magnetic field is very weak and turbulent throughout the volume of the cluster. In this case, though the effects of magnetic pressure and tension are neglected, the anisotropy of the heat and momentum fluxes manifests itself as an effective suppression of the transport coefficients when averaged over a small volume, assuming the magnetic field is isotropically tangled. We may then re-write Equations \ref{eqn:momentum} and \ref{eqn:energy} as:
\begin{eqnarray}
\frac{\partial{(\rho{\bf v})}}{\partial{t}} + \nabla \cdot (\rho{\bf vv} + p{\sf I} + {\sf \Pi}) &=& \rho{\bf g} \label{eqn:momentum_hd}, \\
\frac{\partial{(\rho{e})}}{\partial{t}} + \nabla \cdot \{[(\rho{e} + p){\sf I} + {\sf \Pi}] \cdot {\bf v} + {\bf Q}\} &=& \rho{\bf g \cdot v}\label{eqn:energy_hd},
\end{eqnarray}
and the conductive and viscous flux terms may be written as
\begin{eqnarray}
{\bf Q} &=& -f_c\kappa_{\rm Sp}\nabla{T},\\
{\sf \Pi} &=& -f_v\mu_{\rm Sp}\nabla{\bf v},
\end{eqnarray}
where the ``suppression factors'' $f_c$ and $f_v$ account for the reduced conductivity and viscosity due to the averaging over the random direction of the magnetic field, or other microscale processes that suppress the effective conductivity and/or viscosity below the Spitzer value. This HD limit of the equations is the approximation that has been assumed in a number of simulation works, including some important ones considered in this review. Consideration of this limit is relevant, since the finite-Larmor radius effects on the transport coefficients may often only reveal themselves in a spatially averaged way due to projection effects and the finite resolution of X-ray observations.

It is also important to note that the Braginskii-MHD equations are ill-posed in situations where the absolute value of the pressure anisotropy exceeds certain bounds (on the order of the magnetic pressure), and in this regime finite-Larmor radius effects become important. We will describe these issues further and the consequences of these effects for simulations of the ICM in Section \ref{sec:plasma_instabilities}.

\subsection{Results from Simulations of the Formation and Global Evolution of Cold Fronts}\label{sec:pure_hydro}

The basic physics of the formation and global evolution of cold fronts can be understood in the pure hydrodynamics framework for both the remnant core and the sloshing variety. The results of these studies, in particular their points of divergence from observed cold fronts, indicated the need for additional physics beyond HD that is discussed in subsequent sections.

Stripping of gas from one gravitational potential falling into another, usually larger, gas-filled potential has been simulated in various contexts, ranging from stripping of gas clouds inside galaxy halos \citep[e.g.,][]{mur93,pit10}, to elliptical galaxies in clusters \citep[e.g.,][etc.]{lea76,tak84,por93,ste99,acr03,tak05} to mergers between clusters \citep[][]{roe97,ric01,rit02,poo06,poo07,poo08}. Such simulations routinely produce a remnant core cold front at the upstream side of the stripped atmosphere. Strictly speaking, several of these idealized simulations do not prove that gas stripping {\it forms} an upstream cold front because they already start with a cold front, a contact discontinuity between the cooler to-be-stripped atmosphere and the hotter ambient medium. However, simulations of high mass ratio cluster mergers such as \citet{poo06} and \citet{zuh11a} demonstrate that remnant merger core cold fronts form readily in such situations. \citet{hei03} pointed out an additional effect that may help to form and maintain the upstream cold front. They suggest that the flow of ambient gas around the stripped atmosphere causes a slow forward motion of the gas inside the stripped potential, bringing cooler gas from the center of the infalling atmosphere towards the upstream edge. However, this does not seem to be a universal process and may depend on the depth and steepness infalling potential as well as on the timescale available for the stripping during the cluster infall (\citealt{rod15a}, hereafter R15A). All purely hydrodynamical simulations of the formation of such remnant-core cold fronts that have sufficient resolution predict the growth of KHIs along the sides of the stripped atmosphere.

\begin{figure}
\begin{center}
\includegraphics[width=0.98\textwidth]{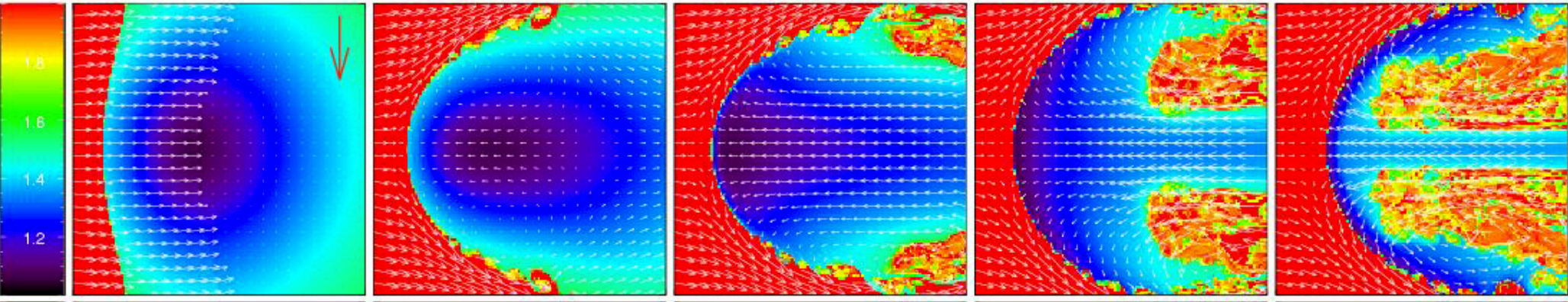}
\caption{Formation of a merger-remnant cold front in a purely hydrodynamic simulation from \citet{hei03}. Slices are of gas entropy, with velocity vectors overlaid. Note the rapid development of KHI.\label{fig:heinz2003}}
\end{center}
\end{figure}

\begin{figure}
\begin{center}
\includegraphics[width=0.48\textwidth]{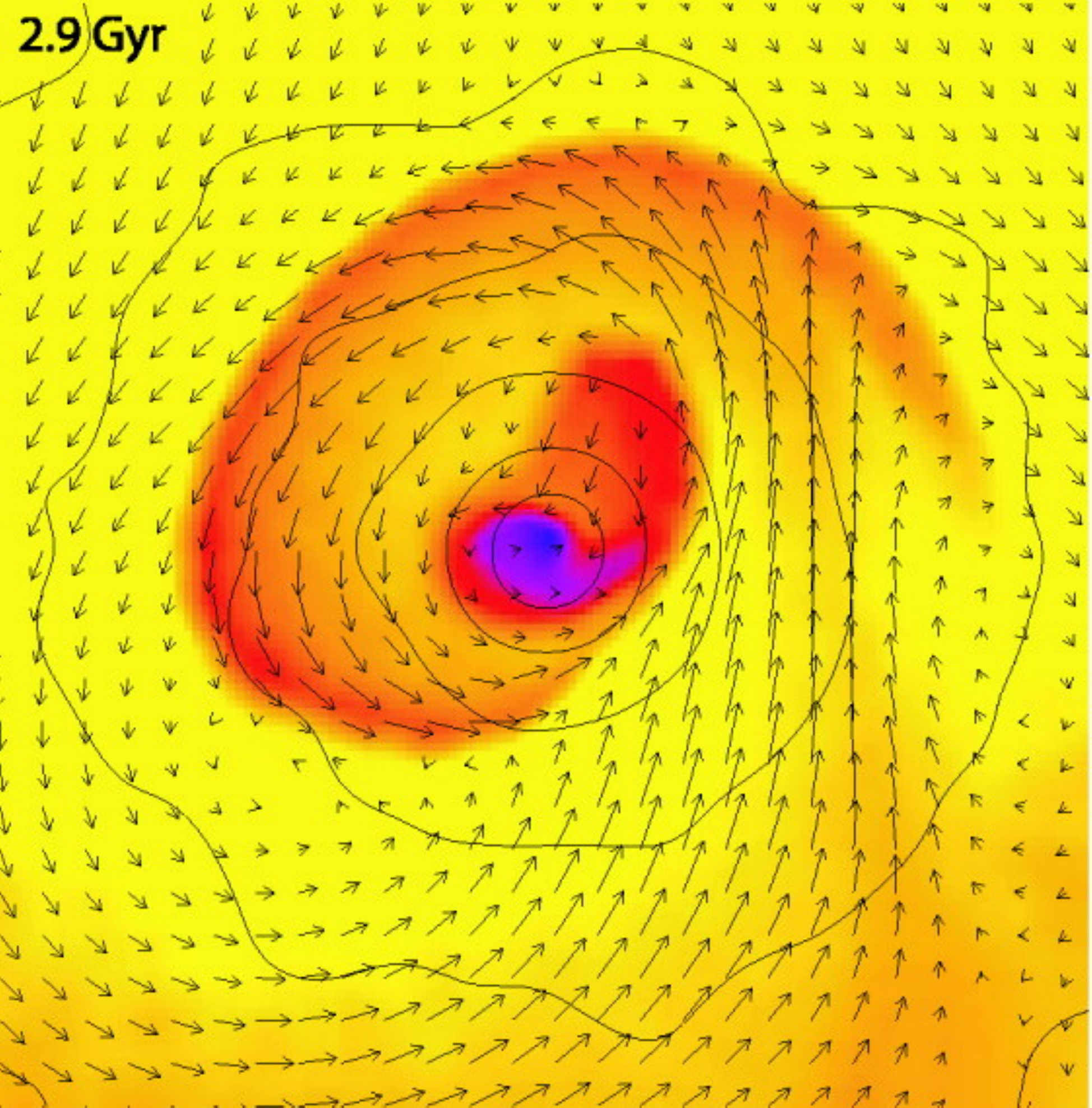}
\includegraphics[width=0.486\textwidth]{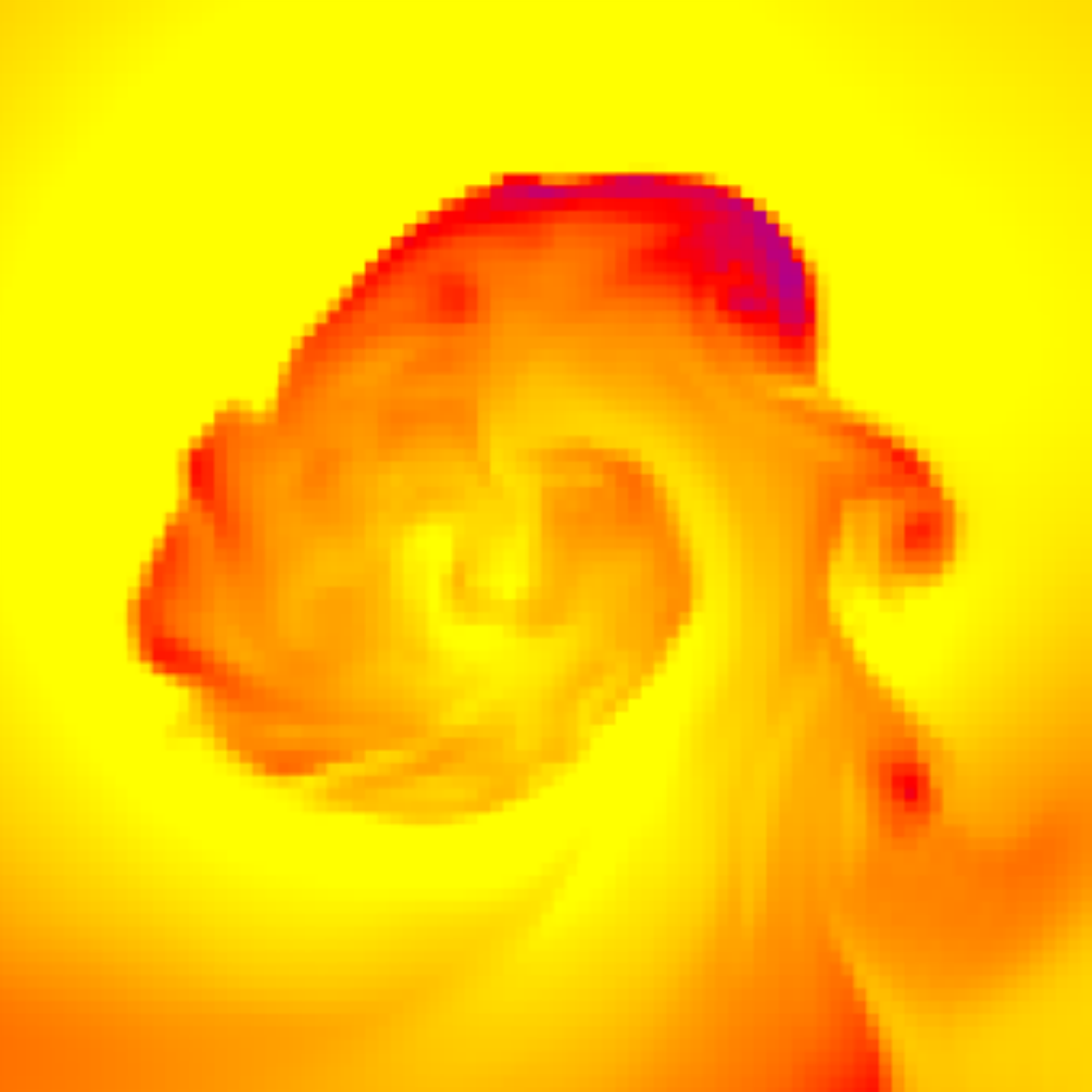}
\caption{Sloshing cold fronts (shown with slices of gas temperature) from an SPH simulation by AM06 (left) and a grid-based simulation of the same cluster setup by \citet{zuh10} (right). The grid-based simulation resolves KHI that are unresolved by the SPH simulation.\label{fig:sph_vs_amr_sloshing}}
\end{center}
\end{figure}

\citet[][hereafter AM06]{AM06} carried out simulations of sloshing cold fronts with the \code{Gadget} SPH code, and first described in great detail the formation and evolution of these features. \citet{rod11} and \citet[][hereafter R12]{rod12} demonstrated that the positions of the sloshing cold fronts and the contrasts in density and temperature across them can be used to determine the merger history of a given cluster. In the simulations of AM06, the cold fronts which formed were stable against KHI in spite of the presence of shear flows along them, since the SPH method employed at the time was unable to resolve the KHIs. Simulations with grid codes, however, routinely predict KHIs at sloshing cold fronts. \citet[][]{zuh10} performed sloshing simulations using identical initial conditions as AM06, except that in their case the simulations were carried out using the \code{FLASH} grid-based code. Though the same cold fronts were formed in both sets of simulations, the \code{FLASH} simulations produced KHI at the interfaces (see Figure \ref{fig:sph_vs_amr_sloshing}). KHIs also developed naturally in the sloshing simulations for the Virgo cluster and A496 \citep[][R12, R13]{rod11}. Their appearance in synthetic X-ray observations is discussed in Section \ref{sec:viscosity}.

Many cold fronts are believed to be free of KHIs. The following sections discuss ICM plasma properties that could suppress them.

\begin{figure}
\begin{center}
\includegraphics[width=0.7\textwidth]{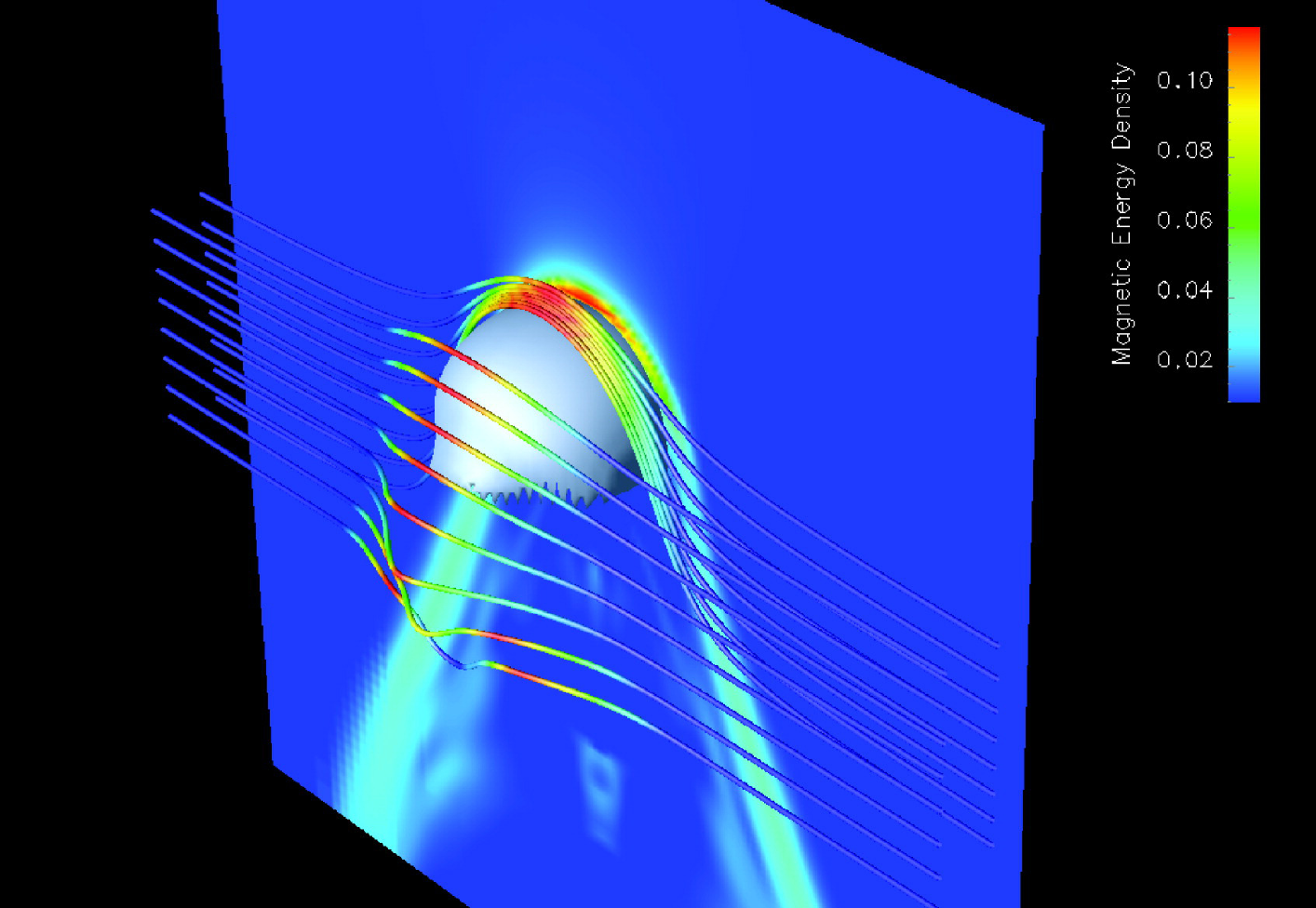}
\caption{The draping of magnetic field lines around a cold cloud in the simulations of \citet{dur08}. The cutting plane is colored by magnetic energy, as are the magnetic field lines. The corresponding animation can be viewed at \url{http://vimeo.com/160767845}.\label{fig:dursi08}}
\end{center}
\end{figure}

\subsection{Magnetic Fields}\label{sec:bfields}

Throughout most of the volume of a galaxy cluster, the magnetic field strength is dynamically negligible. The influence of the magnetic field is typically parameterized by the plasma parameter $\beta = p_{\rm th}/p_B$, the ratio of the thermal pressure $p_{\rm th}$ and the magnetic pressure $p_B = B^2/8\pi$. Faraday rotation measurements (RM) constrain the value of $\beta$ to be as low as $\sim$50 or as high as several hundred, with the magnetic field strength decreasing in radius from the cluster center, and turbulent on scales ranging from 100~kpc down to 100~pc, as determined from RM maps \citep[see][]{car02,eil02,gov06,lai08,bon10,gui08,gui10,fer12}.

However, the existence of the discontinuous changes in density and temperature observed in cold fronts leads to the expectation that the magnetic field in these regions may not be typical of the cluster as a whole. This was first shown for the case of remnant-core cold fronts using analytic arguments by \citet{lyu06}. As a cold front moves through the magnetized medium, the field is is amplified and aligned with the cold front surface as the flow stretches and wraps or ``drapes'' the field lines around the interface. This effect occurs whether the cold front is moving subsonically or supersonically, so long as the velocity is super-Alfv\'enic, which is easily satisfied for the weak magnetic fields in the ICM. The draping layer that forms is very thin compared to the radius of curvature of the cold front, and has a magnetic pressure nearly equal to the surrounding pressure.

Using an analytical stability analysis and simulations, \citet{dur07} showed that a thin magnetized layer located at a cold front-like interface could stabilize interfaces even against long-wavelength perturbations of KHIs and Rayleigh-Taylor instabilities (RTIs). In particular, they showed that given a magnetic field layer with thickness $l$, to stabilize wavelengths $\lambda \approx 10l$ requires that the Alfv\'en speed in the layer is comparable to the relevant destabilizing velocity--namely, the shear velocity in the case of KHI. This was followed up with simulations by \citet{dur08}, who performed a detailed study of the propagation of cold clumps of gas through a magnetized medium. They established that the magnetic field strength in the draping layer that forms is set by a competition between the ``plowing up'' of field lines around the cold front and the slipping of field lines around the sides of the front. An animation showing this effect can be seen in Figure \ref{fig:dursi08}. Another early set of MHD simulations of the formation of remnant-core cold fronts that were explicitly used to investigate the effects of the draping layers were carried out by \citet[][hereafter together referred to as ``Asai-MHD'']{asa04,asa05,asa07}. These simulations used a similar setup to the simulations of \citet{dur08}, but also included gravity and had various initial configurations for the magnetic field lines ranging from uniform to turbulent. The \citet{asa04} simulations were carried out in 2D, and the \citet{asa05} and \citet{asa07} simulations were carried out in 3D.

\begin{figure}
\begin{center}
\includegraphics[width=0.97\textwidth]{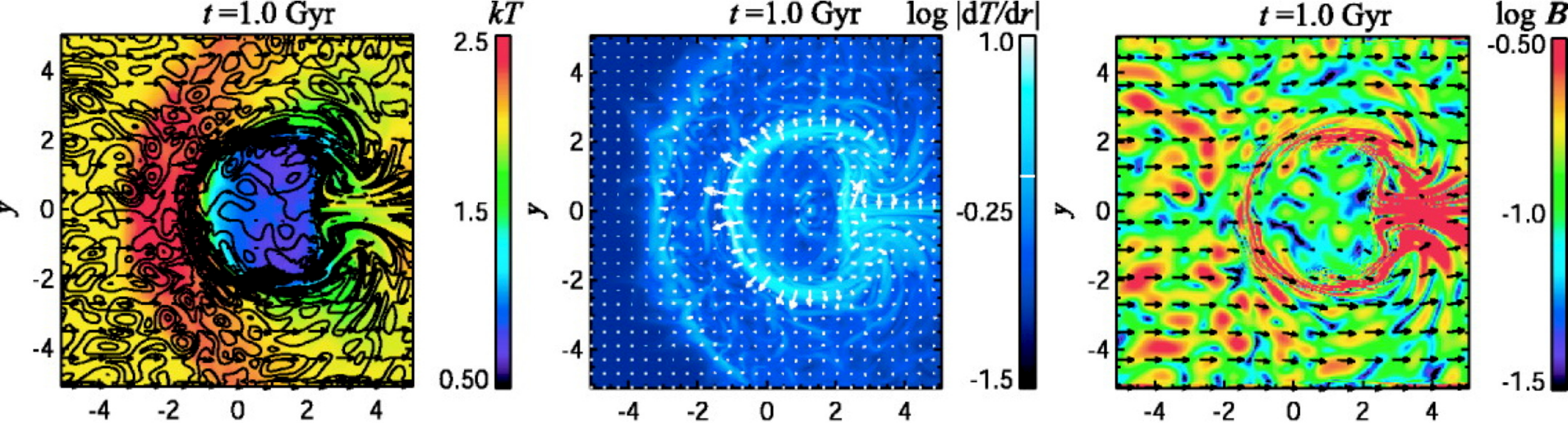}
\caption{Slices through the temperature (left), temperature gradient (middle), and magnetic field strength (right) for the formation of a cold front in a magnetically turbulent medium from \citet{asa07}. The solid curves in the left panel show the contours of the magnetic field strength. Arrows in the left and right panels show the velocity vectors, and arrows in the middle panel show the gradients of temperature. A length unit of ``1'' corresponds to 250~kpc.\label{fig:asai2007_slices}}
\end{center}
\end{figure}

In all of these simulations, the magnetic draping layers that form suppress KHI at the interface (see Figure \ref{fig:asai2007_slices} for examples of the temperature and magnetic field distributions from one of the \citet{asa07} simulations), as predicted by \citet{lyu06}. Importantly, these works showed that such layers also formed regardless of the initial field configuration, whether uniform or turbulent. They also find that the convergence of the flow behind the subcluster amplifies the field in this location as well. The Asai-MHD simulations also included anisotropic thermal conduction, which will be discussed in Section \ref{sec:thermal_cond}.

The MHD cluster merger simulations of \citet{tak08} also produced magnetic draping layers at the cold front surfaces, but these simulations had low resolution and were not used to investigate the properties of cold front stability. It should also be noted that magnetic draping effects may also be responsible for the stability of AGN-blown bubbles; which has been demonstrated by a number of simulation works \citep{rob04,jon05,rus07,don09,one09}. In particular, \citet{rus07} pointed out that the ability of the magnetic fields to stabilize a buoyantly rising bubble depends on the coherence length of the fields. If the latter is smaller than the bubble radius, no useful draping layer can form, and the bubble dissolves by KHIs and RTIs. The simulations of \citet{asa07} indicate that for remnant core cold fronts a draping layer forms quickly also for a turbulent magnetic field with a coherence length smaller than the remnant atmosphere.

\begin{figure}
\begin{center}
\includegraphics[width=0.96\textwidth]{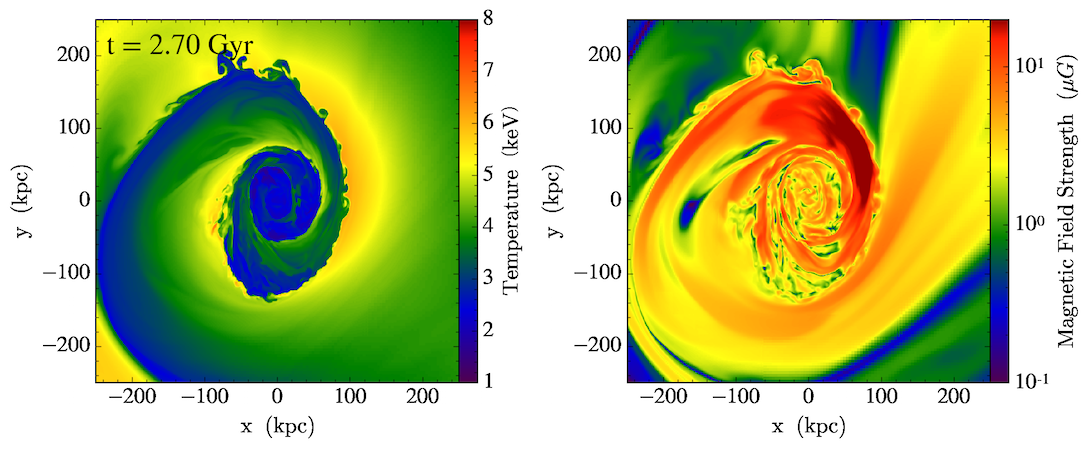}
\caption{Slices of gas temperature (left) and magnetic field strength (right) in an MHD simulation of gas sloshing (ZuHone et al., in preparation). An animation of the development of the cold fronts and the magnetized layers over $\sim$2~Gyr can be found at \url{http://vimeo.com/160771386}. Along the front surfaces, the magnetic field strength is increased, and the magnetic field within the cold fronts remains high as the cold fronts expand.\label{fig:mag_movie}}
\end{center}
\end{figure}

\begin{figure}
\begin{center}
\includegraphics[width=0.97\textwidth]{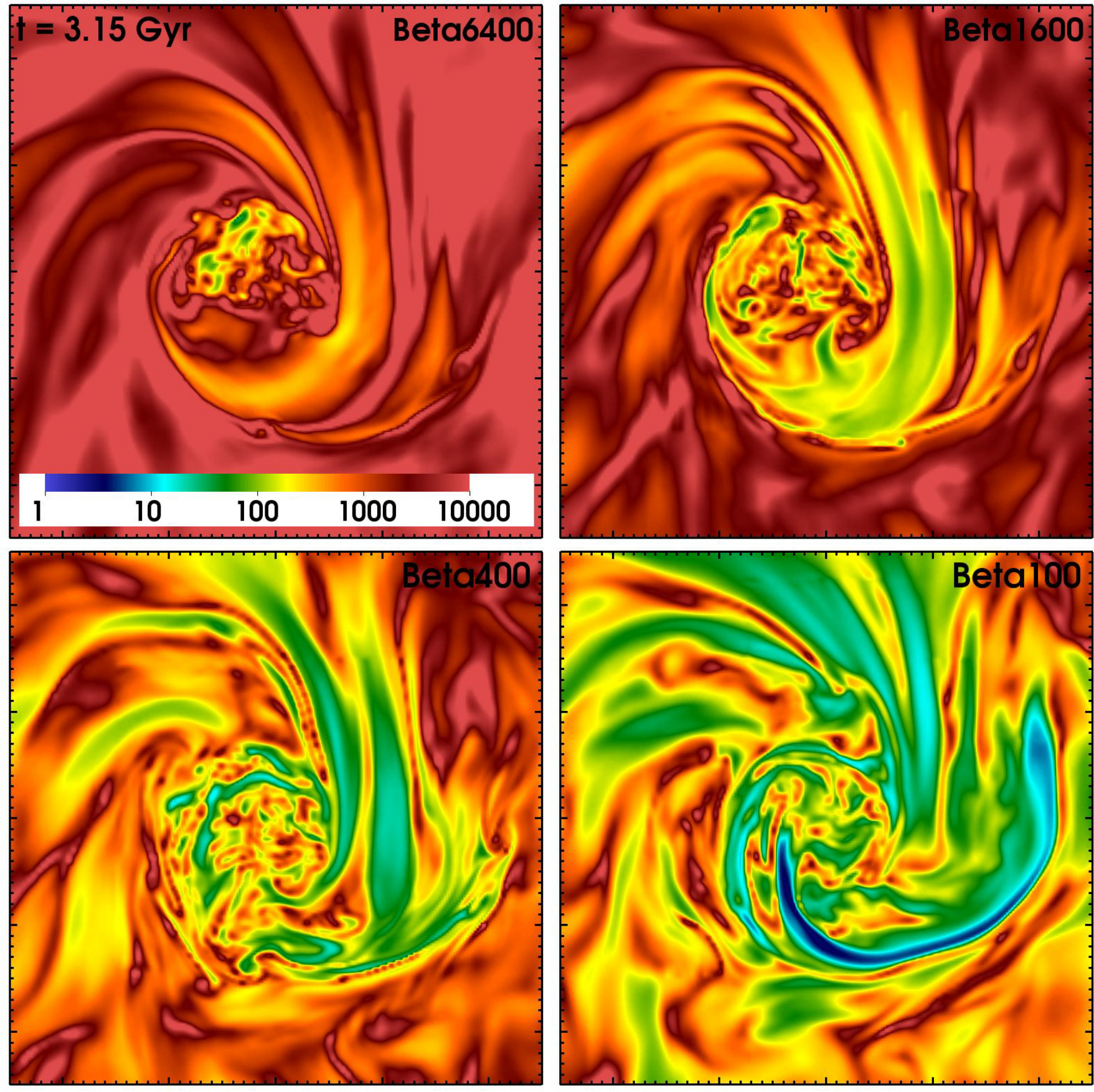}
\caption{Slices through the plasma $\beta$ for the simulations from ZML11 with varying initial $\beta$. Each panel is 500 kpc on a side. Major tick marks indicate 100~kpc distances.\label{fig:diff_beta_beta}}
\end{center}
\end{figure}

\begin{figure}
\begin{center}
\includegraphics[width=0.97\textwidth]{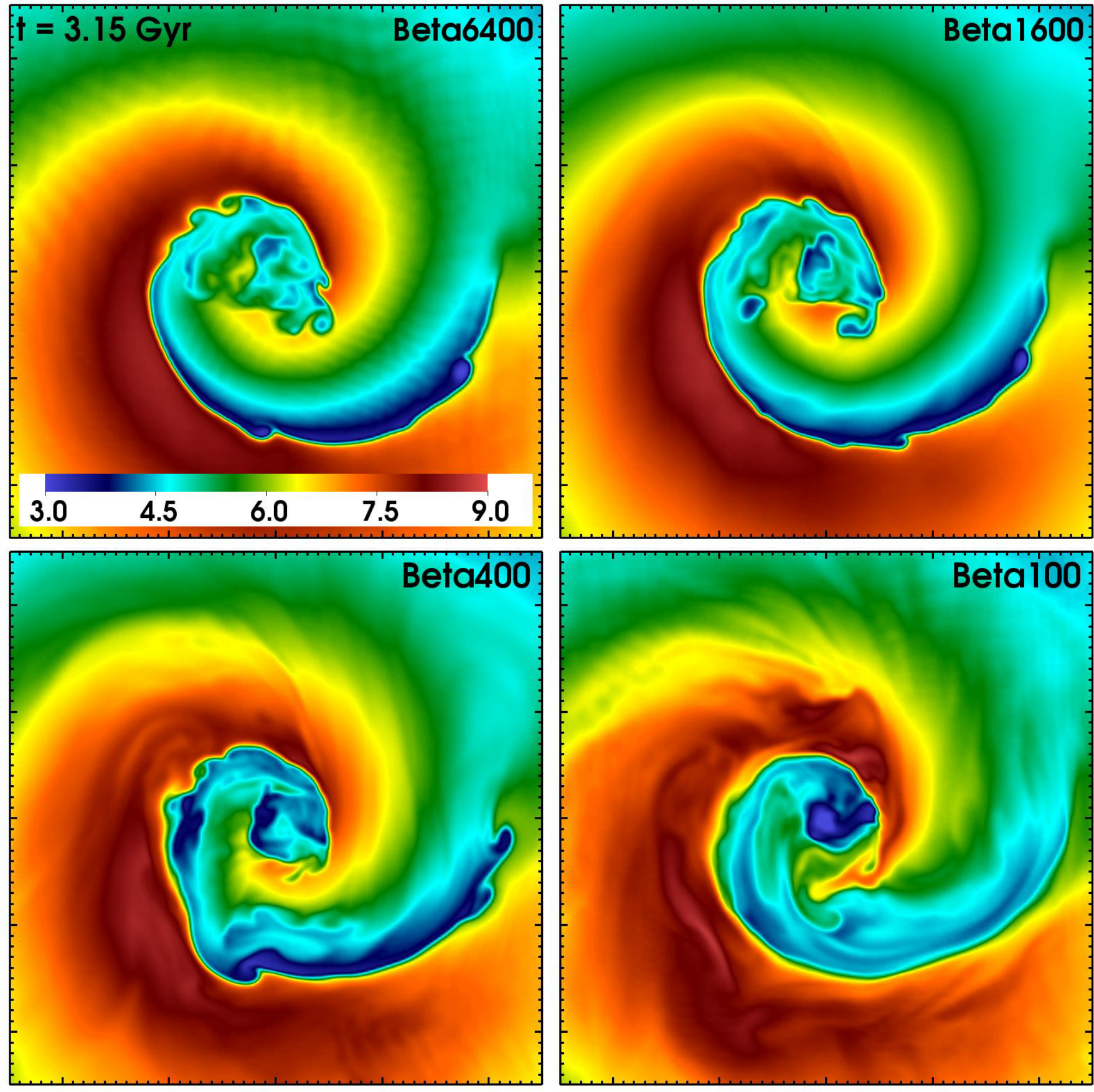}
\caption{Slices through the gas temperature for the simulations from ZML11 with varying initial $\beta$. Each panel is 500 kpc on a side. Major tick marks indicate 100~kpc distances.\label{fig:diff_beta_temp}}
\end{center}
\end{figure}

\begin{figure}
\begin{center}
\includegraphics[width=0.96\textwidth]{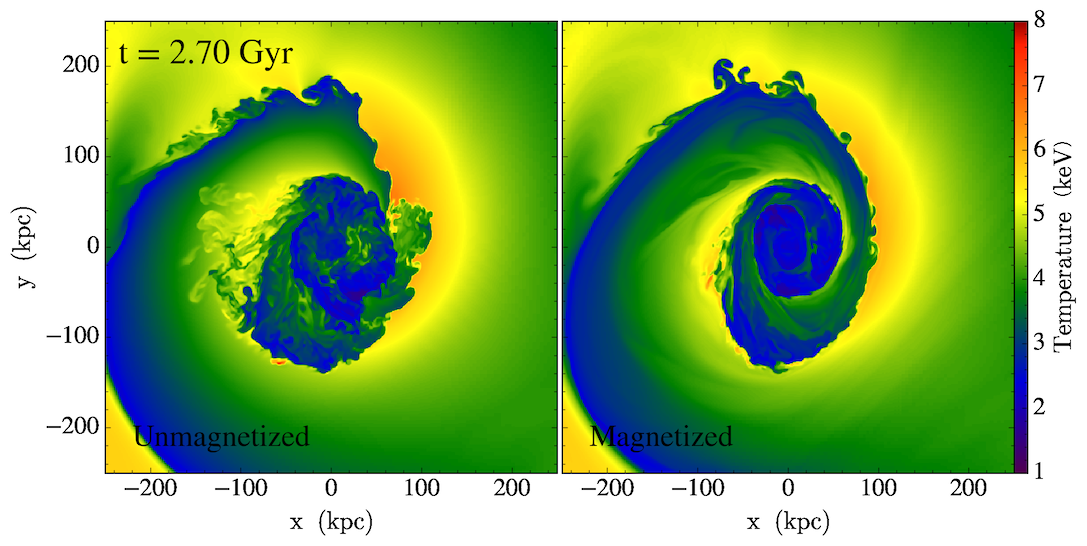}
\caption{Comparison of simulations of gas sloshing with and without magnetic fields. Panels show slices of gas temperature. In the magnetized simulation, $\beta \sim 100$. An animation of the sloshing cold fronts developing over $\sim$2~Gyr can be found at \url{http://vimeo.com/160770528}. KHI develop more readily in the unmagnetized simulation.\label{fig:compare_movie}}
\end{center}
\end{figure}

Though similar magnetic field layers to the remnant-core cold fronts are also expected to appear in sloshing cold fronts, they are also somewhat different. In the remnant-core case, the movement of the cold, dense gas through the surrounding ICM builds up a layer of amplified magnetic field on the {\it outside} of the interface \citep{lyu06}. In the case of sloshing cold fronts, in simulations we often find a velocity shear across the front surface due to the velocity difference between the fast moving cold gas beneath the front, moving tangential to the interface, and the slowly moving hot gas above the front \citep[see also][]{kes12}. This shear amplifies the magnetic field via line freezing, $d({\bf B}/\rho)/dt \simeq [({\bf B}/\rho) \cdot \nabla]{\bf u}$, on the {\it inside} of the interface \citep{kes10}. Figure \ref{fig:mag_movie} shows an animation of sloshing cold fronts using slices of gas density and magnetic field strength from an MHD simulation (ZuHone et al., in preparation).

The first simulations of sloshing cold fronts to include magnetic fields were those of \citet[][hereafter ZML11]{zuh11}. They performed a parameter space exploration over the initial magnetic field configuration. The initial magnetic field was either turbulent or purely tangential, and the average field strength and its coherence length scale were varied. The former was set by enforcing the $\beta$ parameter to be the same everywhere. This produces a declining magnetic field radial profile which is consistent with existing observations \citep{bon10}.

In these simulations, the shear flows initiated by the gas sloshing produce highly magnetized (low-$\beta$) layers just underneath the cold front surfaces. Figure \ref{fig:diff_beta_beta} shows slices of the plasma $\beta$ for the simulations with varying magnetic field strength. These layers are produced in all of the simulations, but the strength of the layer depends on the strength of the initial field. For example, the bottom-right panel of Figure \ref{fig:diff_beta_beta} shows a magnetic layer in the cold front to the southwest which is nearly in equipartition with the thermal pressure, but the corresponding layers in the simulations with initial weaker field (increasing $\beta$) are not nearly as strong. Roughly, the magnetic pressure in the layers is increased by an order of magnitude over the initial field. These strongly magnetized layers are transient, only occurring directly underneath the front surface as the fronts expand. Within the increasing volume defined by the expanding fronts, the magnetic field that develops in the core behind the front is stronger than the initial field by a factor of a few. This field is typically very turbulent, except exactly below the front surface.

Figure \ref{fig:diff_beta_temp} shows slices through the gas temperature for the same simulations at the same epoch. For simulations with stronger initial field, the magnetic field in the layers becomes stronger, and more capable of suppressing KHI. Figure \ref{fig:compare_movie} shows a side-by-side animation of gas sloshing in an unmagnetized vs. a magnetized medium (ZuHone et al., in prep), showing again that the magnetic field suppresses the growth of KHI considerably compared to a case where there is no magnetization. For essentially complete suppression of KHI, the initial magnetic pressure must be at least roughly 1\% of the thermal pressure, or $\beta \simlt 100$. In contrast, ZML11 did not find that the suppression of KHI has a strong dependence on the coherence length of the magnetic field fluctuations or the magnetic field geometry (whether turbulent or tangential), provided that the field strength was the same (Figure 22 of ZML11).

However, they did find that these conclusions were somewhat dependent on resolution--for a simulation with initial $\beta$ = 100 and a finest resolution of 2~kpc, KHI is highly suppressed, but for an otherwise identical simulation with a finest resolution of 1~kpc, small-wavelength KHI were able to grow (Figures 38 and 39 of ZML11). This is likely due to a confluence of two factors: the decreased numerical viscosity associated with higher spatial resolution, and the decrease in the wavelength at which KHI can be suppressed due to the thinning of the magnetized layer which occurs with decreased resolution \citep[see][]{dur07}. Given these facts, and the uncertainties associated with the magnetic field strength in clusters, it is uncertain that magnetic fields provide the observed suppression of KHI in all clusters where such suppression is observed.

\begin{figure}
\begin{center}
\includegraphics[width=0.48\textwidth]{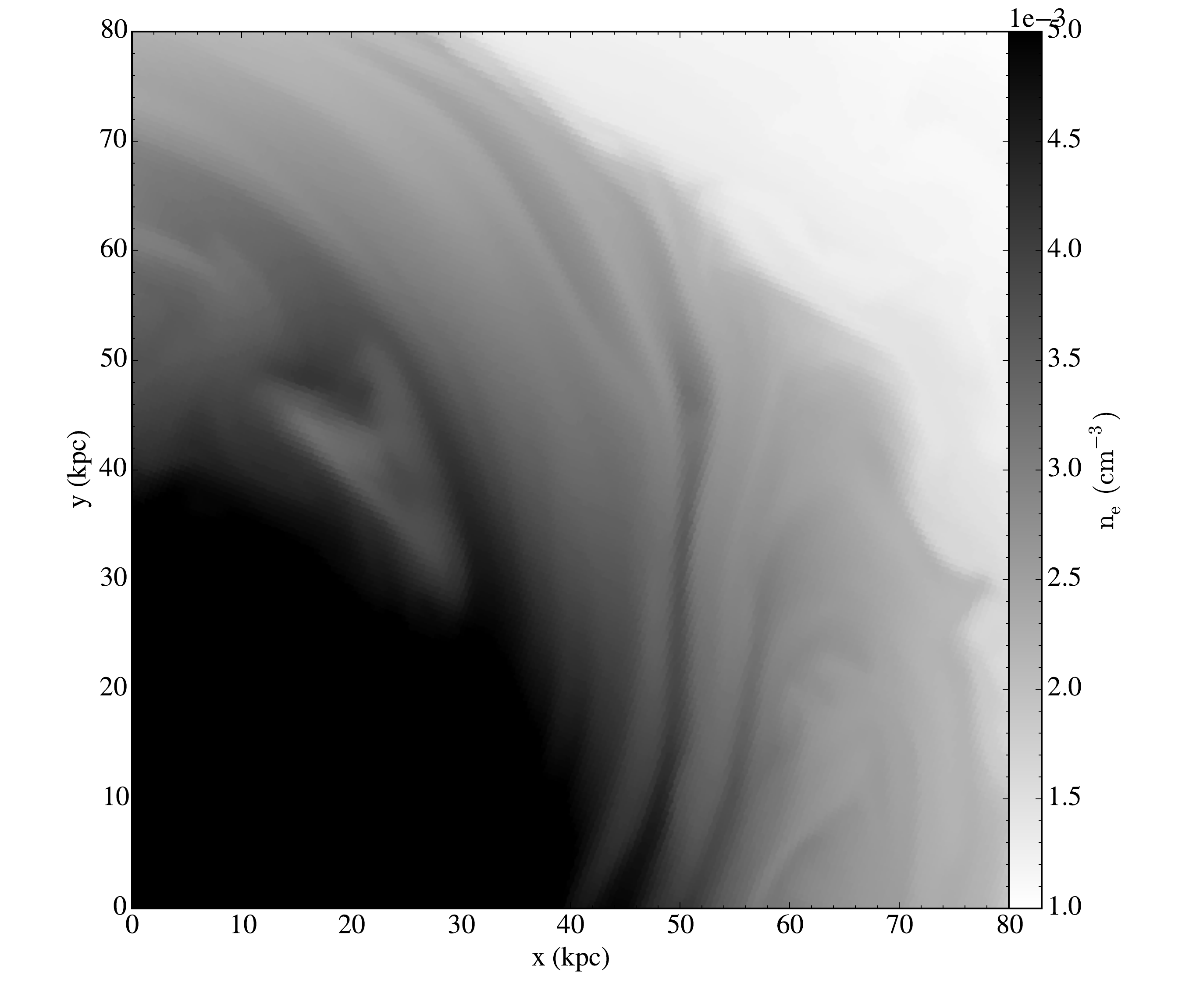}
\includegraphics[width=0.48\textwidth]{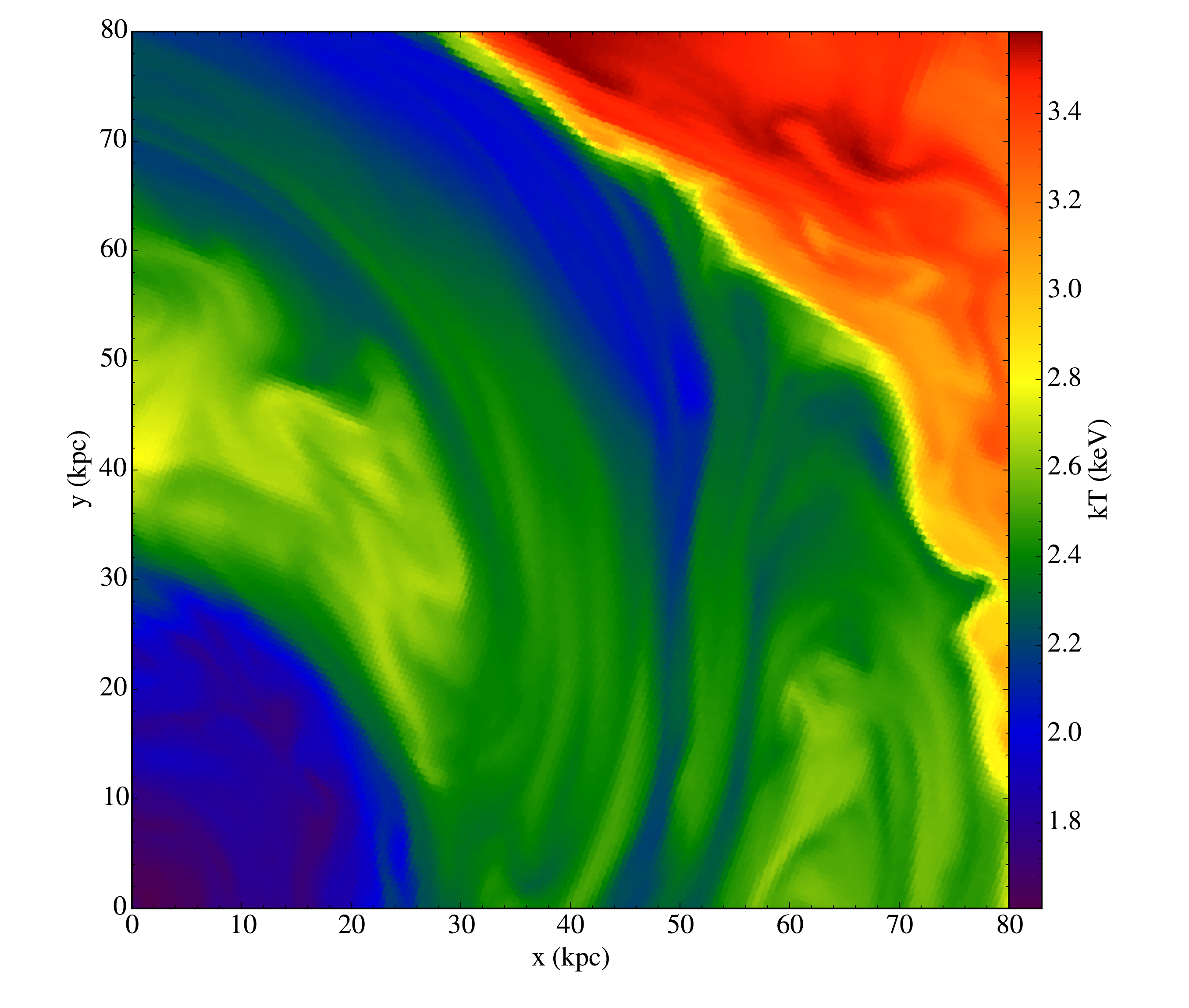}
\includegraphics[width=0.48\textwidth]{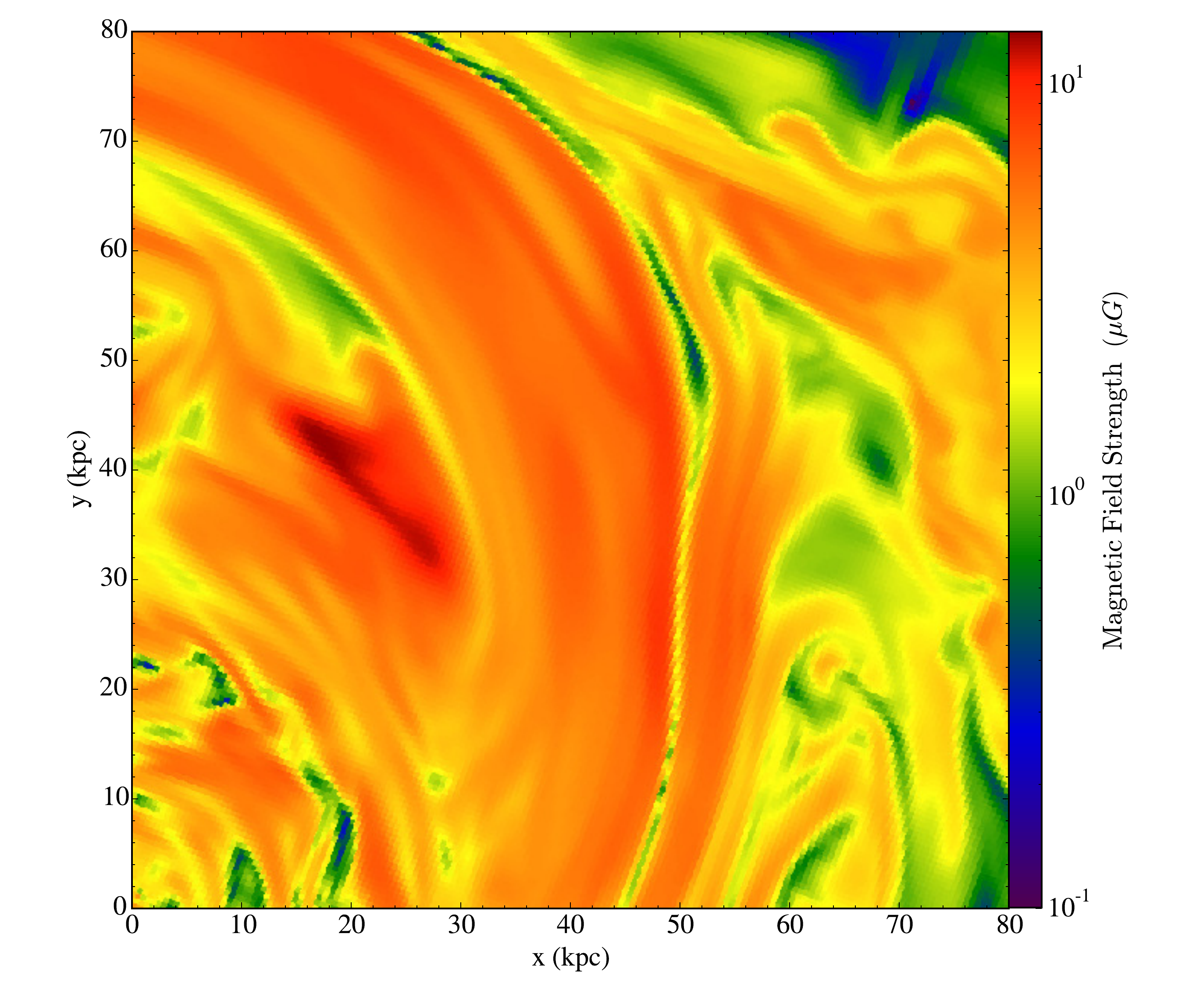}
\includegraphics[width=0.48\textwidth]{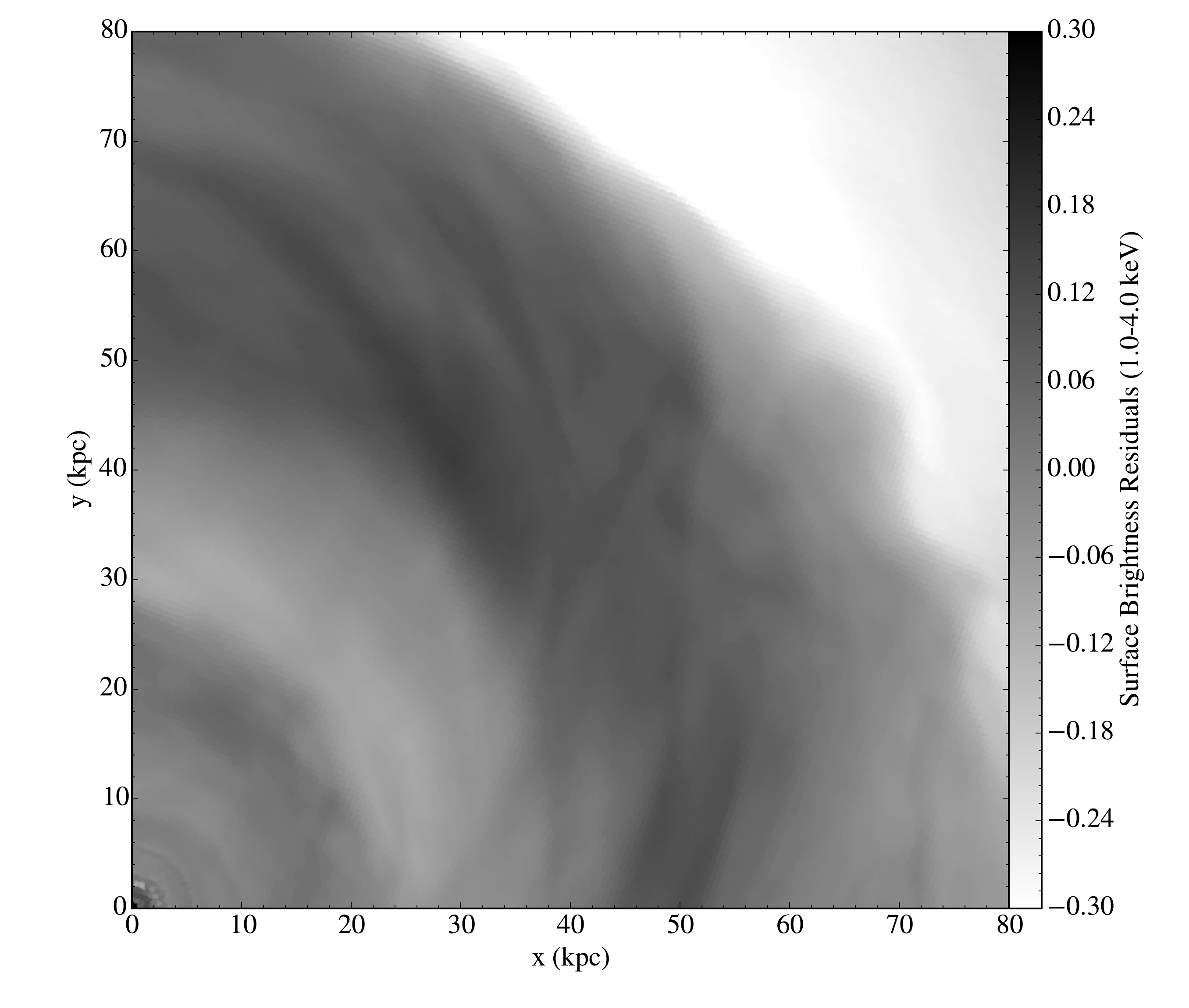}
\caption{Sloshing in a Virgo-like simulated cluster from \citet{wer16}. Top left: Slice through the center plane of electron density. Top right: Slice of gas temperature. Bottom left: Slice of magnetic field strength. Bottom right: Projected surface brightness residuals. \label{fig:virgo_sim_plots}}
\end{center}
\end{figure}

Magnetic fields may also explain the peculiar linear features seen in the observations of the Virgo cold front by \citet{wer16} (Figure \ref{fig:virgo_chandra}). Figure \ref{fig:virgo_sim_plots} shows slices of density, temperature, and magnetic field strength, and residuals of surface brightness for a simulation designed to reproduce the Virgo cluster cold fronts (based on the simulations from \citet{zuh15a}, using the initial conditions from R12. The initial magnetic field strength for the simulation was set by $\beta \sim 100$. The images show a region of the cluster approximately located at the same place where the observations of \citet{wer16} were taken. They show narrow, quasi-linear, dense, cold features, which correlate to narrow channels of weak magnetic field in between wide bands of strong magnetic field which are less dense and hotter. The dense features produce surface brightness features similar to those seen in Figure \ref{fig:virgo_chandra}.

Since the magnetic field in the simulation is turbulent before the sloshing begins, the local magnetic field strength will vary from place to place. When the sloshing motions begin, the field is stretched and amplified in the direction of the sloshing, which erases much of the small-scale structure in the field. However, we may expect that perpendicular to the direction of motion, some memory of the variations in magnetic field strength will be preserved. Since the simulations demonstrate that the final amplified field strength does depend to some extent on the initial field strength, these variations will produce regions of slightly weaker magnetic field sandwiched between regions of stronger field. The total pressure across these bands (magnetic and thermal) will be continuous.

The narrow regions of slightly weaker field are correlated with regions of higher density and lower temperature compared to the surrounding regions with higher magnetic field strength. The anticorrelation between density and temperature is somewhat peculiar, as the simulations exhibiting these features are adiabatic (without sources of heating or cooling), and for an individual fluid element the entropy should be constant. This implies that a decrease in the gas thermal pressure due to an increase the magnetic pressure in the wide bands should correspond to a decrease in both the gas density and temperature in these bands, yet the temperature is higher. This may be explained by low-entropy, low-temperature, high-density gas being advected from the cluster center to larger radii by the sloshing, but it is not completely clear why such gas should be associated with regions of low magnetic field strength. In real clusters, the density enhancement and temperature decrement will be enhanced by gas cooling. A more careful investigation of this phenomenon from simulations is required.

\begin{figure}
\begin{center}
\includegraphics[width=0.7\textwidth]{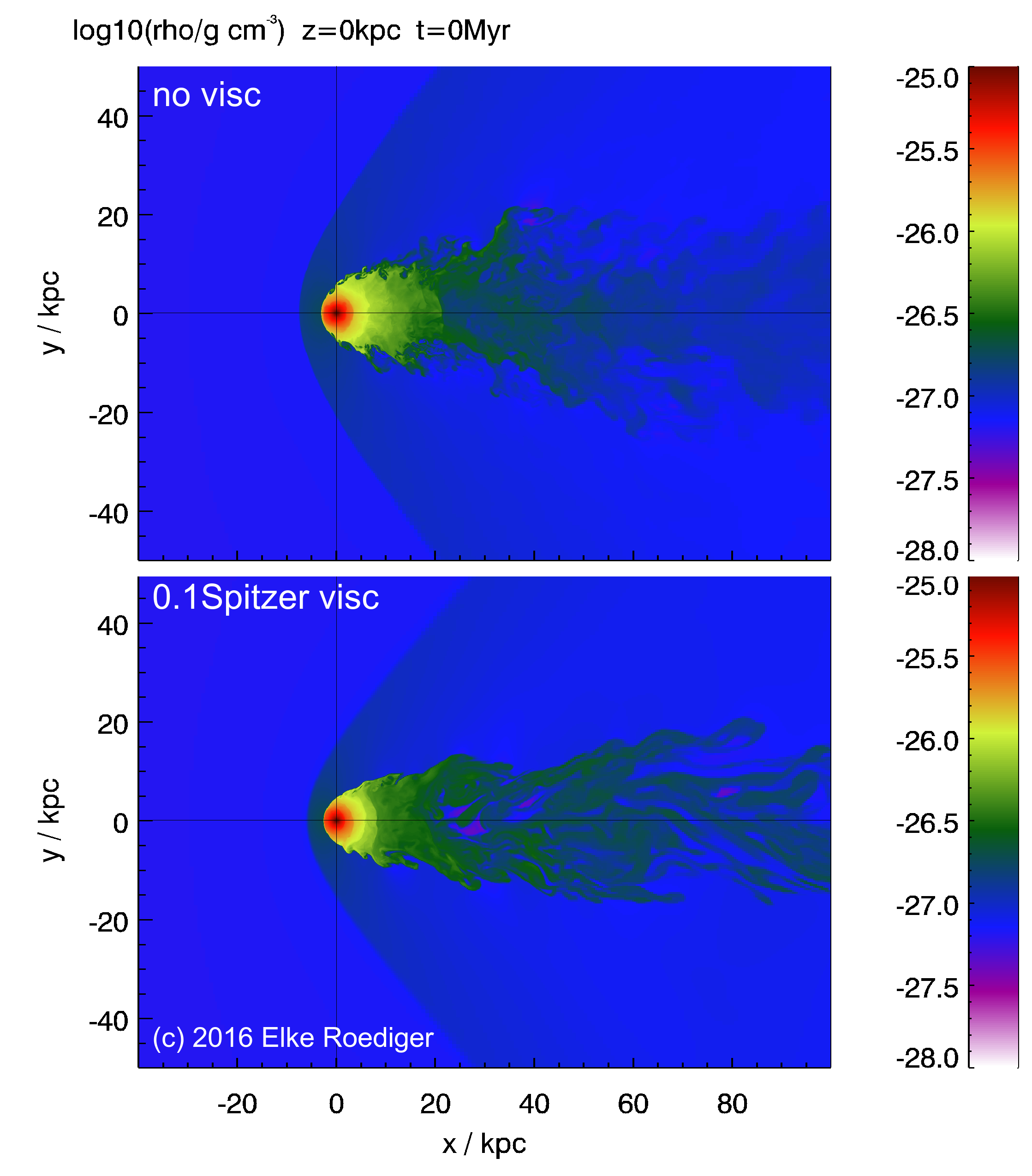}
\includegraphics[width=0.7\textwidth]{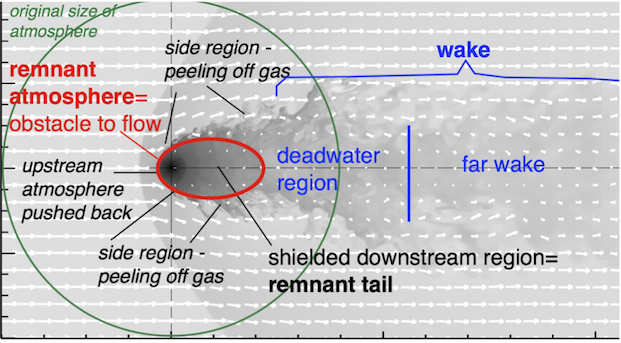}
\caption{Snapshots from simulations of progressive gas-stripping of an elliptical galaxy, from R15A and R15B. The two top panels and the associated animation found at \url{http://vimeo.com/160773185} show density slices through the stripped galaxy for inviscid stripping and with a viscosity of 0.1 of the Spitzer level. During the infall into the host cluster, characteristic flow regions develop, which are labelled in the bottom panel. A remnant-core cold front exists at the upstream side, stretching around to the sides of the atmosphere. Here the interface between the galactic gas and the ICM becomes KHI unstable at sufficiently low viscosity. The downstream atmosphere is shielded from the ambient flow for a long time and takes the appearance of a remnant tail. Only in the wake can stripped gas mix with the ambient gas, unless the mixing is suppressed by viscosity.\label{fig:elkeRPS1}}
\end{center}
\end{figure}

\subsection{Cold Fronts and Viscosity}\label{sec:viscosity}

Small-scale KHIs will be suppressed if the shear velocity across the interface does not change discontinuously, but is smoothed over a certain thickness $\pm d$. In this case, KHIs of length scale $\simlt 10d$ are suppressed \citep[][Section 102]{chandrasekhar}. In a viscous fluid, momentum diffusion across a shear flow interface will establish exactly such a smoothed out shear flow layer. Thus, we can naturally expect that with increasing viscosity KHIs are suppressed at ever larger length scales. Assuming an isotropic viscosity, \citet{rod13b} showed that the suppression of KHIs occurs roughly at Reynolds numbers ${\rm Re} = LU/\nu$ smaller than 30 to 100, where $L$ is the length scale of the KHI, $U$ the shear velocity, and $\nu = \mu/\rho$ is the kinematic viscosity. The detailed value of the critical Reynolds number depends on the nature of viscosity (e.g., constant versus temperature dependent $\nu$) and density contrast across the shear interface.

At full Spitzer viscosity (Equation \ref{eqn:spitzer_viscosity}), the critical Reynolds number for suppressing KHIs is easily reached for both remnant core and sloshing cold fronts.  Furthermore, a sufficient ICM viscosity prevents the mixing of gas stripped from the infalling subcluster or galaxy with the ambient ICM in the wake (\citealt{rod15b}, hereafter R15B). Recent deep {\it Chandra} and {\it XMM-Newton} observations resolve the structure of the cold fronts and of galaxy or subcluster wakes. However, the interpretation of the observations on their own is not straightforward because the appearance of individual objects depends on several other parameters such as the infall and/or merger history, viewing geometry, initial gas contents, or internal perturbations due to, e.g., AGN activity.

R15A and R15B simulated the infall of the elliptical galaxy M89 into the Virgo cluster and followed the progressive gas stripping process as the galaxy is crossing its host cluster. The authors showed that progressive gas stripping of a galaxy or subcluster leads to the formation of a ``remnant tail'' of the atmosphere (Figure \ref{fig:elkeRPS1}). Gas from the galaxy’s atmosphere is removed predominantly upstream and from its sides, but the downstream part of the atmosphere is shielded from the ICM head wind (see animation in Figure \ref{fig:elkeRPS1}). Consequently, the stripped atmosphere develops a head-tail structure, where the tail is simply the unstripped downstream atmosphere. Making the analogy to the flow around a blunt body, the wake of the galaxy starts only downstream of the remnant tail. These global flow patterns, i.e., the upstream edge or cold front of the remnant core, its remnant tail, and the downstream wake in the ICM are independent of viscosity.

\begin{figure}
\begin{center}
\includegraphics[height=6cm]{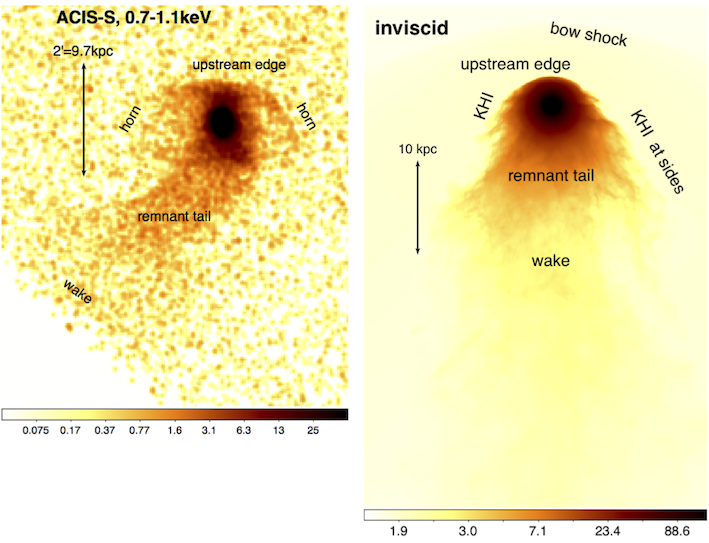}
\includegraphics[height=6cm]{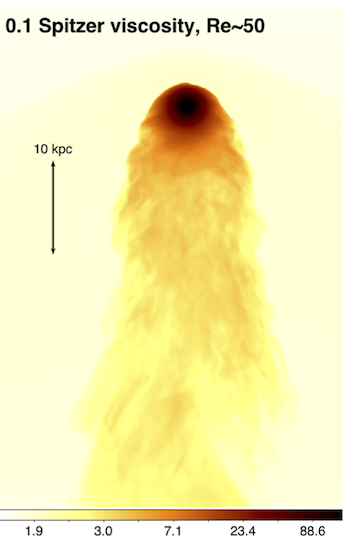}
\caption{{\it Chandra} X-ray image of the stripped elliptical galaxy M89 in the Virgo cluster (left), compared to mock X-ray images of inviscid and a viscous stripping simulations for the same galaxy (middle and right panels). The observation and simulations clearly show the remnant-core cold front, labelled ``upstream edge'' here. Comparing the simulations and the observations identifies the observed near tail of M89 as the remnant tail, and its wake is only found at the edge of the observed field of view. In the inviscid simulation, KHIs peel off galactic gas from the sides of the remnant core. If the Virgo ICM has a viscosity roughly 0.1 of the Spitzer viscosity, KHIs at the sides of M89 would be suppressed, and it should have a cold bright wake out to 10 times the remnant-core radius downstream of the galaxy. Reproduced from R15B.\label{fig:elkeRPS2}}
\end{center}
\end{figure}

R15B focused on observable signatures of an isotropic but Spitzer-like temperature dependent viscosity (Figure \ref{fig:elkeRPS2}). In inviscid stripping, KHIs at the sides of the remnant atmosphere take the shape of horns or wings that are observable in X-ray images, and mixing of stripped colder galactic gas leads to fainter, warmer wakes. Interestingly, already at a level of 1/10 the Spitzer viscosity, KHIs are suppressed, horns or wings at the sides of the remnant core are absent, and the stripped-off colder and denser galactic gas leads to bright, cool, wakes far behind the galaxy, implying that the viscosity is low in systems such as M89 which possess evidence of ``horned'' cold fronts (Figure \ref{fig:elkeRPS2}).

\begin{figure}
\begin{center}
\includegraphics[width=0.98\textwidth]{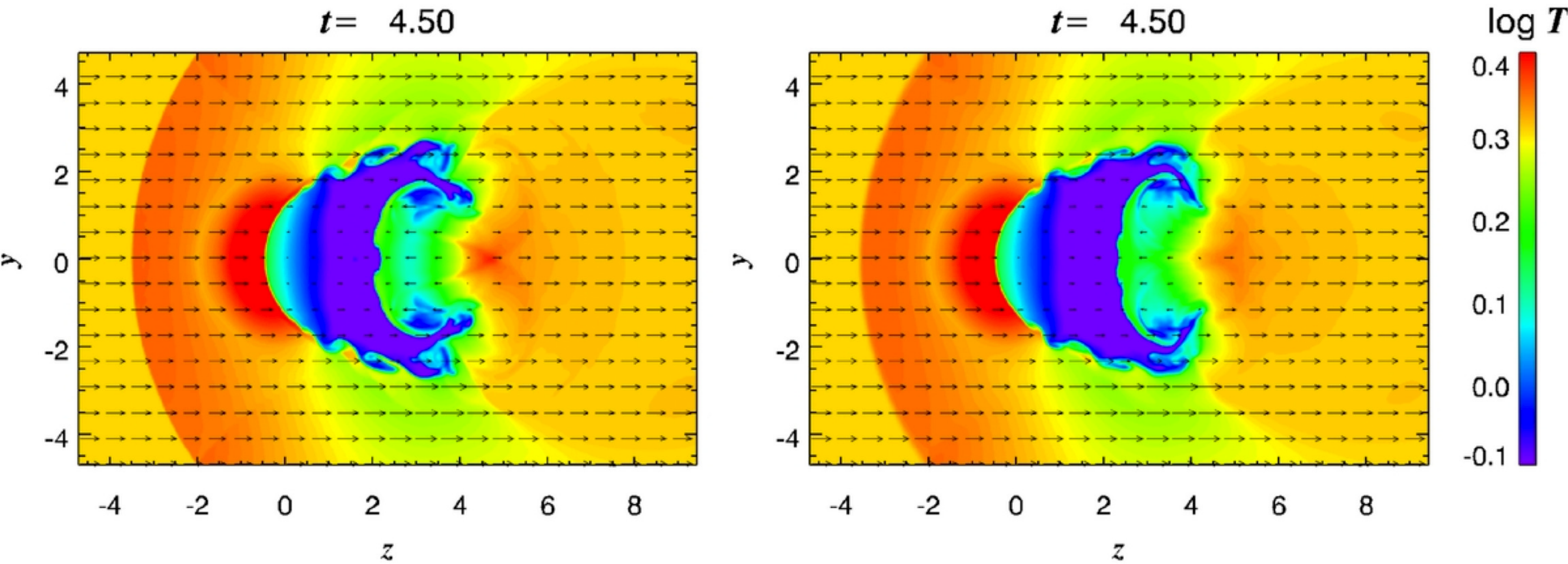}
\includegraphics[width=0.98\textwidth]{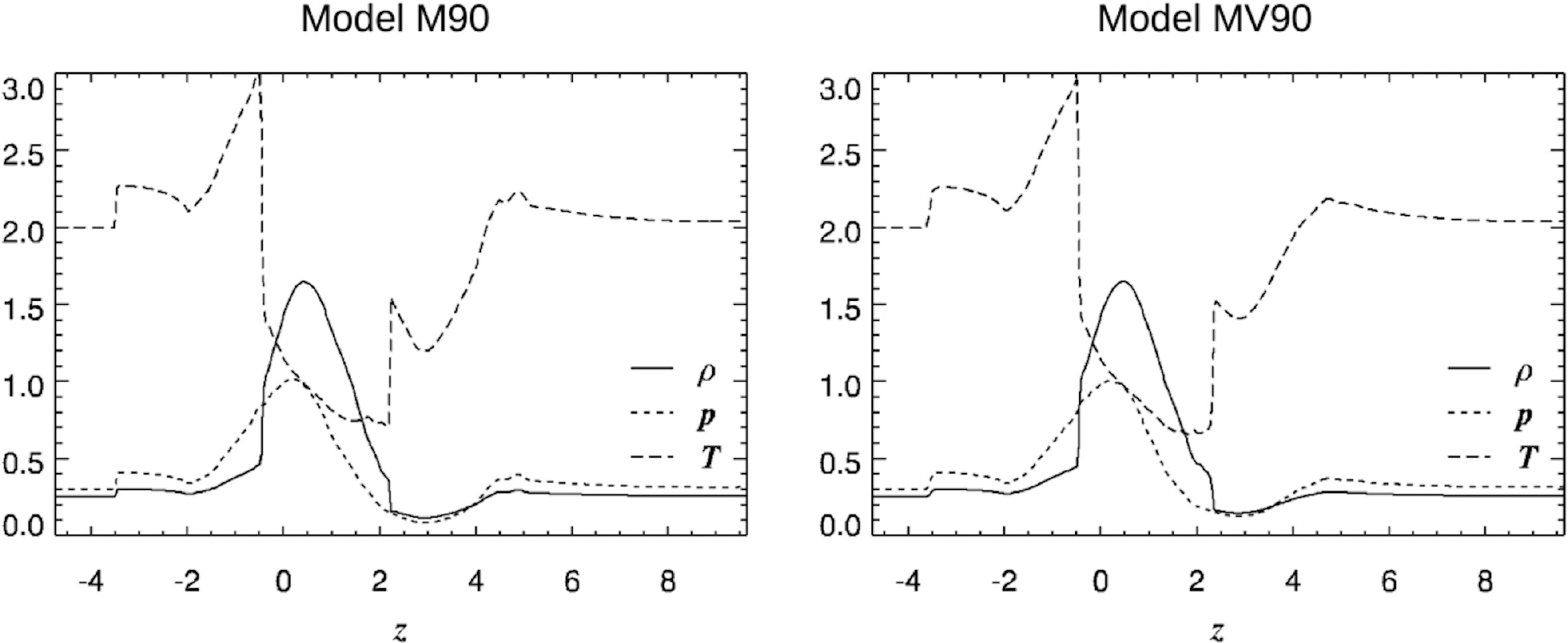}
\caption{The effects of Braginskii viscosity on remnant-core cold fronts, from S13. Top panels: Slices through the gas temperature in the $y-z$ plane from the inviscid (left) and Braginskii viscosity (right) simulations. In these units, 1 = 250~kpc. Bottom panels: Profiles of the gas density and temperature across the cold front interface, from the inviscid (left) and Braginskii viscosity (right) simulations. The cold front interface is located at $z \approx -0.5~(-125~{\rm kpc})$. The growth of KHI in both simulations is similar, demonstrating that in this scenario Braginskii viscosity has only a marginal effect on suppressing KHI.\label{fig:suzuki2013_visc}}
\end{center}
\end{figure}

\begin{figure}
\begin{center}
\hfill\includegraphics[width=0.45\textwidth]{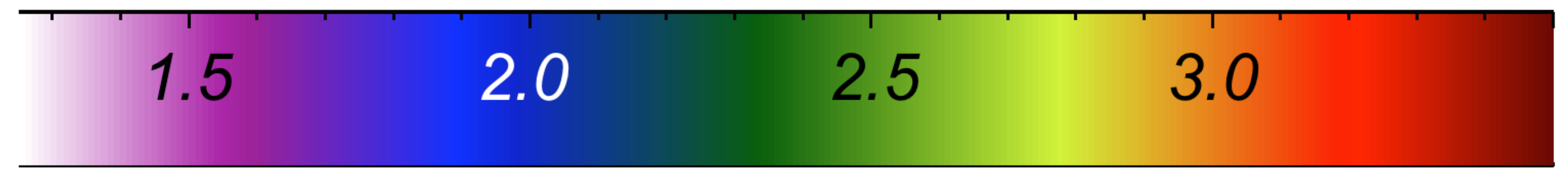}
\hfill\hfill\includegraphics[width=0.54\textwidth]{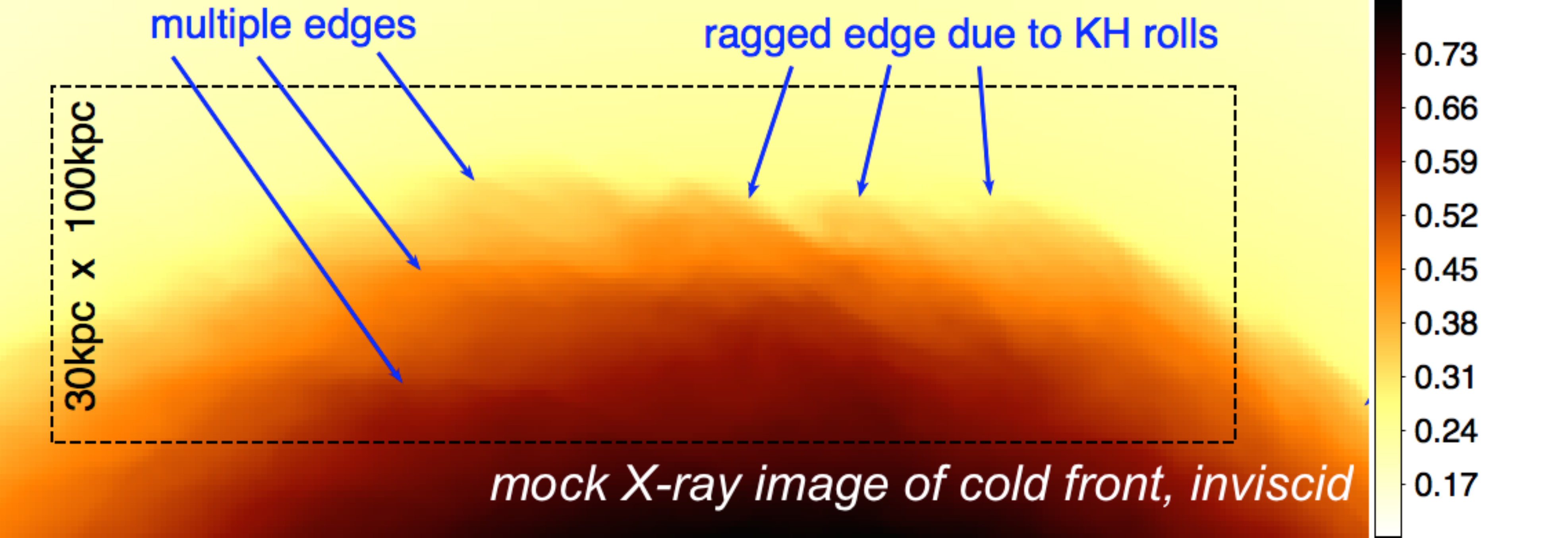}
\includegraphics[width=0.49\textwidth]{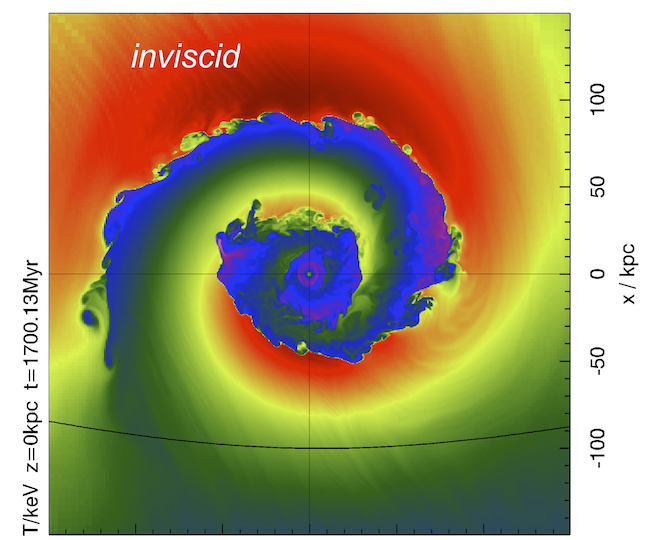}
\includegraphics[width=0.49\textwidth]{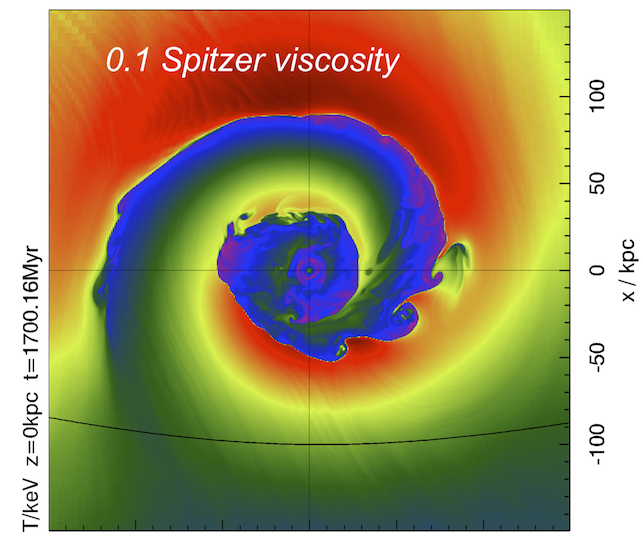}
\includegraphics[width=0.48\textwidth]{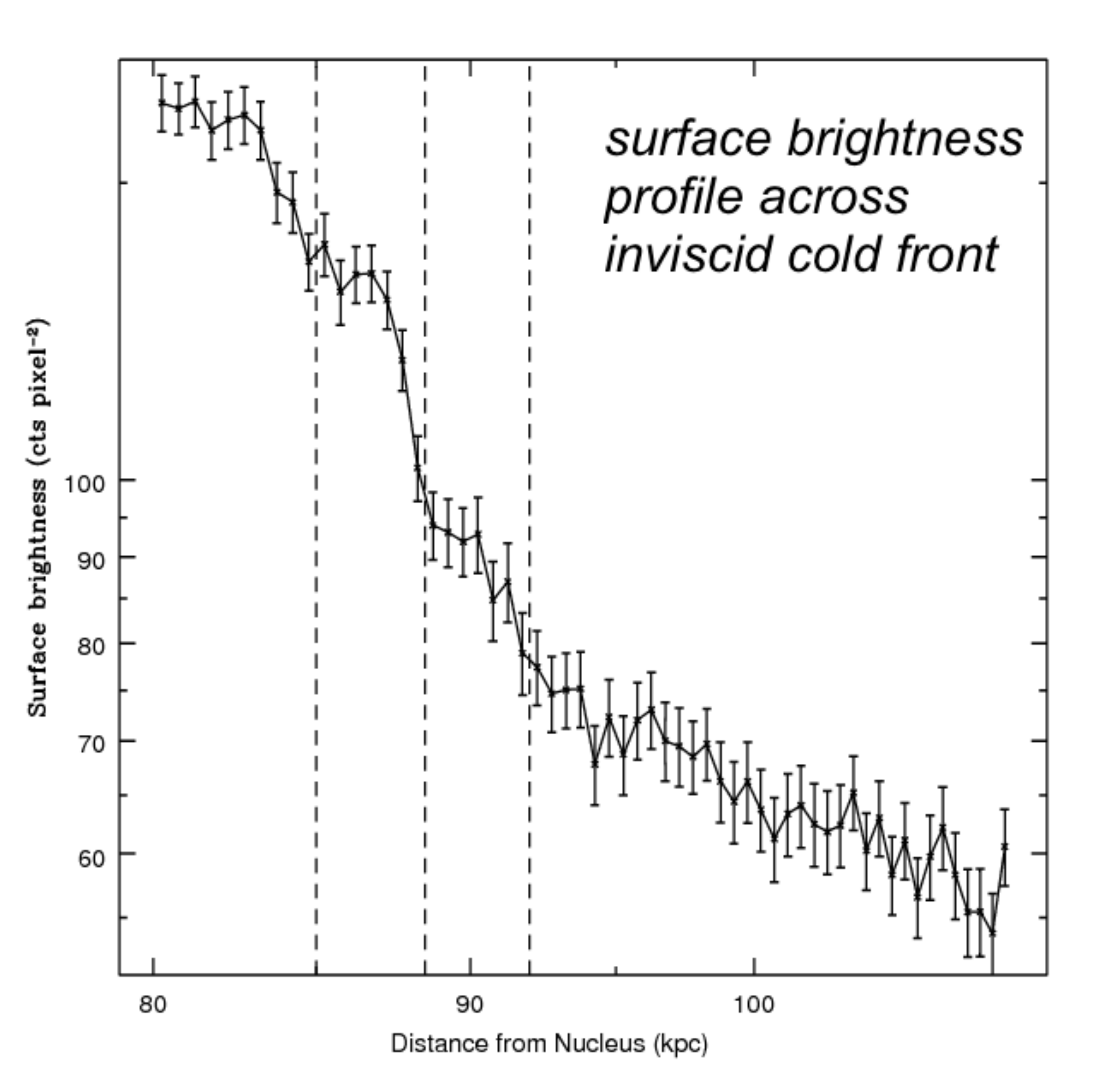}
\includegraphics[width=0.48\textwidth]{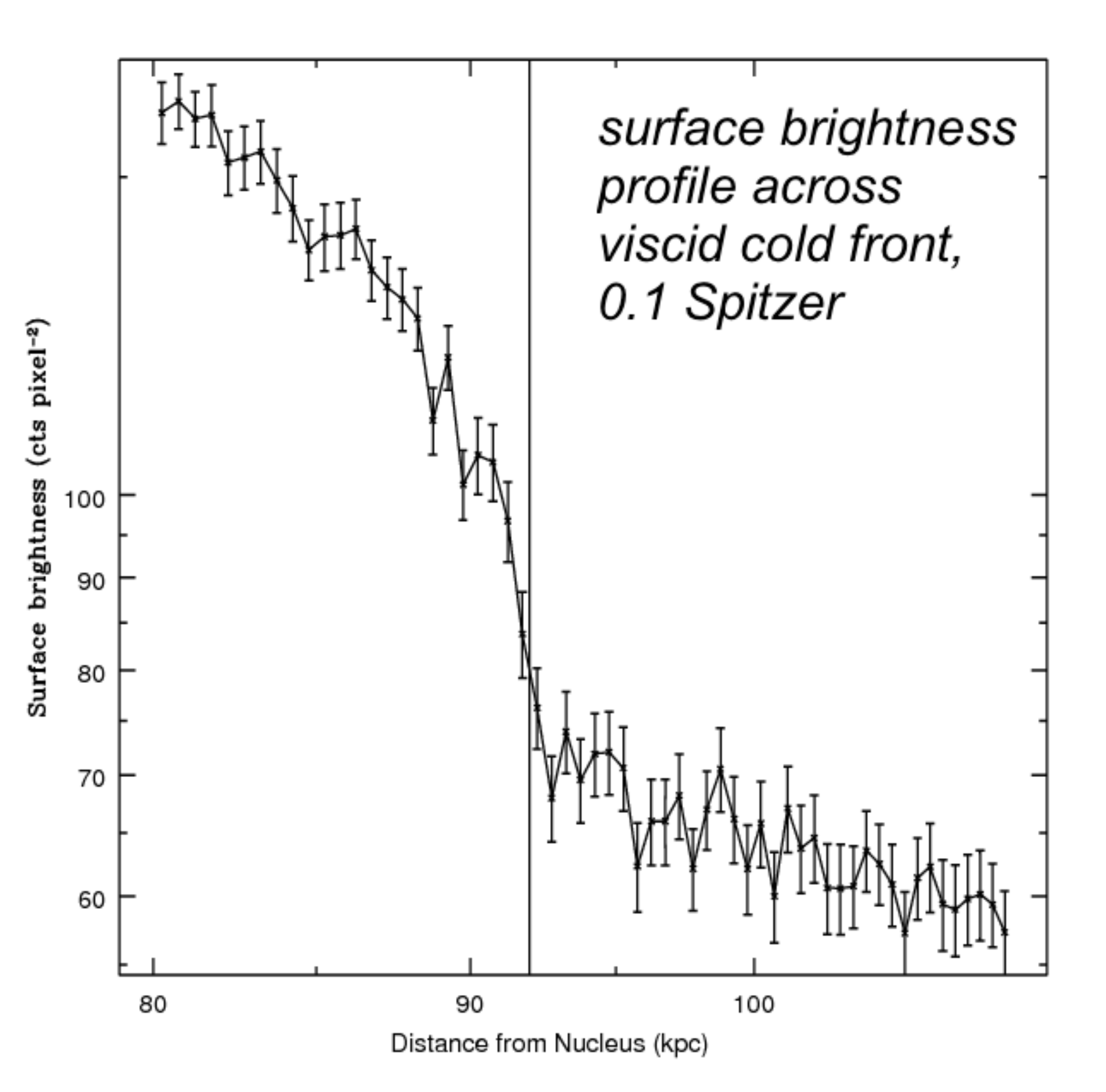}
\caption{Snapshot from a simulation of inviscid and viscid sloshing in Virgo cluster, from R13. An animation of the middle two panels may be found at \url{http://vimeo.com/160772084}. The middle row shows temperature slices through the cluster center in the orbital plane of the perturber. A viscosity at 0.1 of the Spitzer level erases pre-existing KHIs along the northern front, which continued to grow and form in the inviscid run. The top-right panel shows a mock X-ray image of the inviscid sloshing, zoomed in on the northern cold front. In X-ray images, KHIs appear as multiple edges and give the cold front a ragged appearance. The multiple edges also appear in narrow-wedge surface brightness profiles across the cold front (three edges are marked in the profile in the bottom-left panel). At a sufficient viscosity, however, the cold front appears as a single edge in the surface brightness profile (bottom-right).\label{fig:elkeSlosh1}}
\end{center}
\end{figure}

\citet[][hereafter S13]{suz13} carried out the first investigation of remnant-core cold front evolution under the influence of an anisotropic (Braginskii) viscosity. They used weak magnetic fields so that their effect on KHI suppression would be negligible. The magnetic fields were uniform in direction, oriented either perpendicular to the axis of motion of the cold front, or inclined by a 45$^\circ$ angle. They showed that anisotropic viscosity is not as efficient as an isotropic viscosity at suppressing KHIs because the damping effect of viscosity is reduced by the dependence on the field line direction (see Figure \ref{fig:suzuki2013_visc}). In the plane parallel to the initial field line direction, KHI grows very fast, since the field lines are perpendicular to the velocity gradient at the interface. Conversely, in the plane perpendicular to the initial field line direction, the growth rate of KHI was somewhat reduced, due to the fact that the field line had components parallel to the velocity gradient. A similar dependence on the initial field direction on the growth of KHI at the surfaces of AGN-blown bubbles was found by \citet{don09}, who also included Braginskii viscosity in their simulations. This points to the need for future simulations to include more realistic magnetic field configurations.

The first simulation of sloshing cold fronts with an isotropic viscosity was carried out by \citet{zuh10}, who studied how sloshing motions in a galaxy cluster provide a source of heat to offset cooling via mixing of higher-entropy gas from larger radii with the lower-entropy gas from the core. They compared simulations without viscosity and with a constant kinematic viscosity which roughly corresponded to the Spitzer value in the core region. They found that gas mixing is very efficient if the ICM is inviscid, whereas it is negligible if the plasma is viscous. They noted that the viscosity greatly suppressed KHIs along the cold fronts.

\begin{figure}
\begin{center}
\includegraphics[width=0.50\textwidth]{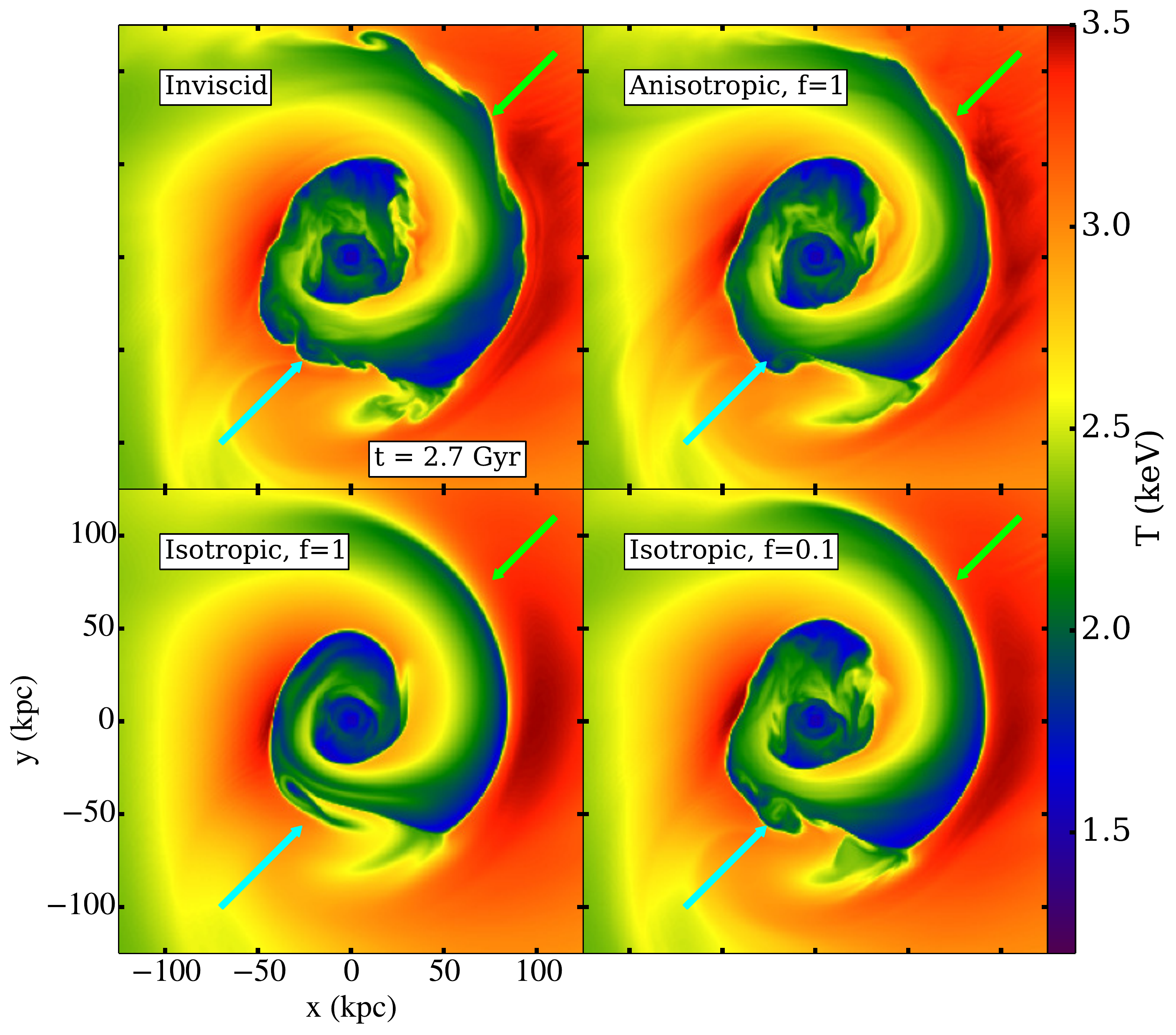}
\includegraphics[width=0.45\textwidth]{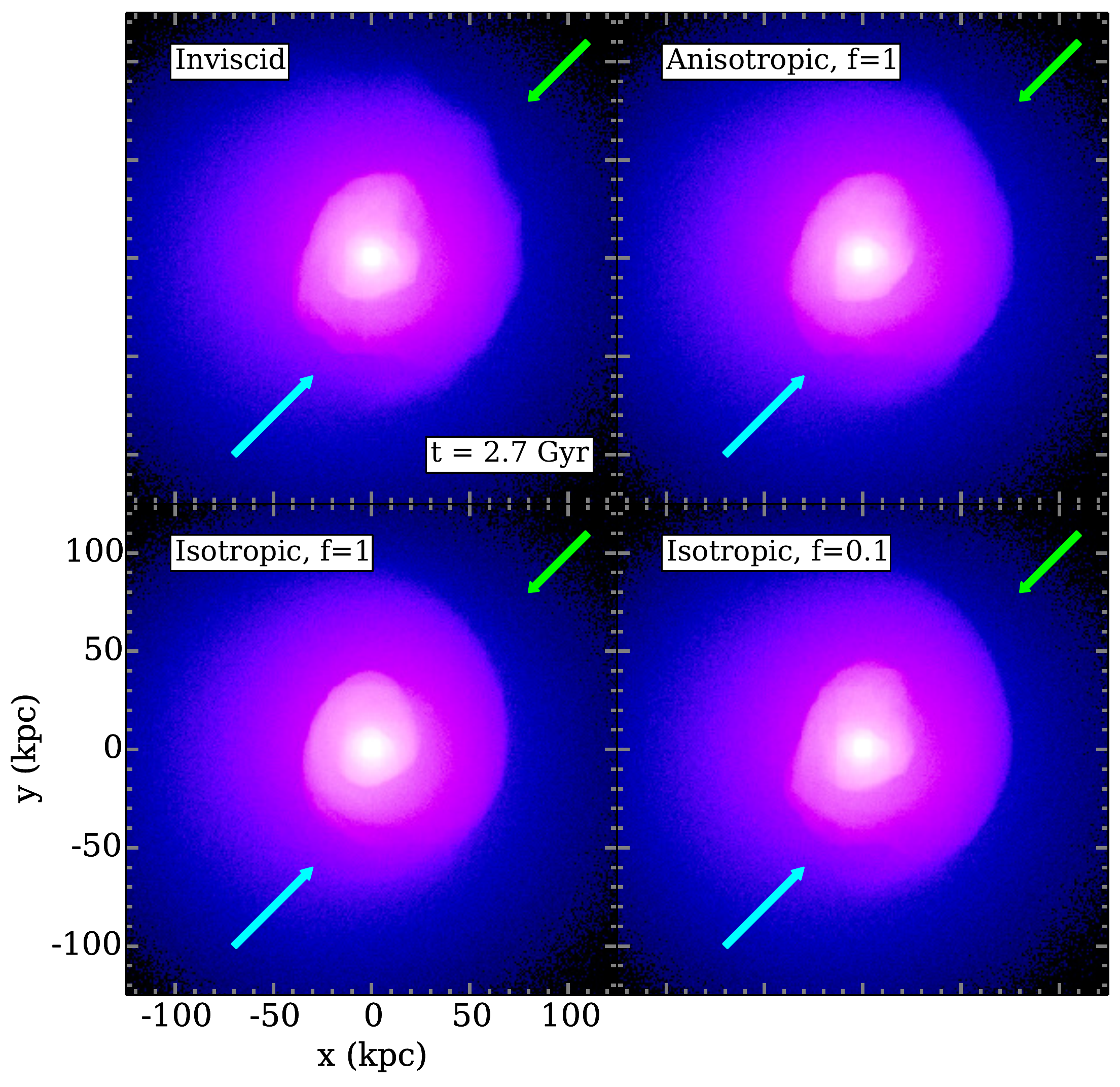}
\caption{Simulations of the Virgo cold fronts with different prescriptions for viscosity, including anisotropic (Braginskii) viscosity. From top-left counterclockwise, simulations are: invsicid, Spitzer anisotropic viscosity, Spitzer isotropic viscosity, 1/10~Spitzer isotropic viscosity. Left panel: Slices through the gas temperature in keV. Right panel: Simulated 300~ks {\it Chandra} observations. Cold fronts with an appearance that has been visibly modified by viscosity are marked with arrows; for a closeup of the NW cold front see Figure \ref{fig:virgo_closeup}.\label{fig:virgo_braginskii}}
\end{center}
\end{figure}

R13 showed that the \textit{presence} of KHIs at sloshing cold fronts leads to characteristic observable signatures such as a ragged appearance of the surface brightness edge and a multi-step structure of the cold front in surface brightness profiles (Figure \ref{fig:elkeSlosh1}). In contrast, suppressing KHIs with an isotropic, temperature-dependent, Spitzer-like viscosity with a suppression factor of $f_v \sim 0.1$, led to smooth sloshing cold fronts and a clean, single step edge in surface brightness profiles. Furthermore, the region below the cold fronts shows much stronger surface brightness fluctuations in the inviscid case than in the viscous one. The deep Virgo observations of \citet{wer16} show both surface brightness fluctuations below the cold front and a multi-step structure of the cold front itself, which may indicate the presence of KHIs and a strongly suppressed ICM viscosity in the Virgo cluster. Distorted or ``ragged'' sloshing cold fronts have also been identified in the merging groups NGC 7618/UGC 12491 (\citealt{rod12a}, Figure \ref{fig:n7618}) and in A496 \citep[][R12]{dup07}.

\begin{figure}
\begin{center}
\includegraphics[width=0.95\textwidth]{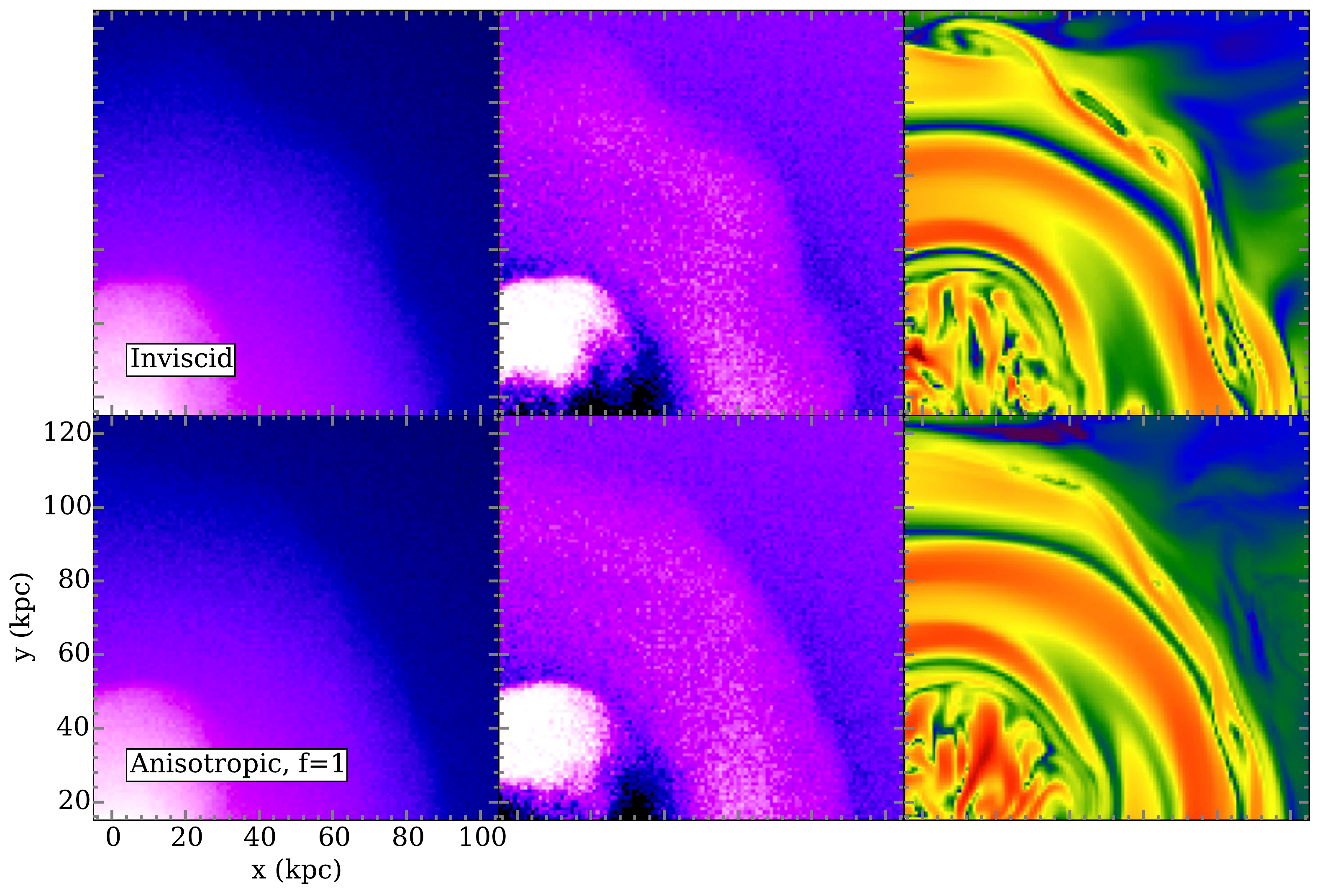}
\caption{Close-up of the NW cold front in Figure \ref{fig:virgo_braginskii} for simulations with inviscid and anisotropic (Braginskii) viscosity. The top panels show the invsicid simulation, whereas the bottom panels show the anisotropic viscosity simulation. From left to right, the panels show the simulated 300~ks exposure, simulated 300~ks residuals, and slice through the magnetic field strength. In the inviscid simulation, KHI distort the cold front and tangle the magnetic field lines.\label{fig:virgo_closeup}}
\end{center}
\end{figure}

The first simulations of sloshing cold fronts to employ anisotropic viscosity were those of \citet[][hereafter Z15]{zuh15a}. They compared magnetized simulations with anisotropic viscosity to those with isotropic viscosity with various suppression factors $f_v$. Similar to S13, they used weak magnetic fields (with $\beta \sim 1000$) to isolate the effect of viscosity on KHI. Using synthetic X-ray observations, they found that anisotropic viscosity is capable of suppressing KHI at cold front surfaces to a degree that is qualitatively similar to an isotropic viscosity with a Spitzer fraction of $f_v = 0.1$ (see Figures \ref{fig:virgo_braginskii} and \ref{fig:virgo_closeup}). They also compared simulations with isotropic viscosity with and without magnetic fields, and noted that even their weak seed field has effects on cold front stability that can yield observational consequences. This demonstrates that it will be essential for future studies to consider the effects of viscosity and magnetic fields together.

\subsection{Cold Fronts and Thermal Conductivity}\label{sec:thermal_cond}

Due to the small electron Larmor radius, thermal conduction in the ICM should be greatly reduced perpendicular to magnetic fields. Given that magnetic fields readily align with cold fronts, we expect thermal conduction across cold fronts to be greatly reduced. \citet{xia07} used pure-HD simulations of remnant-core cold front formation with isotropic thermal conduction and various suppression factors for the thermal conductivity to show that to reproduce the width of the cold front in A3667 the heat flux across the interface must be suppressed by a factor of $\sim$67. They also found that the thickness of the diffusive layer near the stagnation point of the cold front does not depend on the distance along the front from this point, implying that the front does not widen with distance from this point.

\begin{figure}
\begin{center}
\includegraphics[width=0.44\textwidth]{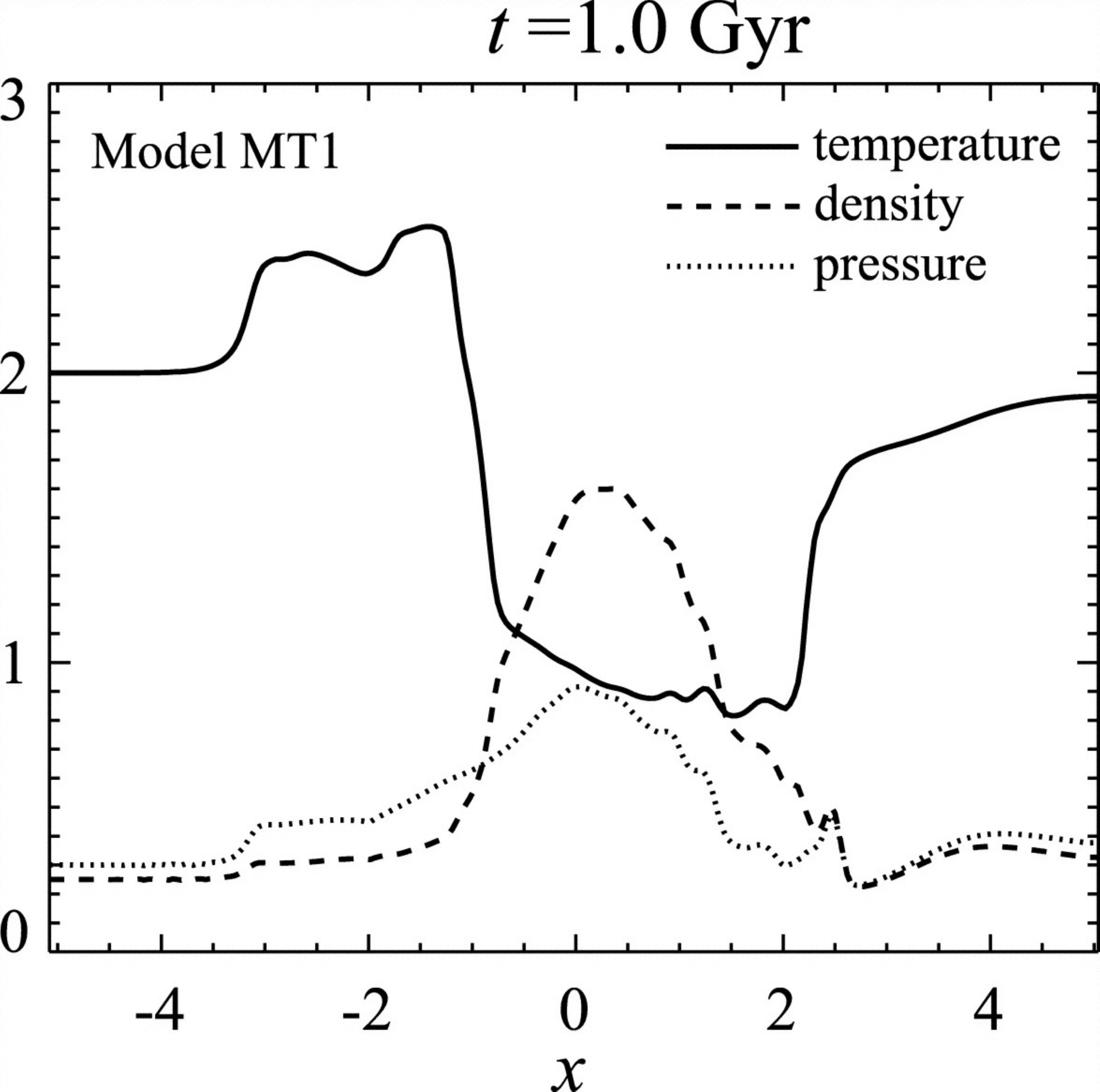}
\hspace{0.3in}
\includegraphics[width=0.44\textwidth]{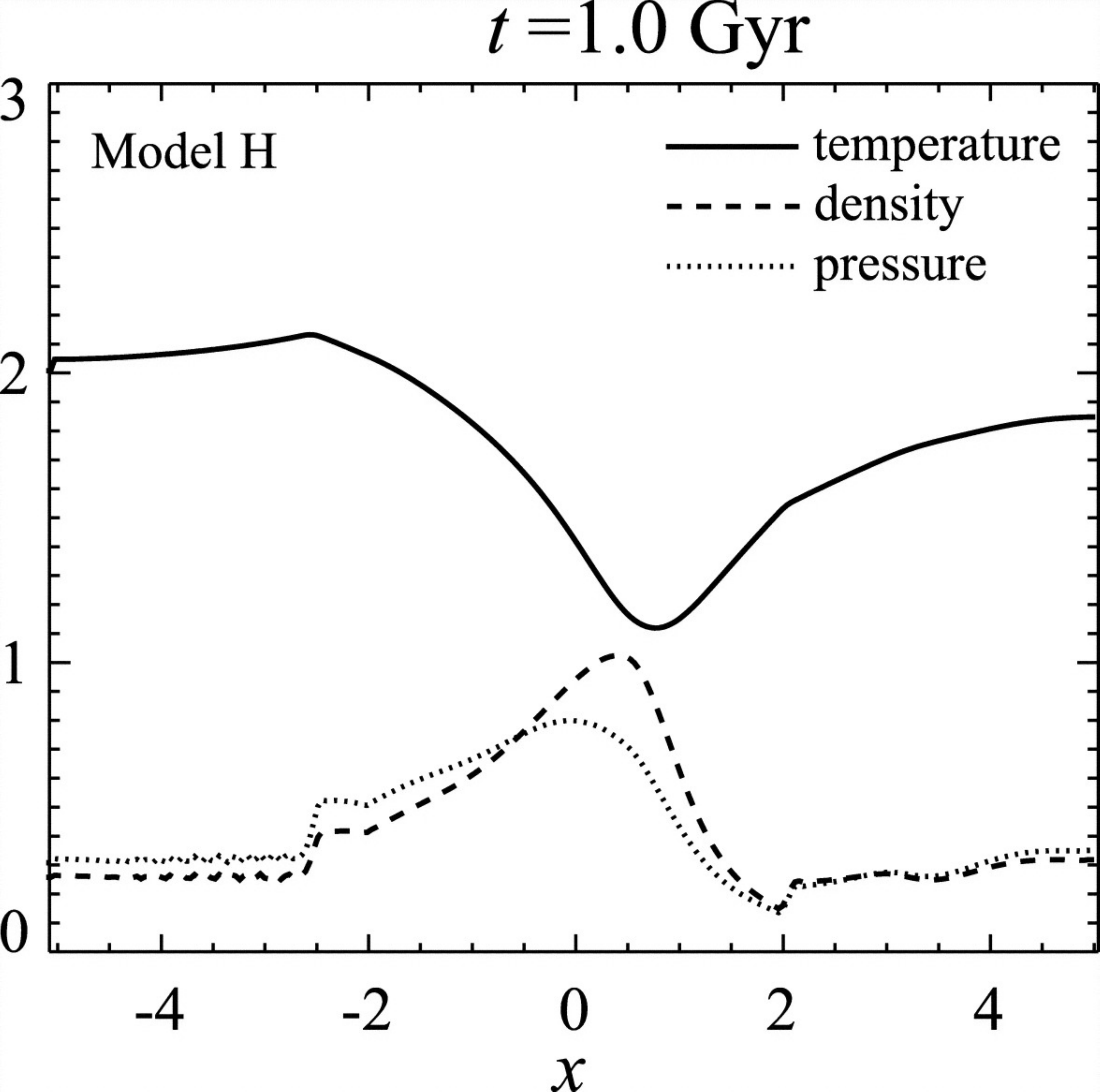}
\caption{Profiles of the gas density, temperature, and pressure across the cold front interface from the simulations of \citet{asa07}. Left panel: Profiles from the magnetically turbulent simulation with anisotropic thermal conduction. Right panel: Profiles from the unmagnetized simulation with isotropic thermal conduction. The cold front interface that is preserved by the magnetic field layer at $x \approx -1~(-250~{\rm kpc})$ in the left panel are wiped out by conduction in the right panel.\label{fig:asai2007_prof}}
\end{center}
\end{figure}

The previously described Asai-MHD simulations were the first to investigate the effects of anisotropic thermal conduction and magnetic fields on a remnant-core cold front. These simulations demonstrated that sufficient magnetic draping would occur around the developing cold front to suppress thermal conduction strongly enough that a sharp cold front edge could be maintained for timescales much longer than that required to smear out the cold front interface by uninhibited conduction, regardless of whether or not the initial magnetic field was uniform or turbulent. Figure \ref{fig:asai2007_prof} shows profiles of gas density, temperature, and pressure across the cold front for models with anisotropic and isotropic thermal conduction, where in the former case the cold front is shielded from conduction by the magnetic field, but in the unmagnetized latter case conduction has completely smoothed out the front surface.

\begin{figure}
\begin{center}
\includegraphics[width=0.98\textwidth]{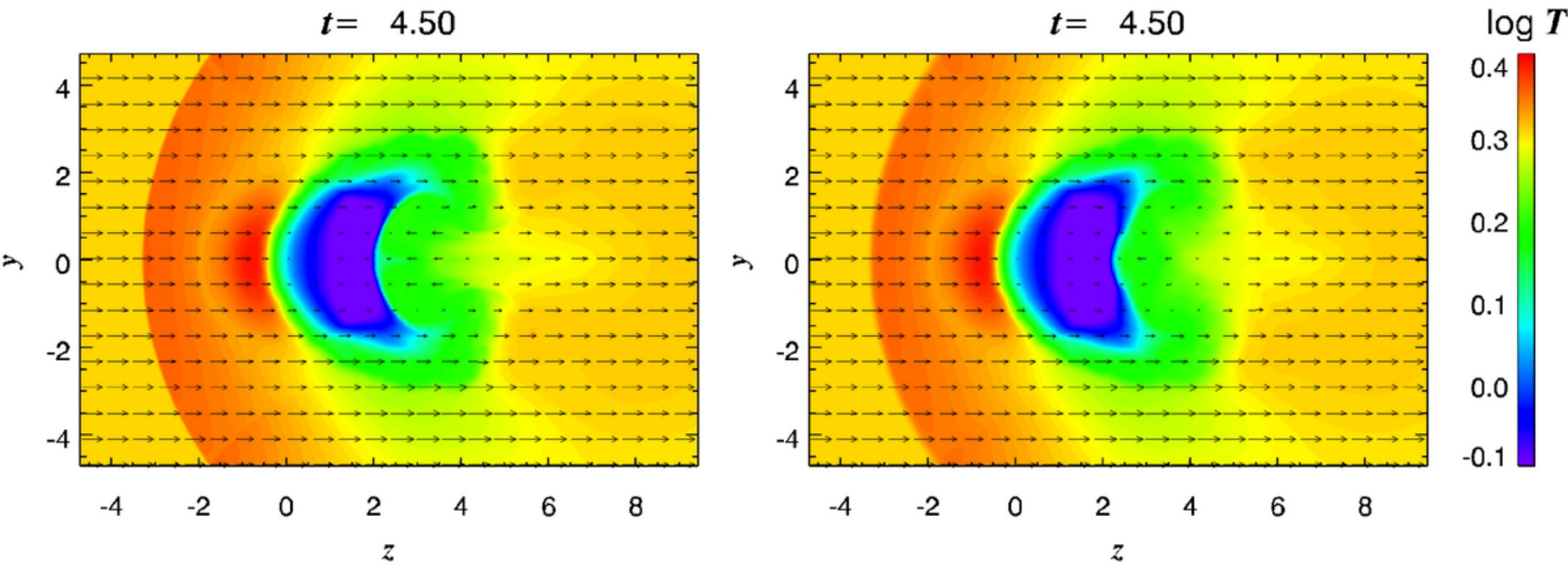}
\includegraphics[width=0.98\textwidth]{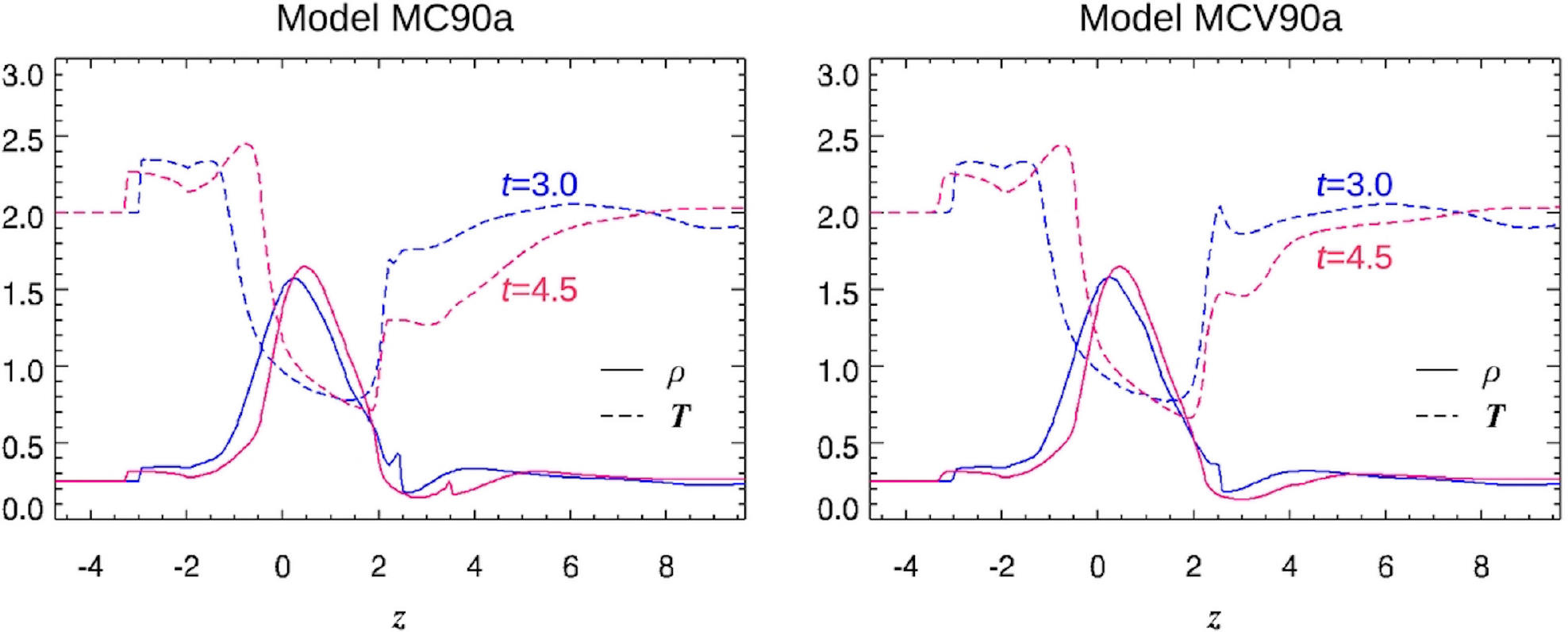}
\caption{The effects of anisotropic thermal conduction and viscosity on remnant-core cold fronts, from S13. Top panels: Slices through the gas temperature in the $y-z$ plane from the inviscid (left) and Braginskii viscosity (right) simulations. In these units, 1 = 250~kpc. Bottom panels: Profiles of the gas density and temperature across the cold front interface, from the inviscid (left) and Braginskii viscosity (right) simulations. The density and temperature jump across the cold front interface at $z \approx -0.5~(-125~{\rm kpc})$ appears to be smeared out by conduction, regardless of the presence or absence of Braginskii viscosity (compare with the sharp jumps in the bottom panels of Figure \ref{fig:suzuki2013_visc}).\label{fig:suzuki2013_cond}}
\end{center}
\end{figure}

S13 also claimed to find that the magnetic field layer also suppressed conduction across the cold front interface for their nearly identical setup, though they noted that some smoothing out of the temperature discontinuity occurred at early times before the magnetic draping layer was able to develop. The top panels of Figure \ref{fig:suzuki2013_cond} show slices through the gas temperature of their magnetized simulations with anisotropic thermal conduction (with and without Braginskii viscosity). The interface does not appear quite as sharp as in their non-conductive simulations (see Figure \ref{fig:suzuki2013_visc}). The bottom panels of Figure \ref{fig:suzuki2013_cond} highlight this fact in more detail, showing the profiles of density and temperature across the cold front surface at $z \approx -0.5$.

\citet[][hereafter Z13]{zuh13} were the first to carry out simulations of the formation of sloshing cold fronts in the ICM with magnetic fields and anisotropic thermal conduction. They used the same set of cluster initial conditions as ZML11, performing simulations with different thermal conductivities and magnetic field geometries, with the initial magnetic field strength set by enforcing that $\beta = 400$ everywhere. The two most important cases were a simulation with full Spitzer conduction along the field lines, and an otherwise identical simulation with 1/10th Spitzer conduction along the field lines, modeling a situation where thermal conduction is suppressed {\it along} the field line, by strong curvature of the field lines, stochastic magnetic mirrors, or microscale plasma effects \citep[][also see Section \ref{sec:plasma_instabilities} and references therein]{cha99,mal01,nar01,sch08}.

\begin{figure}
\begin{center}
\includegraphics[width=0.97\textwidth]{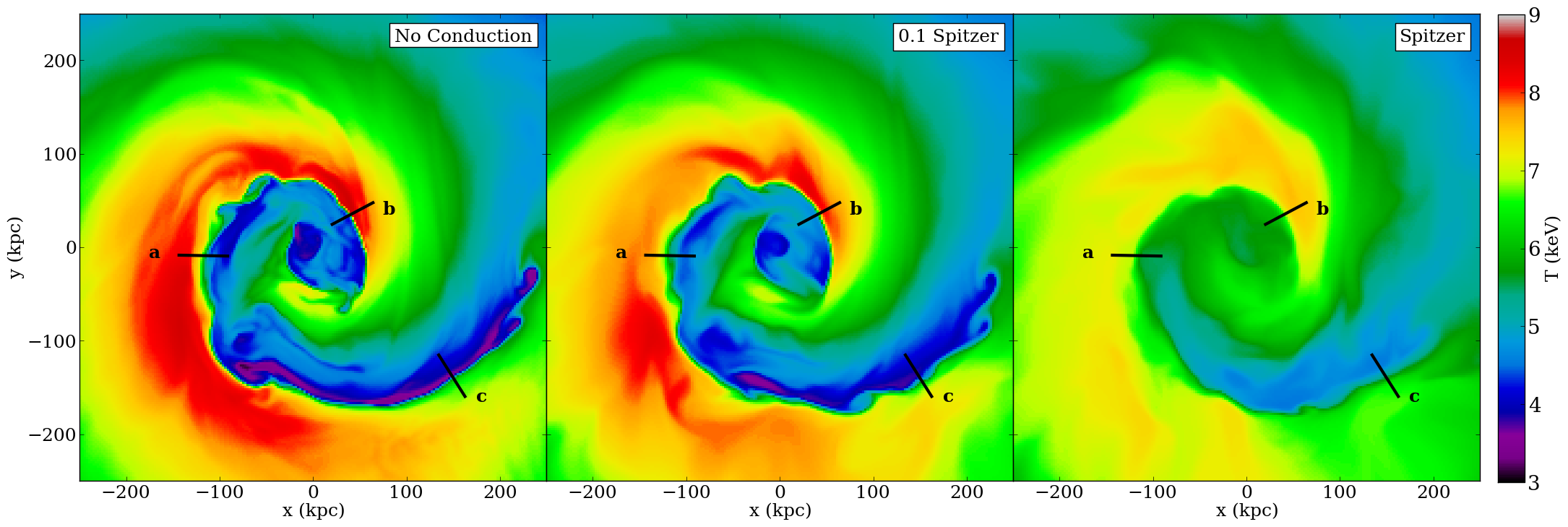}
\includegraphics[width=0.31\textwidth]{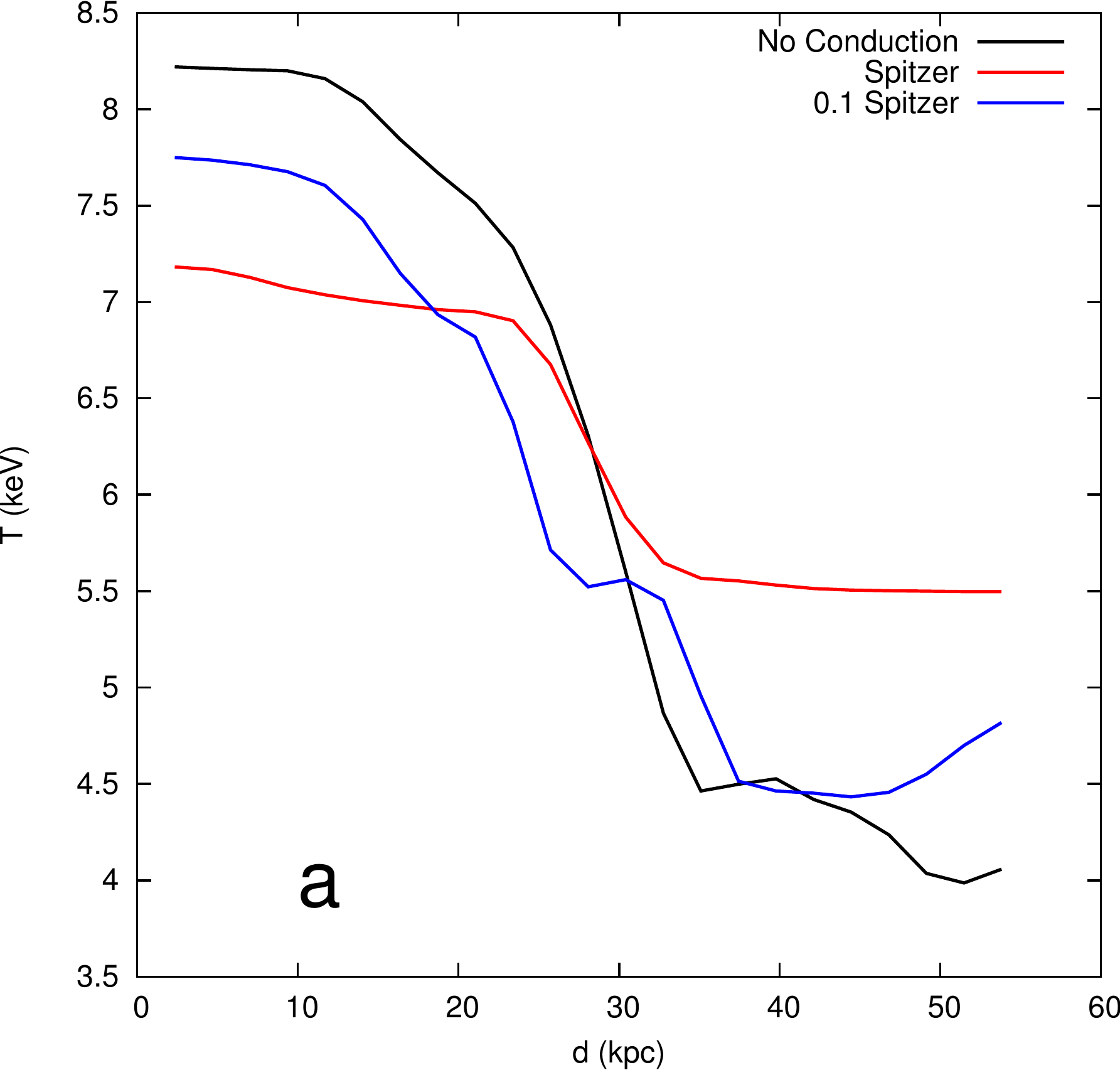}
\includegraphics[width=0.31\textwidth]{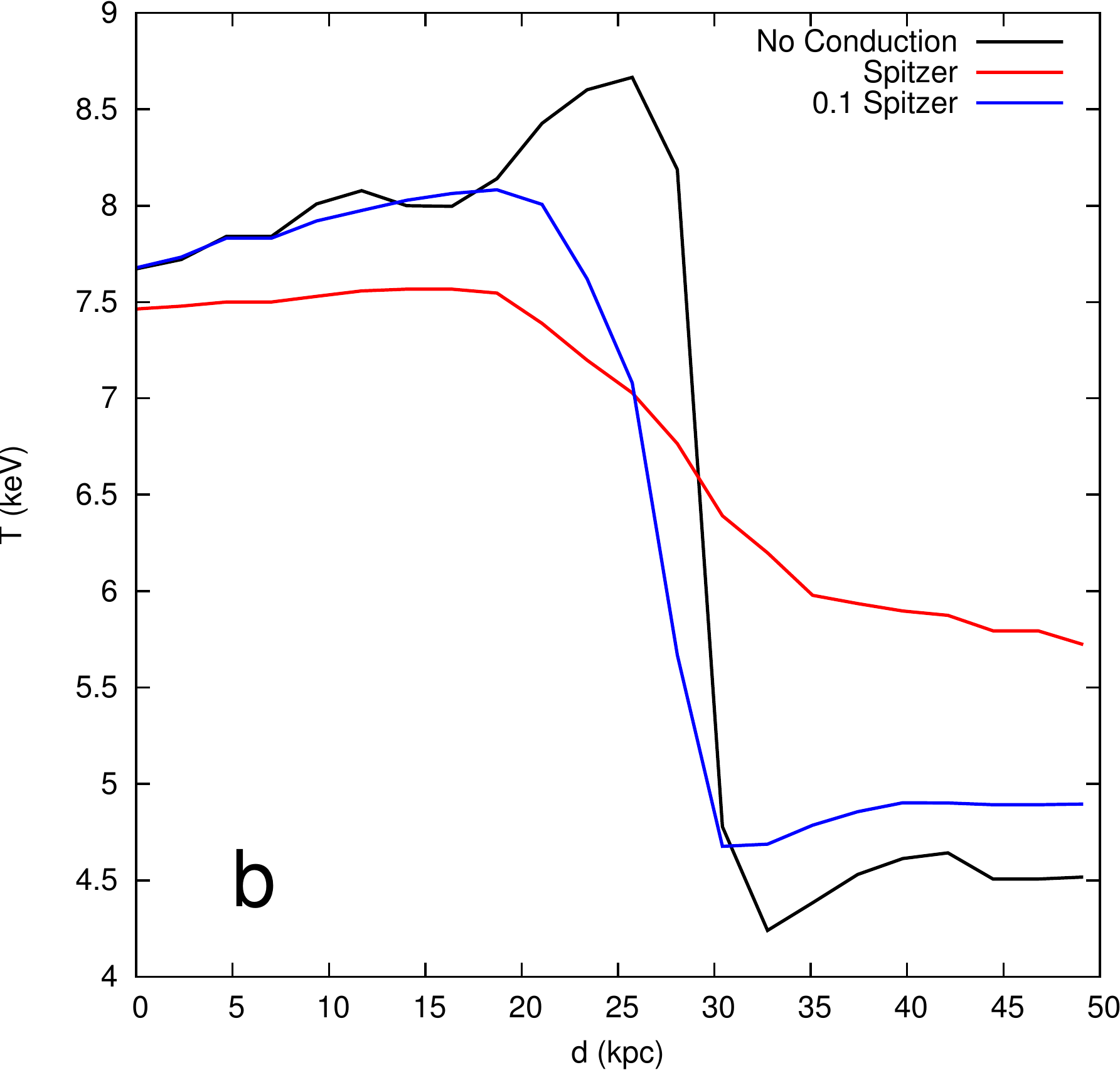}
\includegraphics[width=0.31\textwidth]{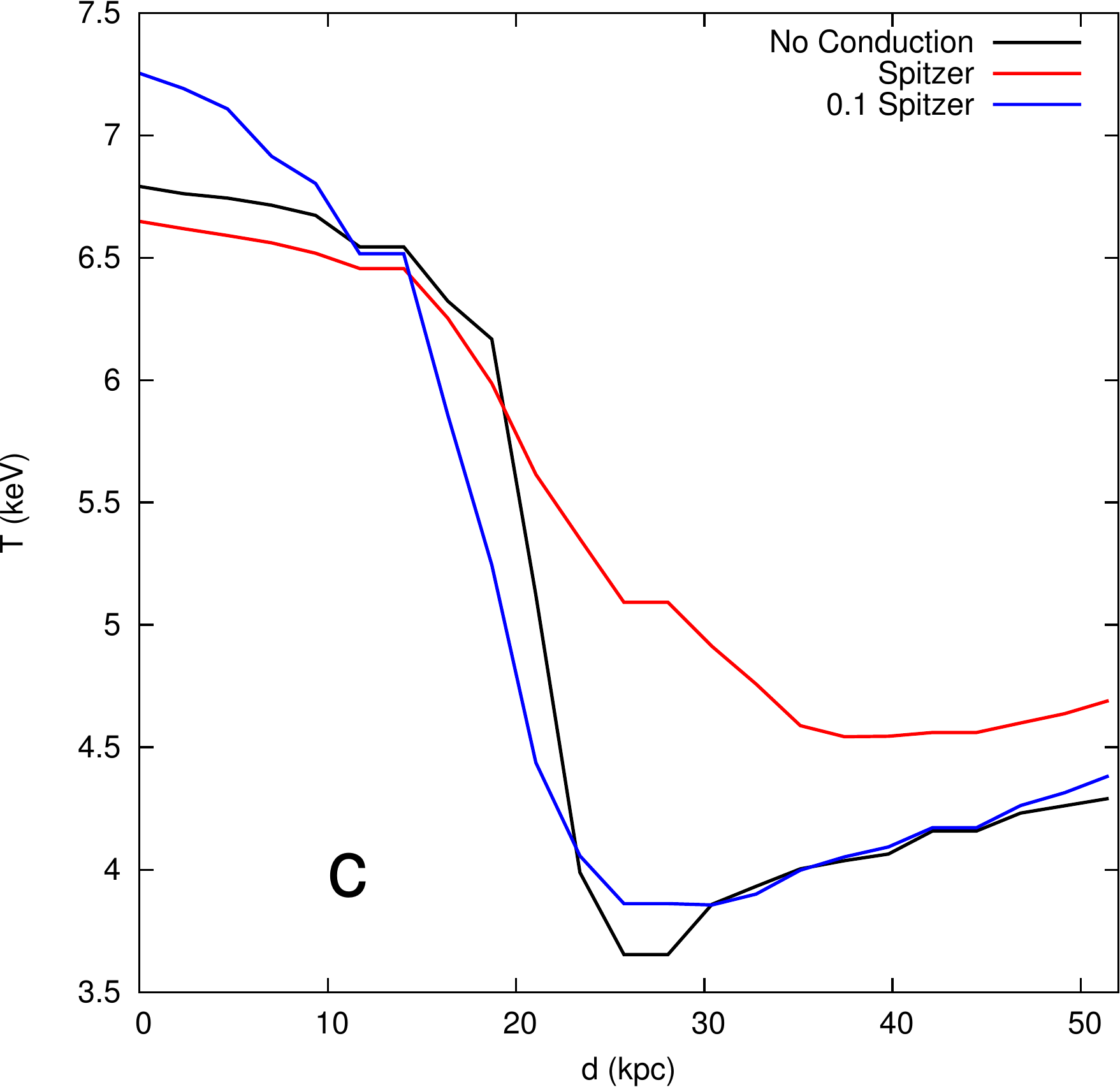}
\caption{The effect of anisotropic thermal conduction on sloshing cold fronts (reproduced from Z13). Top panels: Slices of gas temperature through the center of the domain in the $x-y$ plane from the simulations of Z13 at the epoch $t$ = 3.25~Gyr. Each panel is 500~kpc on a side. Black lines indicate the locations of profiles taken across the cold fronts. Bottom panels: Profiles of the gas temperature taken across the lines shown in the top panels.\label{fig:conduction_slices}}
\end{center}
\end{figure}

\begin{figure}
\begin{center}
\includegraphics[width=0.97\textwidth]{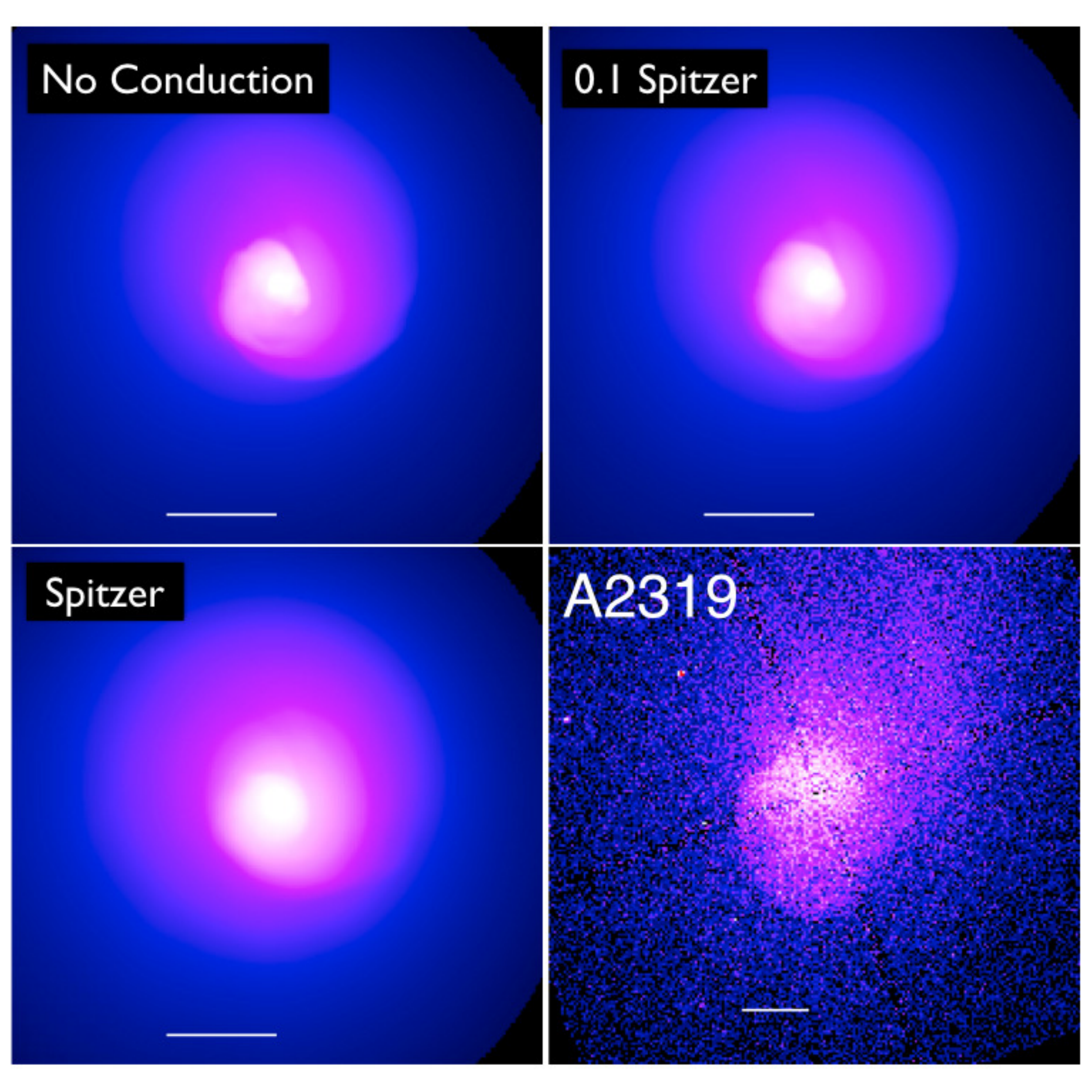}
\caption{Projected X-ray emission along the $z$-axis of the simulation domain for the simulations from Z13 with a {\it Chandra} X-ray image of A2319 included for comparison. White bars indicate 100~kpc distances. Conduction smooths out the surface brightness jumps, making them barely discernable in comparison to the sharp jumps in emission seen in A2319.\label{fig:xray_cond}}
\end{center}
\end{figure}

These simulations revealed an unexpected result: despite the presence of magnetic field layers, thermal conductivity was able to smear out the front surfaces to a certain degree. Figure \ref{fig:conduction_slices} shows slices of temperature through the midplane of the sloshing cold fronts (top panels), along with profiles taken across the cold fronts in particular places (shown in the bottom panels, marked by black lines in the top panels). In the case of full anisotropic Spitzer conduction (top-right panel), the temperature jumps have been smeared out and the interior of the cold fronts have considerably higher temperature than the case of no conductivity. In the simulation with 1/10th Spitzer conduction (top-middle panel), the temperature jumps are essentially identical to those of the case without conduction. Figure \ref{fig:xray_cond} shows that thermal conductivity would have a significant effect on the appearance of cold fronts as seen in X-rays--full Spitzer conduction, even when restricted to the field lines, smears out the interface to such a degree that the cold fronts are essentially eliminated, as shown in Figure \ref{fig:xray_cond}. For 1/10 Spitzer conduction, the cold fronts appear roughly the same as in the case without conduction.

\begin{figure}
\begin{center}
\includegraphics[width=0.97\textwidth]{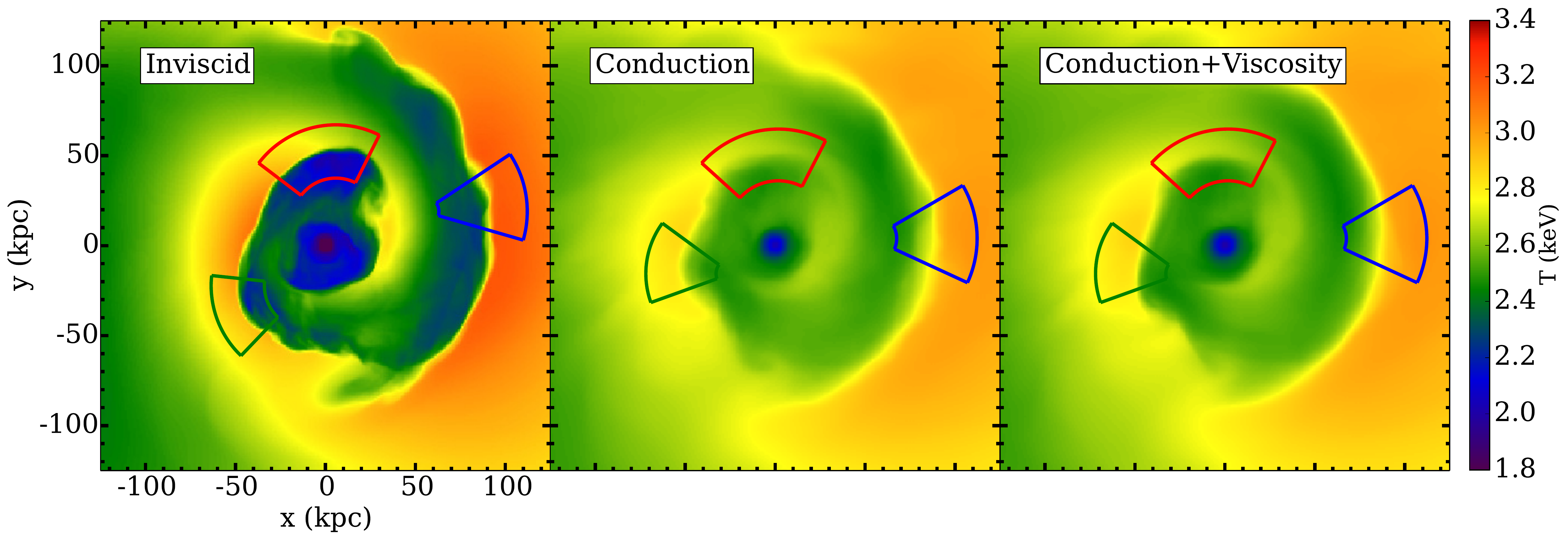}
\includegraphics[width=0.97\textwidth]{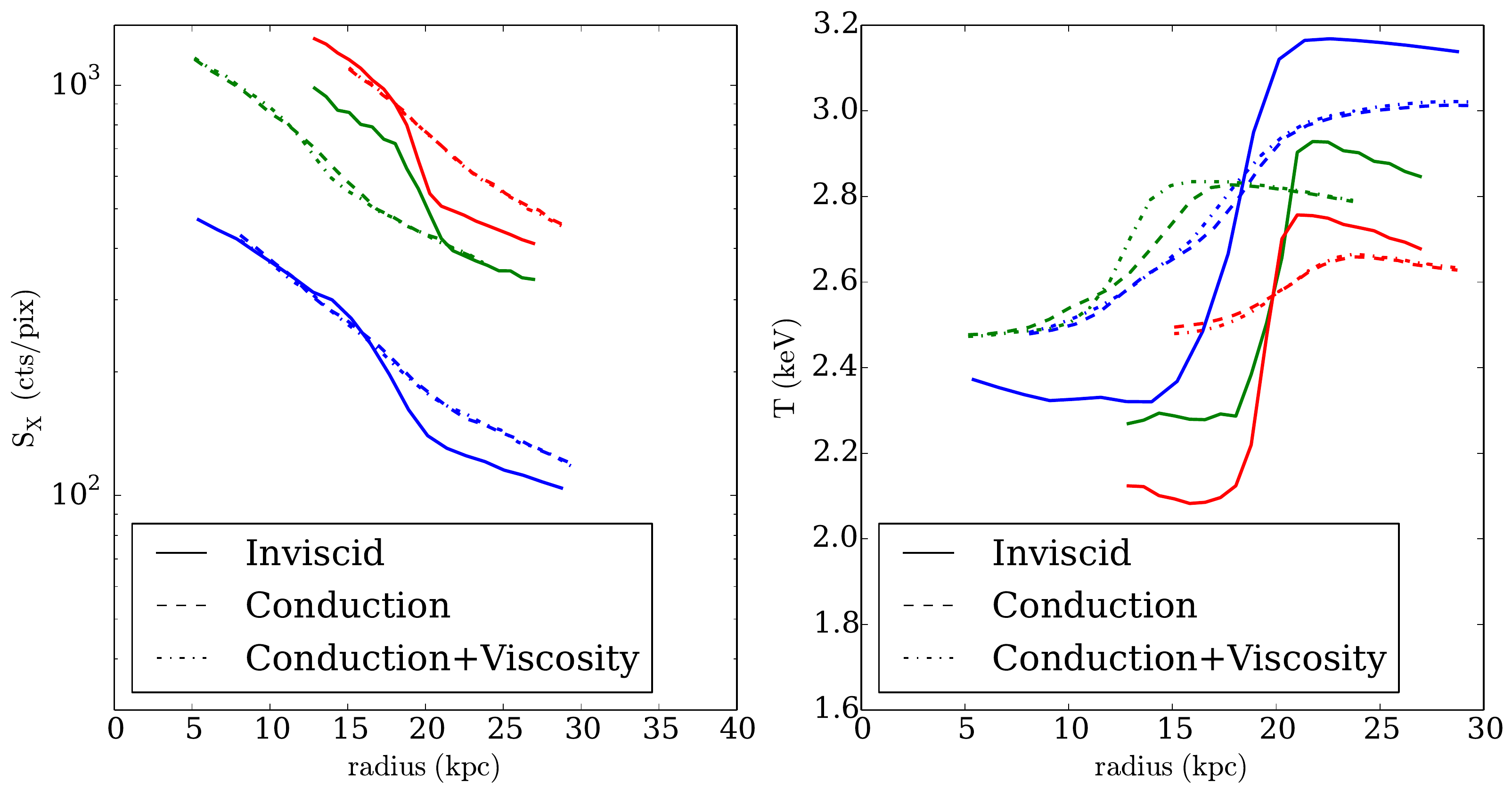}
\caption{The effect of anisotropic thermal conduction on a simulated cluster similar to Virgo, from \citet{zuh15a}. Top panels: projected spectroscopic temperature for three different models. Colored regions show the locations of profiles in the bottom panel. Bottom panel: Profiles of surface brightness and projected temperature across the cold front surfaces at the locations given by the regions in the top panel.\label{fig:virgo_cond}}
\end{center}
\end{figure}

These results are in conflict with the initial expectation that a magnetic layer completely protects the cold front from heat conduction. Not surprisingly, the devil is in the details--the full simulations reveal two reasons why the simple picture, drawn from analytic considerations, breaks down. First, although the magnetic field is stretched tangential to the front surface by the shear flow, KHI may develop rapidly along parts of the front, tangling the field lines at the interface once again, allowing some heat to diffuse between the hot and cold phases (c.f. Figure 8 of Z13). This effect is also seen in the plane-parallel simulations of \citet{lec12}, which showed that perturbations in the magnetic field direction across an interface similar to a cold front can cause a similar smearing of the interface. Only in a scenario where the amplified magnetic field were strong enough to prevent KHI completely would this effect be absent. Secondly, the three-dimensional nature of the problem and the geometry of the cold fronts leave the cold gas underneath the front magnetically connected to hot gas above the front and perpendicular to the sloshing plane, providing yet another avenue for the transfer of thermal energy (c.f. Figures 9 and 10 of Z13). We will discuss the discrepancies between the different simulations of cold fronts with anisotropic thermal conduction in Section \ref{sec:cond_supp}.

In the previously described work by Z15, simulations with anisotropic thermal conduction were also included, with and without Braginskii viscosity. Regardless of whether or not the ICM is viscous, the results from these simulations are the same as in Z13: anisotropic thermal conduction smears out cold fronts to such a degree that they would be unobservable, even by {\it Chandra} (see Figure \ref{fig:virgo_cond}). This implies that these constraints on thermal conduction may apply generally to clusters, as the cluster model resembling the Virgo cluster from \citet{zuh15a} has a much colder average temperature than the more massive cluster model from Z13, and hence much weaker thermal conductivity, due to the strong dependence of the latter on the temperature ($\propto T^{5/2}$).

\section{Discussion}\label{sec:disc}

\subsection{Open Questions}\label{sec:open_qs}

\subsubsection{The Role of Plasma Instabilities}\label{sec:plasma_instabilities}

The Braginskii-MHD equations outlined in Section \ref{sec:physics} do not have an explicit dependence on the Larmor radius; it is simply assumed to be much smaller than the mean free path. However, a number of recent works have shown that the effects of a finite-Larmor radius may be relevant on larger scales in the ICM. In particular, when the pressure anisotropy violates the approximate inequalities
\begin{equation}\label{eqn:firehosemirror}
-\frac{B^2}{4\pi} \simlt p_\perp - p_\parallel \simlt \frac{B^2}{8\pi} ,
\end{equation}
a situation which is expected to occur readily in the cluster plasma \citep{sch05,lyu07,kun11}, rapidly growing Larmor-scale instabilities (namely, the firehose instability on the left side of the equation and the mirror instability on the right side) are triggered and act to regulate the pressure anisotropy back to within its stability boundaries. The effect of these microscale instabilities on the pressure anisotropy and, conversely, the transport of momentum, is an open question.

Collisionless kinetic simulations of the driven firehose instability \citep[e.g.,][]{mat06,ht08,kun14} have shown that the collisionality of the plasma, supplemented by the anomalous scattering of particles off the microscale fluctuations, adjusts to maintain a marginally firehose-stable pressure anisotropy. This effectively reduces the parallel viscosity of the plasma, making it possible for KHI to develop more easily at surfaces like cold fronts. In addition, with the pressure anisotropy microphysically pinned at the firehose stability threshold (the left inequality of Equation \ref{eqn:firehosemirror}), the resulting viscous stress would effectively cancel the magnetic tension, again easing development of KHI. In the case of the mirror instability \citep[see][]{kun14,riq14}, the pressure anisotropy appears to be regulated not just by anomalous particle scattering but also by an increasing population of resonant particles becoming trapped in magnetic mirrors where the pressure is naturally less anisotropic.

Consideration of these plasma effects requires choices to be made when employing the Braginskii-MHD equations in simulations of cold fronts. In such simulations, when the pressure anisotropy exceeds the bounds of Equation \ref{eqn:firehosemirror}, these instabilities will be triggered, but since the Larmor radius is severely underresolved, they will occur on the smallest scale available, that of the cell size. The effects of the instabilities will then be damped out by the numerical diffusivity associated with the grid scale. Alternatively, as suggested by the kinetic simulations, the pressure anisotropy may be limited within the bounds suggested by Equation \ref{eqn:firehosemirror}, which amounts to an increase of the parallel Reynolds number of the ICM.

In regions near cold fronts, these effects may not be of significant concern, as the magnetic field is amplified to near-equipartition scales, and the boundaries in Equation \ref{eqn:firehosemirror} will be more difficult to transgress. However, future simulation studies of cold fronts using the Braginskii-MHD equations should investigate the potentially observable effects on cold fronts of limiting the pressure anisotropy or not.

\subsubsection{Distinguishing Between the Effects of Magnetic Fields and Viscosity}\label{sec:viscosity_or_bfield}

All of the cold front simulations we have detailed in this review have shown that magnetic fields and viscosity are both capable of suppressing KHI at cold front surfaces, provided that their effects are strong enough. The question then remains as to which effect is most responsible for this suppression in the cold fronts that lack evidence of KHI. Does one effect dominate over another, or are both effects important?

The major difficulty in answering this question are the lack of strong independent constraints on either effect apart from the lack of KHI at cold fronts itself. From the observational side, we do not currently have an independent check on the strength of the magnetic field in the layer at the cold front surface. In principle, an estimate of the field strength along the line of sight could be made using RM observations in the vicinity of the front. However, the strength of this component may or may not be comparable to the orthogonal component in the plane of the sky where the cold fronts are being observed, especially if the latter component is increased by shear amplification and/or draping effects at the front surface.

The most promising way to constrain the magnetic field strength in the layer would be to observe a deficit in the thermal pressure at a front surface, providing an essentially direct measurement of the magnetic pressure at that location, since the sum of the thermal and magnetic pressures should be continuous, provided the measurement is made on the side of the front where the magnetized layer is located. This would be easiest to check in sloshing cold fronts, which are isobaric across the front surface. Recently, \citet{rei14} used deprojected pressure profiles across a number of sloshing cold fronts from the literature and found that many of the fronts in their sample have pressure discontinuties consistent with the existence of enhanced magnetic pressure underneath the front surface. The average ratio of thermal to total pressure in these regions was $\xi \simeq 0.8$, which implies $\sim$20\% of the total pressure is in the form of magnetic fields. The average value is statistically significant, though for many of the cold fronts in their sample the significance is modest (up to a few $\sigma$). They also did not examine the characteristics of each individual front in detail to determine if the presumed magnetization in that front would be sufficient to prevent the development of KHI. However, these initial results are promising, and deserve future study and longer exposures by {\it Chandra}.

For the viscosity, if we assume its origin is tied to the ion collisions in the ICM, it will have the Spitzer form (Equation \ref{eqn:spitzer_viscosity}), which can be easily determined from the plasma properties. However, the effective viscosity within a small region will be dependent on the magnetic field direction, and as noted in the previous section there are still large theoretical uncertainties regarding the effect of microscale plasma instabilities on the collisionality of the plasma.

To break this apparent degeneracy between magnetic fields and viscosity, simulations should investigate other aspects of the cold fronts and their near vicinity and make predictions for what observable effects may arise from either aside from the suppression of KHI. For example, R13 showed in their unmagnetized simulations of the sloshing cold fronts in Virgo that in the absence of viscosity, the cold gas underneath the front surface should exhibit multiple edges in surface brightness as seen in projection. These edges are absent if the plasma is viscous. It is not yet known whether the same features would be present in a cold front with a magnetized layer. In the context of remnant-core cold fronts, the characteristics of the edges of the cold front and the stripped tail and wake of the subcluster explored in detail by R15A and R15B may be compared to otherwise identical simulations with magnetic fields to determine which effect best matches the observations.

\subsubsection{How Efficient is Thermal Conduction in the ICM?}\label{sec:cond_supp}

As detailed in Section \ref{sec:thermal_cond}, MHD simulations of cold fronts with anisotropic thermal conduction have given mixed results as to whether or not magnetic fields can effectively protect the sharp density and temperature jumps from being smeared out by an efficient thermal conductivity. In the simulations reviewed here, magnetic fields seem to largely prevent the sharp temperature and density gradients of remnant-core cold fronts from becoming smeared out due to conduction, but cannot do so in the case of sloshing cold fronts (though we again note as we did in Section \ref{sec:thermal_cond} that the cold front jumps {\it do} appear to be somewhat smeared out by conduction in the S13 simulations; see Figures \ref{fig:suzuki2013_visc} and \ref{fig:suzuki2013_cond}). Why is this the case?

Several factors may be at play. As noted above, the behavior of magnetic fields at the cold front surface is rather different in the case of remant-core cold fronts and sloshing cold fronts: in the former, the field is ``draped'' on the {\it outside} of the front, in the latter, the field is amplified by shear amplification on the {\it inside} of the front. The draping layers on the outside of the fronts tend to be more long-lasting in the simulations than the layers on the inside of the fronts, which are more transient (see Section \ref{sec:bfields}). The geometry of the cold front is also an important consideration. In the Asai-MHD and S13 simulations, the magnetic field layer appears to wrap essentially all the way around the cold front, strongly restricting conductive heat flux across the interface. As noted above, in the sloshing simulations of Z13 and Z15, tangling of field lines along the front surface by KHI and the presence of other regions of hot gas which are connected to the cold gas by magnetic fields permit a flux of heat to the cold phase. However, the KHI in these simulations were able to develop due to the initial magnetic field strength being set by $\beta = 400/1000$ (Z13/Z15); in a more strongly initially magnetized simulation ($\beta \simlt 100$), the draping layers may have been more effective.

It should also be noted that the remnant-core cold front simulations from the Asai-MHD and S13 works have poorer resolution ($\sim$7-10~kpc, in the 3D simulations) than the sloshing cold front simulations by ZuHone ($\sim$1-2~kpc). The lower-resolution simulations have higher numerical dissipation, preventing some KHI modes from growing and limiting the development of turbulence, both of which could re-tangle magnetic field lines if they were not strong enough to prevent the growth of either effect in the first place. For these reasons, the remnant-core scenario with anisotropic thermal conduction should be re-examined at higher resolution.

The question of whether or not the narrow widths of cold front interfaces are even compatible with anisotropic thermal conduction is important for the study of the thermodynamic properties of the ICM as a whole. A considerable amount of attention in recent years has been devoted to the effect of anisotropic thermal conduction on the stability properties of the ICM. In the classical picture, the ICM is convectively stable since the entropy in clusters is stratified with $dS/dr > 0$. However, a number of studies \citep[][]{bal00,bal01,par08a,par08b,qua08,bog09,par09} have demonstrated that with efficient thermal conductivity in the ICM, the entire cluster atmosphere is unstable to convection provided that $dT/dr \neq 0$, which is observed to be true for nearly all clusters, especially those with cool cores (though other works have emphasized that these instabilites are affected strongly by the inclusion of Braginskii viscosity; see \citealt{kun11,kun12}). Other works have emphasized the possible role of thermal conduction in setting the temperature profiles of clusters \citep[][]{zak03,guo08,mcc13}, partially offsetting cooling in cores \citep{ber86,bre88,rus02,kim03,voi04,par10,rus10,rus11}, and influencing the condensation of multiphase gas in cluster cores and giant elliptical galaxies \citep[][]{sha12,mcc12,wag14,voi15a,voi15b,yan15}. Therefore, determining if cold fronts demonstrate that the thermal conductivity in the ICM is very low would have a significant impact on our understanding of the thermodynamics of the cluster core. Such a suppression of thermal conductivity has been suggested by previous theoretical studies, either due to strong curvature of magnetic field lines at small scales or by microscale plasma instabilities \citep[][]{cha99,mal01,nar01,sch08}.

\subsection{Future Observations}\label{sec:future_observations}

More observations are needed to increase the number of cold front observations with high-quality, long-exposure images and spectra for simulations to compare against. To best constrain the plasma properties, observations of nearby clusters with {\it Chandra} are most useful. For instance, other portions of the largest Virgo cold front than the one observed by \citet{wer16} should also be observed with similar exposure ($\sim500$~ks). Simulations predict (R13) that the velocity shear may change along the length of the front, indicating that the magnetic field strength in the layer and/or the presence of KHI may be different at other locations along the front.

An important leap forward in the use of cold fronts for studying the physics of the cluster plasma will be provided by the next generation of X-ray observatories, which will have effective areas that far surpass what is currently available. There is no doubt that the increased effective area of {\it Athena} \citep{nan13}, with over an order of magnitude improvement in area over that of current instruments, will greatly improve the statistics of measuring the temperature on either side of cold fronts, but this improvement will be blunted by the poorer spatial resolution of {\it Athena} ($\sim$5") with respect to {\it Chandra}'s superb sub-arcsecond resolution. The mission concept {\it X-ray Surveyor} \citep{wei15} would provide both a similar increase in effective area, coupled with {\it Chandra}-like spatial resolution.

The next generation of X-ray observatories will also possess high resolution microcalorimeters which will be able to measure the properties of the velocity field of the ICM from the shifting and broadening of emission lines in the X-ray spectrum. The first such mission will be {\it Astro-H} \citep{tak14}, to be launched in early 2016. A recent study carried out by \citet{zuh15b} used synthetic observations to show that {\it Astro-H} will be able to measure the basic kinematic properties of sloshing cold fronts, namely the shift of the spectral lines produced by the bulk motion of the sloshing gas (provided that our line of sight is at least partially within the sloshing plane) and the line broadening produced by the variation in this bulk motion and sloshing-driven turbulence. Unfortunately, they also showed that due to {\it Astro-H}'s poor spatial resolution of $\sim$1', it will be unable to place any meaningful constraints on the ICM viscosity from studies of cold fronts. Meaningful constraints on the cluster plasma from kinematics will require a mission with {\it both} high spatial and spectral resolution, such as {\it Athena}, or (more likely) {\it X-ray Surveyor}, both of which will have calorimeters with spectral resolution equivalent to or higher than {\it Astro-H}.

\section{Summary}\label{sec:summary}

Cold fronts are important probes of plasma physics in galaxy clusters. We presented a review of recent simulation works, along with examples of relevant observations, which have provided the strongest constraints on the properties of the fronts themselves as well as the ICM physics implied by these features. Many insights have been gained from these simulation studies, but a number of open questions remain. To summarize:

\begin{itemize}
\item Recent simulations conclusively demonstrate that the surfaces of cold fronts should be locations of strong magnetic field layers oriented parallel to the front surface, whether on the outside of the front (as in remnant-core cold fronts) or the inside of the front (as in sloshing cold fronts). Such layers may suppress KHI, explaining the fact that KHI do not appear to be present in many observations of cold fronts. However, whether or not this is the case depends strongly on the initial magnetic field strength of the surrounding medium.
\item In the absence of strong magnetic fields, viscosity may play an important role in explaining the lack of evidence of KHI in many cold fronts. Recent simulations have demonstrated that even a small viscosity may be enough to explain the observed features of some cold fronts. On the other hand, a number of cold fronts {\it do} show evidence of KHI, in the form of distortions such as the ``horns'' seen in the cold front of the remnant-core cold front in M89, and the ``ragged'' appearance of some cold fronts as seen in NGC~7618, UGC~12491, and A496. In these systems, such features imply a low viscosity for the cluster plasma. More detailed comparisons between observations and simulations for a variety of systems are necessary to determine what range of Reynolds number is permissible in the ICM.
\item Where KHI suppression is observed in cold fronts, it is not clear which effect is likely to be more responsible--magnetic fields or viscosity? The most straightforward way to break this degeneracy is to use long-exposure {\it Chandra} observations of cold fronts to attempt to measure the thermal pressure deficit at the interface arising from the presence of a strongly magnetized layer. More quantitative comparisons of the observable signatures of cold fronts predicted from simulations with magnetic fields and viscosity are also needed.
\item Whether or not magnetic fields suppress thermal conduction across cold fronts appears to depend on the type of cold front. Simulations of remnant-core cold fronts suggest that the highly magnetized layers effectively prevent the flow of heat to the colder side of the front. In contrast, for sloshing cold fronts, simulations predict that the regions above and below the cold fronts are connected with magnetic fields along which sufficient heat conduction can occur, erasing the sharp cold fronts. Though there are geometrical differences between the two types of cold fronts which may be responsible for this difference, the remnant-core simulations of this effect performed so far have low resolution. More MHD simulations of remnant-core cold fronts with anisotropic thermal conduction are needed to determine if these differences between these two types of cold fronts persist at higher resolution or is numeric in nature.
\item The focus of future simulations should be to employ the most detailed and computationally feasible physical models available. This includes the full modeling of the Braginskii-MHD equations, with possible Larmor-radius-scale corrections as provided by future kinetic simulations of the cluster plasma. From these, more accurate hydrodynamic approximations may be derived. The role of magnetic reconnection near cold front surfaces should also be investigated.
\item In future comparisons between simulations and observations of cold fronts, careful attention needs to be paid to predicting the distinct observable signatures of cold fronts under different assumptions for the underlying physics, since only if we can recognize and characterize these signatures accurately can they be used as a probe of the ICM physics. The most accurate comparisons to the observations are facilitated by synthetic X-ray observations of cold fronts, which include the effects of Poisson noise and instrumental responses.
\item A more accurate determination of the plasma properties of the ICM from cold fronts will be provided by the next generation of X-ray telescopes with high-effective area and high spectral resolution, such as {\it Athena} and {\it X-ray Surveyor}, though the latter will be essential for examining the cold front interfaces with the same spatial resolution as {\it Chandra}.
\end{itemize}

\acknowledgements
The authors would like to thank the reviewers, Eugene Churazov and Chris Reynolds, for their reading of this manuscript and the helpful comments. JAZ acknowledges support from NASA though subcontract SV2-8203 to MIT from the Smithsonian Astrophysical Observatory, and through Chandra Award Number G04-15088X issued by the Chandra X-ray Center, which is operated by the Smithsonian Astrophysical Observatory for and on behalf of NASA under contract NAS8-03060.

\appendix

\section{Table of Simulations}\label{sec:appendix}

Table \ref{tab:sim_info} lists the major simulations of cluster cold fronts highlighted in this work, including the type of cold front simulated, the method of simulation (HD or MHD), whether or not viscosity and conductivity were included and whether they are isotropic or anisotropic, the dimensionality of the simulation, and the resolution.

\input tab2.tex

\end{document}

%% file: tab1.tex
\begin{table*}
\caption{Relevant Length Scales in the ICM$^*$\label{tab:length_scales}}
\begin{center}
\begin{tabular}{lll}
\hline
\hline
Length Scale & \hspace{1.5cm} & Value \\
\hline
Debye Length &     & 100~fpc \\
Electron Skin Depth & & 2000~fpc \\
Electron Larmor Radius & & 0.05~npc \\
Ion Skin Depth & & 0.08~npc \\
Ion Larmor Radius & & 2~npc \\
Mean Free Path & & 1~kpc \\
Cluster Core Radius & & 100~kpc \\
Cluster Virial Radius & & 1000~kpc \\
\hline
\multicolumn{3}{p{.6\textwidth}}{$^*$Keeping the astrophysical length unit kpc as the most practical reference, we express smaller lengths in nanoparsec (npc) and femtoparsec (fpc). We assume ICM conditions of $n_e \sim n_p = 0.01$~cm$^{-3}$, $kT = 5$~keV, and $B$ = 1~${\rm \mu{G}}$, except the last two scales, which are global.}
\end{tabular}
\end{center}
\end{table*}

%% file: tab2.tex
\begin{landscape}
\begin{table*}
\caption{Simulation Information\label{tab:sim_info}}
\begin{center}
\begin{tabular}{llllllll}
\hline
\hline
Reference & & CF Type$^1$ & Method & Viscosity$^2$ & Conductivity$^2$ & Dimensionality & ${\Delta{x}_{\rm min}}^3$ \\
\hline
\citet{asa04} & (Asai-MHD) & R & HD/MHD & N & I/A & 2D & 8.3 kpc \\
\citet{asa05} & (Asai-MHD) & R & HD/MHD & N & I/A & 3D & 8.9 kpc \\
\citet{asa07} & (Asai-MHD) & R & HD/MHD & N & I/A & 3D & 9.8 kpc \\
\citet{AM06} & (AM06) & S & HD (SPH) & N & N & 3D & 5 kpc \\
\citet{dur08} & & R & MHD & N & N & 3D & 0.016 R$^4$ \\
\citet{hei03} & & R & HD & N & N & 2D cyl & 2 kpc \\
\citet{rod11} & & S & HD & N & N & 3D & 0.5 kpc \\
\citet{rod12} & (R12) & S & HD & N & N & 3D & 1 kpc \\
\citet{rod13a} & (R13) & S & HD & I & N & 3D & 0.5 kpc \\
\citet{rod15a} & (R15A) & R & HD & N & N & 3D & 0.1 kpc \\
\citet{rod15b} & (R15B) & R & HD & I & N & 3D & 0.2 kpc \\
\citet{suz13} & (S13) & R & HD/MHD & I/A & A & 3D & 9.1 kpc \\
\citet{zuh10} & & S & HD & I & N & 3D & 5 kpc \\
\citet{zuh11} & (ZML11) & S & HD/MHD & N & N & 3D & 2 kpc \\
\citet{zuh13} & (Z13) & S & MHD & N & A & 3D & 2 kpc \\
\citet{zuh15a} & (Z15) & S & HD/MHD & I/A & A & 3D & 1 kpc \\
\hline
\end{tabular}
\begin{tabular}{l}
$^1$R = Remnant-core, S = Sloshing \\
$^2$N = None, I = Isotropic, A = Anisotropic \\
$^3$Minimum resolution of simulation, for the ``fiducial'' simulation in the study.\\
$^4$Simulation performed in arbitary units, $R$ is the radius of the core.\\
\end{tabular}
\end{center}
\end{table*}
\end{landscape}